\documentclass[12pt]{article}
\usepackage{graphicx} 
\usepackage{float}
\usepackage{rotating, graphicx}
\usepackage{booktabs}
\usepackage{multirow}
\usepackage{longtable}
\usepackage{array}
\usepackage{placeins}
\usepackage{caption}
\usepackage{subcaption}
\usepackage[a4paper,left=2.54cm,right=2.54cm,top=2.54cm,bottom=2.54cm]{geometry}
\usepackage[natbibapa]{apacite}
\usepackage{multibib}
\usepackage{xcolor}
\usepackage{url}
\usepackage{etoolbox}
\usepackage{appendix}
\usepackage{amsmath}
\usepackage{mathrsfs}
\usepackage{amssymb}
\usepackage{etoc}
\usepackage{pdflscape}

\begin{document}

\begin{titlepage}

    \begin{center}
        \LARGE
        \textbf{A Perfect Storm: First-Nature Geography and Economic Development$^*$} \\
        
        \vspace{0.5cm}
        \large
        Christian Vedel, \\ University of Southern Denmark,\\
        \small
        
        \vspace{0.75cm}
    
        \large
        \textbf{Abstract} \\     
    \end{center}

    \small
    \begin{changemargin}{1cm}{1cm}
    First-nature geography shapes the location of prosperity. I provide evidence by investigating the effects when it suddenly changes. In 1825 a storm breached the Agger Isthmus. This connected Denmark's west Limfjord Region to the North Sea. I demonstrate that trade followed. Prosperity relocated with it: population rose 27.0 percent within a generation — an elasticity of 1.6 relative to market access — with occupational shifts toward fishing and manufacturing. Fertility, not migration, drove the expansion. A mirror experiment, the waterway's closure circa 1086–1208, caused symmetric declines in medieval coin and building finds.

    \vspace{0.05cm} 
    \end{changemargin}

    \vfill
    
    \footnotesize
    $^*$This project was made possible through generous funding by the Independent Research Fund Denmark (DFF – 6109-00123). I want to thank Paul Sharp, Christian Møller-Dahl, Casper Worm Hansen, Max Schulze, Gregory Clark, James Fenske, Mathias Barding, Torben Johansen, Nadja van't Hoff, Kerstin Enflo, Neil Cummins, and former PhD student-colleagues at SDU and at LSE for valuable feedback. This project also benefitted from research assistance of Andreas Slot Ravnholt. Comments from participants at workshops and conferences, including DGPE (2021, 2022), EHES (2022, Groningen), SSEH (2022, Gothenburg), LSE Economic History Graduate Seminar (2022), EHS (2023), University of Copenhagen End-of-Semester workshop (2023), Annual CITP (Nottingham), the Danish Society of Transport Economics, History of Population and Social Structure (CamPop, Cambridge), and SDU PhD seminars are gratefully acknowledged. This paper also benefitted from improvements suggested by ChatGPT and Claude. Important early inspiration came from the podcast 'Kongerækken' and Poulsen (2019). Finally, I want to thank Maria Fay Courtney Bohr. On our vacation during the COVID-summer of 2020, we stumbled over some fascinating local history resulting in the present paper.  \\
    Code and replication materials: \url{https://github.com/christianvedels/A_perfect_storm}.

\end{titlepage}
\newpage

\section{Introduction} \label{intro}

First-nature geography — coastlines, waterways, the physical shape of the land — shapes where people settle and who they trade with. Whether it \textit{causes} prosperity, rather than merely correlating with it, is harder to establish cleanly. Geography is entangled with the institutions and culture that grew on top of it, making cause and coincidence hard to tell apart. A clean test requires a moment when geography changed by nothing other than nature.

On the night of February 3rd, 1825, a storm breached the Agger Isthmus in north-western Denmark — washing away a narrow strip of land and permanently connecting an isolated fjord to the North Sea. The breach was an act of weather, not a decision about where trade should flow. It provides a clean test. Difference-in-differences estimates comparing affected parishes to other Danish coastal areas show that the channel raised trade immediately and, within a generation, lifted population by 27.0 percent — a market access elasticity of 1.59.\footnote{Which in relative terms is similar to the Panama Canal \cite{rauch2022a}.} Higher fertility drove the growth and match exactly the pattern of population growth observed; migration either declined or was unchanged. The expansion was intrinsic, not reallocated from elsewhere. Fishing and manufacturing both grew. First-nature geography lifted a region's prosperity permanently, working through multiple channels simultaneously.\footnote{The analysis draws on ship-level records from the Sound Toll Registers \citep{soundtoll_data} for trade, and micro-level Danish census data covering 1,589 parishes \citep{mathiesen2022linklives} for population, fertility, and occupational structure. Occupations are standardized using a new language model for historical occupational titles \citep{dahl2024breaking}.} In this paper, I present the evidence for this claim.

The conclusion is sharpened by a reverse experiment. The best way to rule out confounders is to repeat the experiment in a society that shares the geographical setup and nothing else --- if the result still holds, the confounders that differ across the two societies cannot be the cause. As a matter of coincidence, the same channel, in the same location, closed sometime between 1086 and 1208, in a society separated from nineteenth-century Denmark by seven centuries of change in culture, religion, technology, and institutions. Archaeological coin and building finds show the same pattern in reverse.\footnote{This draws on a recent strand of literature using archaeological data to infer economic activity \citep{Barjamovic2019, Bakker2021Phonecians, Allen2023, boehm2024}.} The sign flipped with the geography. First-nature geography shaped the location of prosperity similarly across radically different societies. What the pair of experiments identifies is not a fact about Denmark in the 19th century. It is a fact about the influence of geography.

\citet{Diamond1997} argued that geomorphology shaped the fate of entire civilizations. The available causal evidence on first-nature geography mostly comes from the ancient world: \citet{Seror2020Random} on Yellow River shifts in China, \citet{Allen2023} on river shifts in Mesopotamia, \citet{Bakker2021Phonecians} on Phoenician trade routes.\footnote{\citet{Matranga2024} on the invention of agriculture belongs to the same tradition but studies climate seasonality rather than geographic access.} These settings predate the integrated markets in which agglomeration economies are strongest. \citet{Henderson2018} find that first-nature effects on economic activity are systematically \textit{weaker} in more recent economies. It remains unclear how much first-nature geography continued to actively shape the location of prosperity once initial development patterns were established — whether it was still a live force in the last two centuries, or merely a distant seed. I provide direct evidence on this question from 19th-century Denmark.

Specific first-nature features have been causally linked to development - coal deposits \citep{ORourkeCoal2021}, soil suitability \citep{HeavyPlough2016}, ocean fish productivity \citep{Dalgaard2020} - but each channel is single and well-defined. Geomorphology is different. A change in coastal access affects trade, agglomeration, structural change, and fertility simultaneously. That breadth is why its effects exceed what any single-channel mechanism predicts, and why it is harder to identify: geography rarely moves. \citet{Bosker2022} calls for natural experiments to settle the question. I provide one.

The closest parallel in the literature is \citet{Ahlfeldt2015}, who use the Berlin Wall as a natural experiment to measure the effect of \textit{second-nature} geography on prosperity. This paper provides the first-nature counterpart. The identification logic is the same in kind - an abrupt, exogenous change in access - but the shock is natural, not political, and a reverse experiment in a radically different society replaces the Wall's second demolition. The gains from market access are well established for constructed infrastructure: railroads,\footnote{\cite{Donaldson2016, Hornbeck2019, Atack2008, Atack2010, Hornung2015, Berger2017, Berger2019a, Bogart2022, gibbons2024, gorges2025tracksmodernity}} canals,\footnote{\cite{Turnbull1987, Bogart2019, rauch2022a, Feyrer2021}} and ports. Constructed infrastructure goes where political and economic incentives direct it. A storm does not.

The remainder of the paper is organized as follows. The following section provides an overview of the historical background; Section 3 describes the data; Section 4 outlines the empirical strategy. Sections 5 and 6 present the main results on trade and population, respectively. Section 7 presents results for fertility and occupational structure. Section 8 examines the effect of the 12th‑century channel closure, and Section 9 concludes.

\begin{figure}[ht]
\centering
\caption{Map of Denmark and the Event in 1825}
\includegraphics[width=0.8\textwidth]{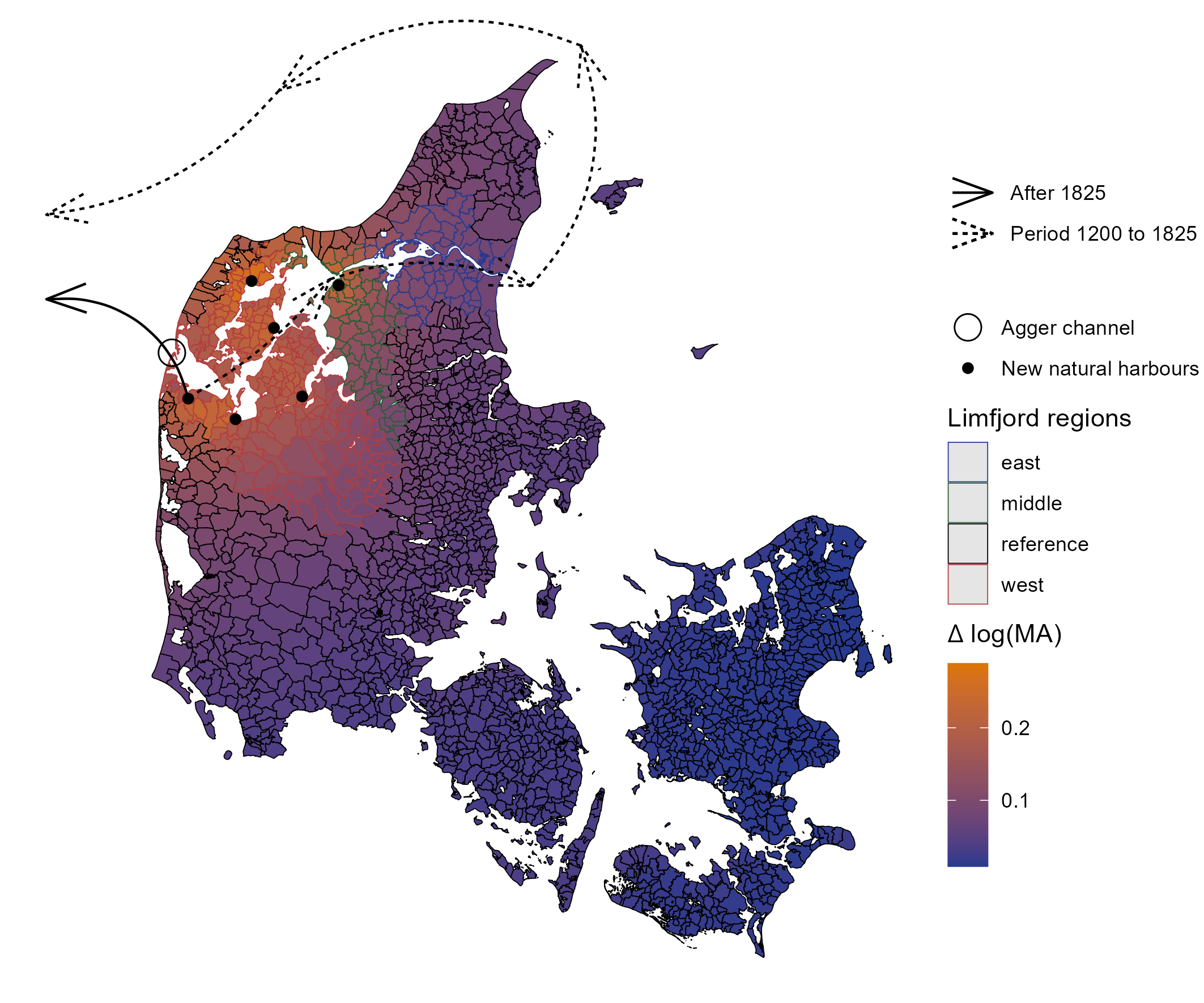}
\parbox{0.9\textwidth}{
\caption*{\footnotesize \textit{Notes:} The map illustrates the effect of the 1825 Agger Isthmus breach on shipping routes. Improved market access is shown in the fill color, while the Limfjord regions are delineated by border color. The arrows indicate shipping routes before (dashed) and after (solid) the breach. \\ \textit{Source:} Parish borders from www.digdag.dk}
}
\label{fig:main_map}
\end{figure}

\section{Historical background} \label{hist}

In the early 19th century Denmark was among the poorest countries in Europe, on the cusp of a rapid economic transformation driven in large part by expanding trade connections with Britain \citep{Lampe2015DanesUK, milkandbutter}. The western Limfjord was cut off from those connections. The 1825 breach changed that.

A fjord is a navigable inlet carved by glacial activity, and in Scandinavia fjords functioned as natural trade highways --- the Old Norse word shares its root with ``port'' \citep{EtymFjord}. Stockholm, Oslo, Bergen, and Aalborg all grew where they did because of fjord access. The Limfjord, cutting across the northern tip of Jutland, was no exception: during the Viking Age it had both eastern and western openings, making it a protected shortcut between the North Sea and the Kattegat \citep{Rasmussen1966}. The western outlet silted shut sometime between 1086 and 1208.\footnote{Established by cross-referencing \citet{saxo} [book XIII, section 5], \citet{Mortensen2018}, and geological evidence in \citet{Christensen2004}.} The Limfjord became a dead end. The closure turned the eastern fjord near Aalborg brackish --- ideal conditions for herring, an important Hanseatic commodity. Aalborg, commanding the only eastern exit, grew \citep{sildeboom2022}. The western parishes stagnated. Section~\ref{arch} documents what this closure left in the archaeological record.

Seven centuries of stagnation followed. Western Limfjord market towns could reach external markets only by sailing east through the shallow passage at Løgstør and then onward through Aalborg. Aalborg's merchants held effective legal control over the region's trade, and multiple 16th- and 17th-century court cases document western towns' unsuccessful attempts to break free \citep[pp.~78--89]{ThistedLokalhistorie1974}. By 1672, Aalborg was the largest Danish market town after Copenhagen \citep{Degn1989}. In 1800, Thisted --- the main western market town --- shipped 6,993 barrels of barley and 6,832 barrels of oats, with 31 and 47 percent of barley and oats respectively reaching markets via the dangerous open coast, and the remainder routed through Aalborg \citep{Aagard1802, Christensen1735}. No Danish province was as unfortunately placed for the sale of its products as Thy: surrounded by sea and fjord yet without a single natural harbor, ultimately dependent on Aalborg for any trade beyond a trickle \citep[p.~381]{Dioerup1842Thisted}.

On the night between February 3rd and 4th, 1825, a storm breached the Agger Isthmus and connected the Limfjord to the North Sea. The new channel gave all western and middle Limfjord market towns a direct outlet to open water, bypassing Aalborg entirely. The economic impact unfolded in three stages. First, altered salinity drove brackish-water fish into narrow rivers where they were easily caught, producing a windfall from 1825 to 1828 \citep{Poulsen2007}. Catches then collapsed, several species went nearly extinct, and the population endured years of starvation until fishermen adapted to North Sea conditions \citep{Poulsen2019}. Second, once the channel became navigable in 1834, trade followed immediately: ship arrivals at Thisted rose from 6 in 1834 to 62 by 1876, and exports grew 354 percent for barley and 997 percent for oats \citep[pp.~153--159]{ThistedLokalhistorie1974}. Overall Limfjord ship traffic grew from 19 vessels in 1835 to nearly 2,000 in 1855 \citep{Svalgaard1977}. Third, ships had sailed west without formal rights since 1834 — geography had already ended Aalborg's monopoly in practice — and in 1841 the King formalized this reality, granting international trading rights to all Limfjord market towns and prompting new customs offices and port facilities across the region. A 1842 county description called it the opening of ``a new and easier path for trade to the great waters'' \citep[p.~381]{Dioerup1842Thisted}.\footnote{A more detailed historical account is given in Appendix~\ref{app:hist}.}

\section{Data}

For trade data, the analysis relies on the Sound Toll Registers from Elsinore, north of Copenhagen, which recorded detailed accounts of all ships taxed from 1420 to 1857—amounting to about 1.8 million entries \citep{gobel2010oresundstolden, soundtoll_data}. These records include information on the origin and destination of each ship, with ports conveniently labeled with geographic coordinates. Based on this, the ships traveling to or from 126 ports in present-day Denmark are counted, and this in turn is used as a measure of levels of trade. The extract covers the period 1750–1855. This dataset is further supplemented by archival data on the number of ships passing the newly formed channel from 1834 \citep{Svalgaard1977}.

Census data were obtained from Link Lives \citep{mathiesen2022linklives} and include individual-level information on occupation, age, gender, and parish of residence. For the analysis, these data are aggregated into parish-level population counts for the years 1787, 1801, 1834, 1840, 1845, 1860, 1880, and 1901. I manually constructed a crosswalk between the census parish names and the historical parish borders (available in the project's public repository), retaining only the 1589 parishes that appear in all censuses (out of 1783). In addition to population counts, the census provides occupational descriptions that have been standardized into HISCO codes using \textit{OccCANINE} \citep{dahl2024breaking}.\footnote{The full micro-level HISCO-coded census data are publicly available from \citep{dk_hisco_data}} 200 random HISCO codes from the algorithm were manually and diligently checked to be 94 percent accurate.

Table \ref{tab:desc_pop} presents summary statistics for the main census data variables used in the analysis. "Population" denotes the number of people in each parish for a given census year. "HISCO Agricultural" counts individuals with HISCO codes beginning with 6, while "HISCO Manufacturing" counts those with HISCO codes starting with 7, 8, or 9. "Born in different county" records the number of residents born outside the county in which they later lived.

"Child–women ratio" measures fertility as the number of children aged 1–5 divided by women aged 15–45, calculated at the parish level. This captures net fertility—births and early childhood survival combined. The measure suits historical contexts where vital registration is incomplete. Children aged 1–5 were born after treatment began, while excluding infants avoids confounding from high infant mortality. Higher values indicate either higher birth rates, lower child mortality, or both—all signaling improved living standards in a post-Malthusian regime \citep{Jensen2022, Klemp2016}.

\begin{figure}
\centering
\caption{Variable distributions}
\label{fig:bal}
\includegraphics[width=1\textwidth]{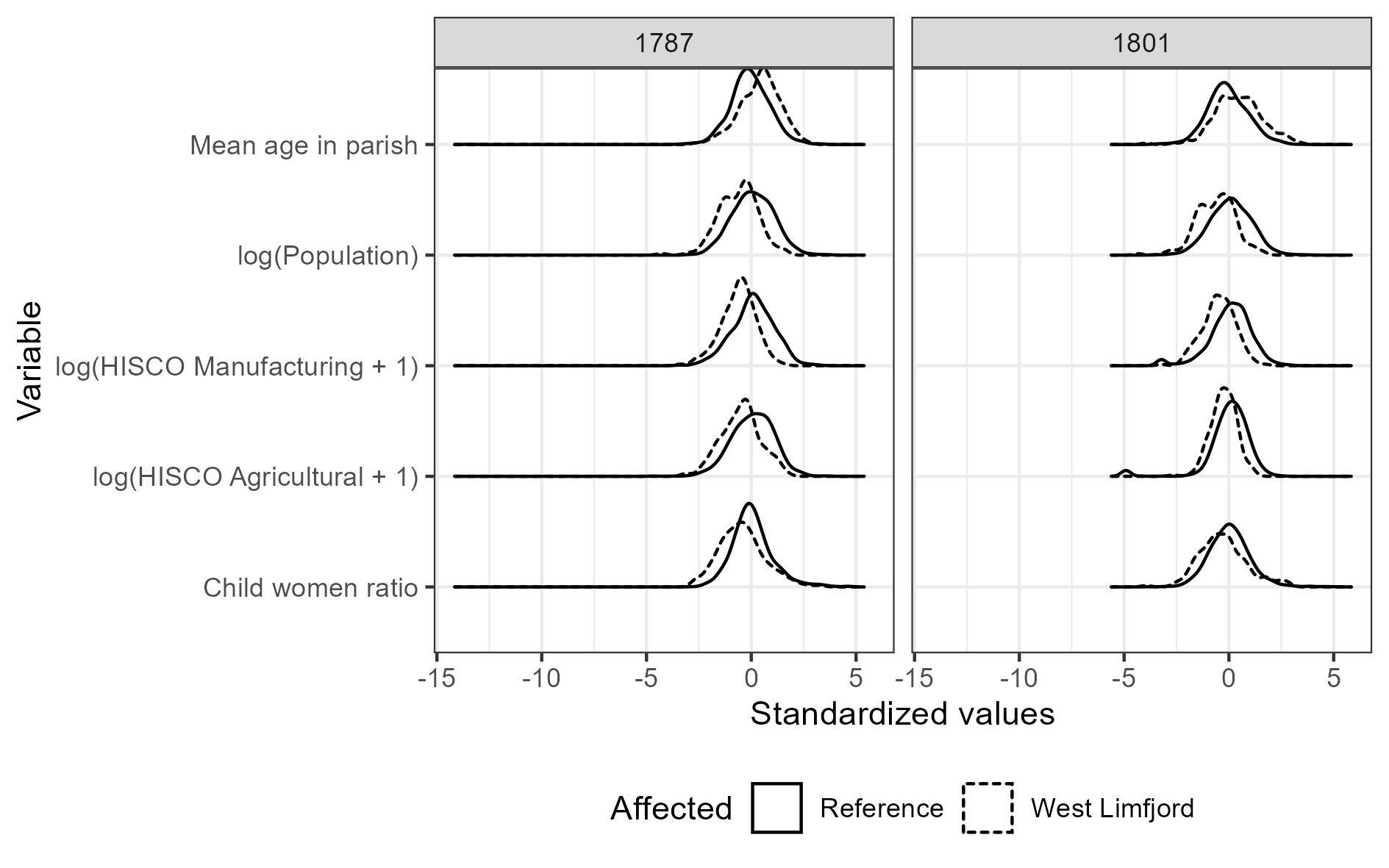}
\parbox{0.9\textwidth}{
\caption*{\footnotesize \textit{Notes:} This figure shows the distribution of variables of interest before the 1825 breach in the West Limfjord and the rest of the country (excluding other Limfjord parishes). Overall, the distributions overlap substantially, though the West Limfjord generally exhibits lower population density and fertility. \\ \textit{Source: Danish census data}}
}
\end{figure}

Figure \ref{fig:bal} displays the standardized distributions of these variables before the Agger channel breach (with the exception of "Born in different county," which is only available from 1845). Note that while there is substantial overlap between the distributions before the event, pre-event differences are also evident; these differences are addressed in Section 7 and further in Appendix \ref{dr_estimates}.

Archaeological data were obtained from a public registry of Danish archaeological findings.\footnote{"Fund og fortidsminder". See https://www.kulturarv.dk/fundogfortidsminder/.} This database contains geo-referenced records of sites that are categorized by type and dated to specific intervals. Each site's coordinates were matched to a parish using the parish borders described above, and the dating information was used to construct a panel of economic activity. Details of this are outlined in the relevant Section \ref{arch}.\looseness=-1

\FloatBarrier

\begin{table}
\centering
\caption{Summary statistics for parish level census data}
\label{tab:desc_pop} 
\footnotesize
\begin{tabular}{lcccccc}
\toprule
  & Observations & Mean & SD & Min & Median & Max\\
\midrule
Population & 14301 & 652.02 & 505.88 & 27.00 & 526.00 & 13087.00\\
Affected: West Limfjord & 14301 & 0.11 & 0.32 & 0.00 & 0.00 & 1.00\\
Affected: $\Delta log(MA_i)$ & 14301 & 0.07 & 0.06 & 0.01 & 0.05 & 0.27\\
HISCO Agricultural & 14301 & 145.67 & 109.22 & 0.00 & 118.00 & 1565.00\\
HISCO Manufacturing & 14301 & 69.83 & 79.21 & 0.00 & 49.00 & 2881.00\\
Born in different county & 7945 & 253.13 & 378.91 & 0.00 & 72.00 & 4299.00\\
Child-women ratio & 14274 & 0.47 & 0.11 & 0.00 & 0.46 & 1.36\\
\bottomrule
\end{tabular}

\parbox{0.9\textwidth}{
\caption*{\footnotesize \textit{Notes:} This table contains summary statistics for the variables used, which are ultimately sourced from the census data. For most variables, there are 14,301 observations corresponding to 1,589 parishes and 9 census years. The child–women ratio is defined as the ratio of children aged 1–5 to women aged 15–45; 27 observations are missing from this variable since no women in the relevant age group are observed. 'Born in different county' is only observed from 1845 onward. \\ \textit{Source: Danish census data}}
}
\end{table}

\FloatBarrier

\section{Empirical Strategy} \label{emp_strat}

The 1825 storm breached the Agger Isthmus because of meteorological
conditions, not because of economic conditions in the affected parishes.
There is no plausible story under which parishes about to grow faster
were more likely to be connected to the North Sea by a winter storm.
Selection into treatment is not a concern.

The estimation follows a standard TWFE specification: 
\begin{equation}
\label{eq:eq501}
\log(y_{it})= \gamma_t + \gamma_i + \textit{Affected}_{i}\,\beta_{t} + \varepsilon_{it},
\end{equation}

\noindent
where $y_{it}$ is the outcome for parish $i$ in year $t$, $\gamma_i$
and $\gamma_t$ are parish and year fixed effects, and
$\textit{Affected}_i$ measures the degree to which parish $i$ was
reached by the new channel. The coefficient $\beta_t$ traces the
year-by-year effect; the last pre-breach year is the reference period.
All treated parishes receive the same shock at the same time, so
treatment is not staggered in a way that creates
the heterogeneous-effects problems \citep{roth2023trendingDiD}.

The identifying assumption is parallel trends: absent the breach,
outcomes in West Limfjord parishes would have followed the same
trajectory as the rest of Denmark. Section~\ref{hist} establishes why
this is plausible --- the western Limfjord was structurally isolated
from Denmark's early-nineteenth-century take-off, not differentially
positioned to benefit from it. I test the assumption directly:
pre-breach coefficients for 1787 and 1801 are near zero and
statistically insignificant.

$\textit{Affected}$ is measured two ways. The first is a dummy:
parishes are classified as West Limfjord if they lie closer to the
Limfjord than to any other coast and fall west of the
northwest--southeast line defined by the coordinates [57.044185,
9.186837] and [56.958951, 9.275585], with a 20~km buffer designating
the Middle Limfjord.\footnote{The coordinate pair defines the boundary
line between West and Middle Limfjord.} West Limfjord parishes are the
treated group; Middle and East Limfjord parishes also enter the specification as controls.

The second measure is the change in market access implied by the
breach. For each parish, I compute the cost-weighted sum of reachable
ports before and after 1825 and take the log difference. Parishes that
could suddenly reach the North Sea cheaply by sea gained the most;
those already connected to other coasts changed little
(Figure~\ref{fig:main_map}). Formally, market access for each parish
is:
\begin{equation}
\label{eq:ma_eq}
\textit{MA}_p = \sum_{h \in \mathscr{H}} \left[\textit{CostDist}(p, h;\, \alpha) + 1\right]^\theta,
\end{equation}

\noindent
where $\mathscr{H}$ is the set of available ports,
$\textit{CostDist}(p, h;\, \alpha)$ is the cost of traveling from
parish $p$ to port $h$ --- with overland travel weighted at $\alpha =
10$ times the cost of sea travel --- and $\theta = -1$ is the distance
elasticity \citep{Harris1954, rauch2022a}.\footnote{The $\alpha = 10$ 
ratio follows \citet{Marczinek2022} and \citet{Bakker2021Phonecians}, 
who apply the same parameter. See also the robustness checks across $\alpha \in [1, 50]$ and $\theta \in [-16, -1]$ in Appendix~\ref{pop_multiverse}.} Treatment is
$\Delta\log(\textit{MA}_p) = \log(\textit{MA}_{p,\text{after}}) -
\log(\textit{MA}_{p,\text{before}})$, where $\mathscr{H}$ expands
after the breach to include the newly accessible ports. 
Minimum-cost paths are computed using Dijkstra's algorithm over a 
combined land-sea cost grid; Appendix~\ref{details_ma} gives full 
details.

\FloatBarrier
\section{The Effect on Trade} \label{trade}

\begin{figure}
\begin{center}
  \caption{Number of ships - sum of inbound/outbound} \label{fig:Sound_toll}
  \includegraphics[width=1\textwidth]{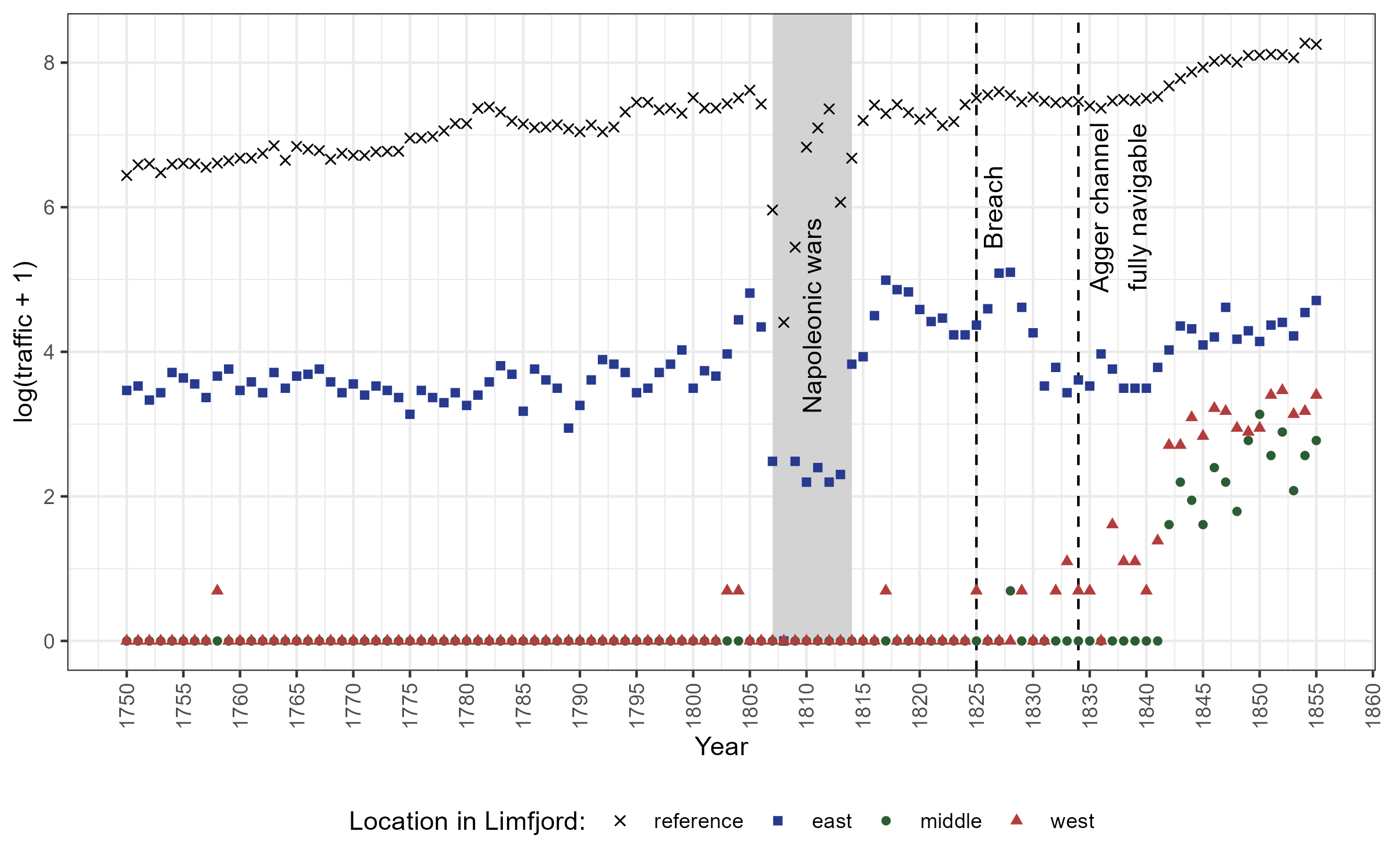}
  \parbox{1\textwidth}{
  \caption*{\footnotesize \textit{Notes:} This shows the log-transformed sum of traffic to and from ports in Denmark as captured by those ships that passed Elsinore. \\ \textit{Source: Sound Toll Registers.}}
}
\end{center}
\end{figure}

\begin{figure}[ht]
\begin{center}
  \caption{Number of ships passing the Agger channel} \label{fig:channel}
  \includegraphics[width=1\textwidth]{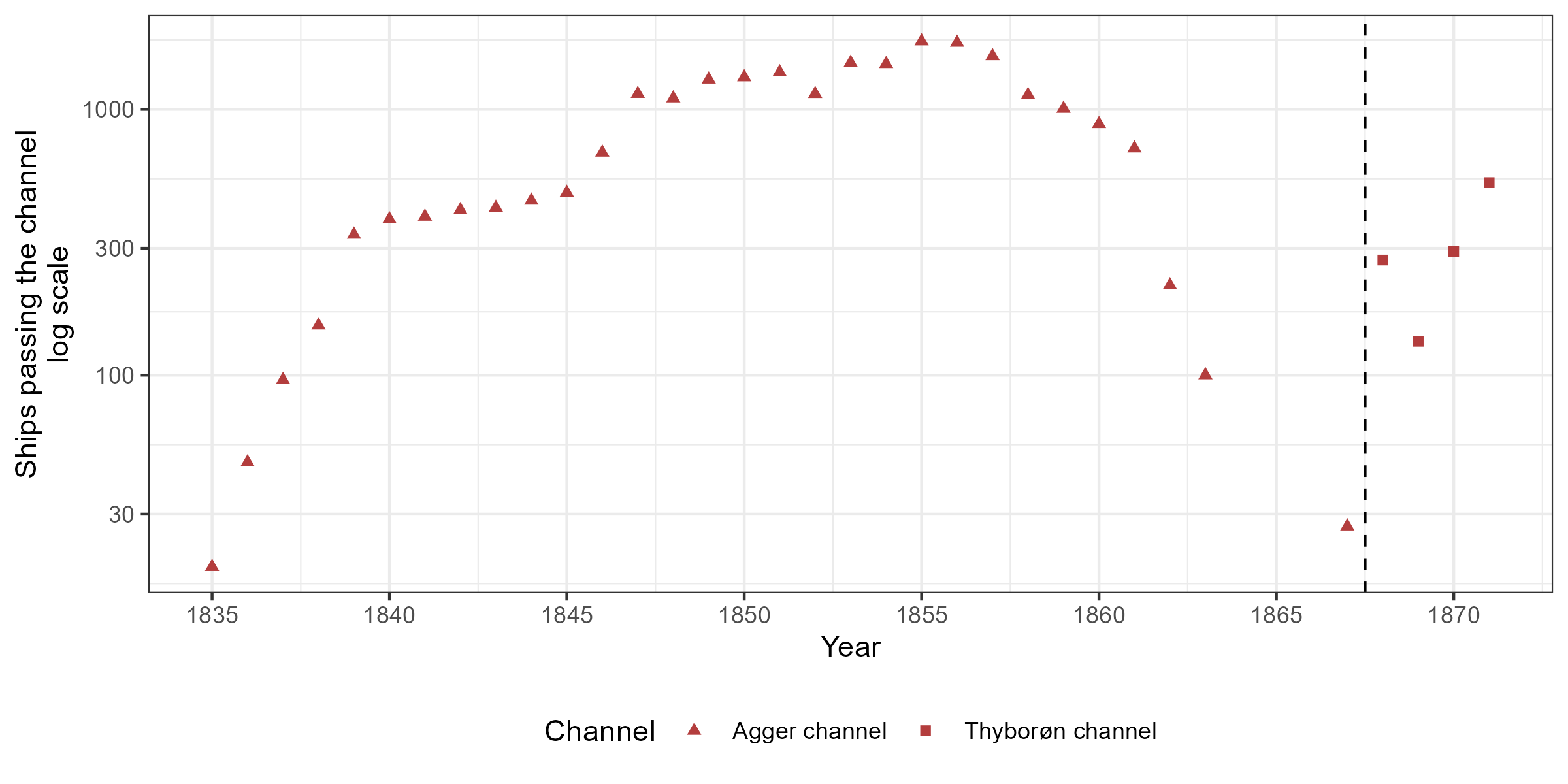}
  \parbox{1\textwidth}{
    \caption*{\footnotesize \textit{Notes:} This shows the number of ships passing the Agger channel and later Thyborøn channel. \\ \textit{Source: Svalgaard (1977). The single observation from 1867 is from Ravn (1993).}}
  }  
\end{center}
\end{figure}

Figure~\ref{fig:Sound_toll} shows log traffic at Danish ports from the
Sound Toll Registers. West Limfjord ports were near-silent before 1825;
trade surged after the channel became navigable in 1834 and stayed
elevated. The Napoleonic Wars (1807--1814) caused a temporary
disruption across all ports \citep{Feldbaek2015}, but the post-1834
divergence is sharp and persistent. Figure~\ref{fig:channel} shows the
same pattern from the other direction: ships passing the Agger channel
grew from 19 in 1835 to nearly 2,000 by 1855 --- matching the
archival record described in Section~\ref{hist}.

To formally test the effect of the channel on trade, the simple dummy
measure of being affected (defined in Section~\ref{emp_strat}) is used
in a difference-in-differences framework. In this specification, the
amount of trade at port $i$ in year $t$ is modeled as a function of
time fixed effects, port fixed effects, and an interaction between a
post-1834 indicator and the affected dummy variable. The estimation is
conducted using the Poisson Pseudo Maximum Likelihood (PPML) estimator,
which has become the standard for this type of regression which
inherently has many zeros in the outcome \citep{Silva2006, Marczinek2022,
SantosSilva2022}. To address potential distortions from the Napoleonic
Wars, results are also reported where these years are excluded. Additional 
robustness checks include estimates using alternative transformations ($\log(y+1)$ 
and the inverse hyperbolic sine) as well as an analysis of the extensive 
margin (i.e., whether any trade occurred).
\FloatBarrier

\begin{table}
\centering
\caption{Channel Introduction and Trade} \label{tab:reg_trade}
\footnotesize
\centering
\begin{tabular}{lccccc}
   \tabularnewline \midrule \midrule
   Dependent Variables: & \multicolumn{2}{c}{traffic} & log(traffic+1) & asinh(traffic) & $1[\text{traffic}>0]$\\
                         & (1)            & (2)           & (3)            & (4)           & (5)\\  
                         &  Poisson       & Poisson       & OLS            & OLS           & OLS\\  
   \midrule
   Post $\times$ east    & -0.4896$^{**}$ & -0.4817$^{*}$ & -0.0029        & -0.0233       & 0.0115\\   
                         & (0.2172)       & (0.2489)      & (0.0915)       & (0.1451)      & (0.1741)\\   
   Post $\times$ middle  & 5.557$^{***}$  & 11.76$^{***}$ & 0.9672$^{***}$ & 1.215$^{***}$ & 0.3811$^{***}$\\   
                         & (0.2151)       & (0.2471)      & (0.0584)       & (0.0700)      & (0.0233)\\   
   Post $\times$ west    & 4.129$^{***}$  & 4.698$^{***}$ & 0.3821         & 0.5062        & 0.3194$^{**}$\\   
                         & (0.5090)       & (0.4714)      & (0.2635)       & (0.3288)      & (0.1263)\\   
   1807--1814 excl.      & No             & Yes           & No             & No            & No\\  
   1825--1833 excl.      & No             & Yes           & No             & No            & No\\  
   \midrule
   \emph{Fit statistics}\\
   Observations          & 13,356         & 11,214        & 13,356         & 13,356        & 13,356\\  
   \midrule \midrule
\end{tabular}
\parbox{0.95\textwidth}{
\caption*{\footnotesize \textit{Notes:} Cluster-robust standard errors (clustered at the port level) are reported in parentheses. Signif. Codes: ***: 0.01, **: 0.05, *: 0.1. Column (1) shows Poisson estimates; Column (2) excludes years 1807–1814 and 1825–1833; Columns (3) and (4) use OLS with $\log(y+1)$ and arcsinh transformations to address zero outcomes; Column (5) analyzes the extensive margin (whether any trade occurred). \\ \textit{Source: Sound Toll Registers Online}}
}
\end{table}

Table~\ref{tab:reg_trade} reports the estimates. West Limfjord trade
increased by 4.1 log points after 1834 (column 1), implying roughly a
62-fold increase ($e^{4.1} \approx 62$); the Middle Limfjord by 5.6
log points ($e^{5.6} \approx 259$). Both are large and significant in
the Poisson specifications. Column (2) excludes two sets of disruption years: 1807--1814
(Napoleonic Wars) and 1825--1833 (the channel had breached but was not
yet navigable). The West coefficient is stable under these exclusions;
the Middle rises sharply to 11.8 log points
($e^{11.8} \approx 133{,}000$). This reflects the Middle Limfjord recording 
only one or two ships in the pre-period, so any post-breach traffic 
produces an enormous estimated effect. Both regions had near-zero trade 
before the breach, making the column~(1) log-point estimates the more meaningful
comparison. In the OLS specifications (columns 3--4),
the West Limfjord coefficient is not statistically significant,
consistent with the sensitivity of log and arcsinh transformations to
zero-heavy distributions; the Poisson estimator is preferred when
zeros dominate \citep{SantosSilva2022}. The East Limfjord shows a
significant negative coefficient in column (1) ($-0.49$, $p<0.05$),
consistent with trade being diverted away from Aalborg following the
breach --- exactly what the historical narrative of Section~\ref{hist}
predicts. The effect holds on the extensive margin (column 5):
the channel brought previously inactive ports into regular use.

\FloatBarrier
\section{The Effect on Population Density}

\begin{figure}
    \centering
    \caption{Effect of the Agger channel on population size}
    \begin{subfigure}[b]{0.8\textwidth}
        \centering
        \caption{Dummy approach}
        \includegraphics[width=0.7\textwidth]{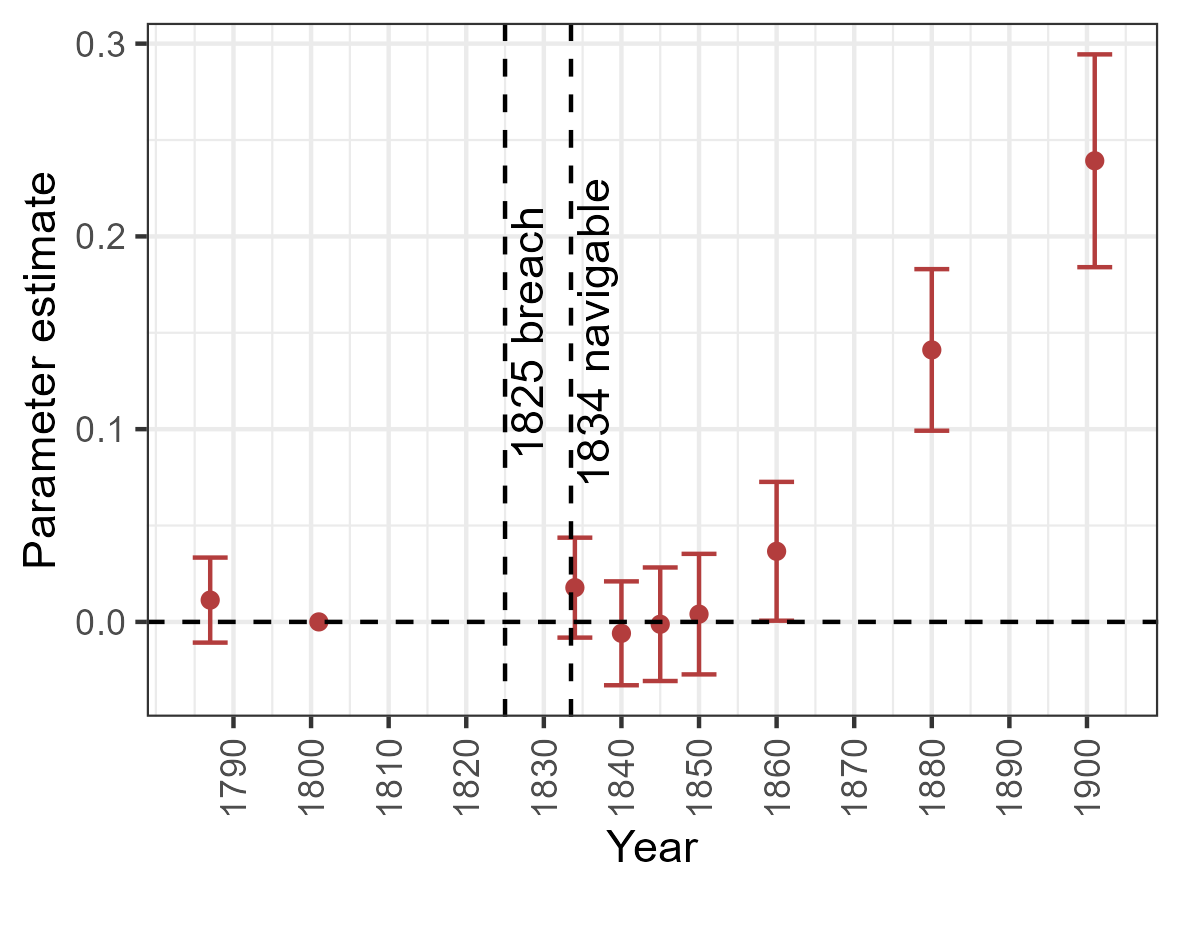}
    \end{subfigure}
    \vspace{1cm}
    \begin{subfigure}[b]{0.8\textwidth}
        \centering
        \caption{Market access approach}
        \includegraphics[width=0.7\textwidth]{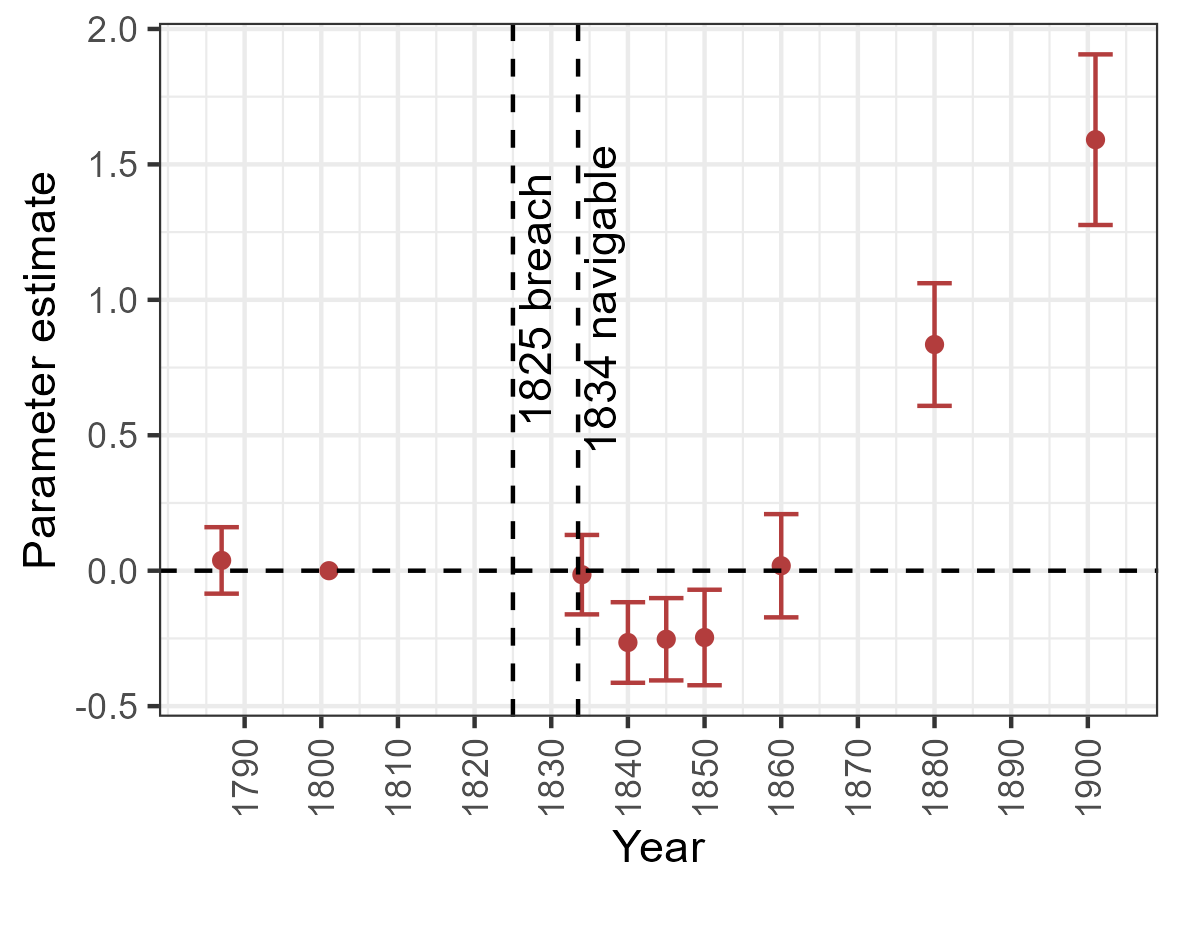}
    \end{subfigure}
    \parbox{1\textwidth}{
    \caption*{\footnotesize \textit{Notes:} Effect of the Agger channel on population size in affected areas. The error bars represent 95\% confidence intervals based on cluster-robust standard errors. Panel (a) uses a dummy measure (being in the West Limfjord), and panel (b) uses a measure based on the estimated improvement in market access. The year 1801 is the reference. All parameter estimates are reported in Appendix \ref{pop_reg_table}. \\ \textit{Source: Danish census data}}
    }
    \label{fig:pop1}
\end{figure}

By 1901, affected parishes were 27.0 percent more populous than
comparable Danish areas --- a 0.239 log-point increase (Figure
\ref{fig:pop1}, panel a). The market access specification yields an
elasticity of 1.59 (panel b) --- comparable in magnitude to the Panama
Canal \citep{rauch2022a}.

The effect was not immediate. Under the market access specification,
the coefficients for 1840, 1845, and 1850 are large, negative, and
statistically significant,
indicating a meaningful population shortfall in treated parishes
during the two decades following the breach. This is consistent with
the documented salinity shock and food shortages described in
Section~\ref{hist}: the sudden influx of salt water disrupted
freshwater fishing and reduced agricultural yields before the new
trade connections could generate lasting income gains. Population
growth becomes discernible only from 1860, and the full effect
materialises by 1901: 76 years after the breach. This mirrors the three-stage adjustment process in
Section~\ref{hist} and is also to some extent what a Malthusian model predicts --- trade
expands first, then fertility responds, then population accumulates.

Figure~\ref{fig:pop1} shows no differential pre-trends: coefficients
for 1787 and 1801 are near zero and insignificant in both panels. This
is consistent with the historical record --- the West Limfjord entered
1825 structurally isolated, not differentially positioned to grow.

Robustness is tested against alternative control groups and parameter
choices ($\alpha$, $\theta$); the multiverse in Appendix
\ref{pop_multiverse} shows no pre-trends in any specification. The
hardest identification threat is pre-existing differences in
observable characteristics between West Limfjord and control parishes.
This is addressed using the doubly-robust estimator of
\citet{Callaway2021did}, adjusting for age, occupation, and fertility
(Appendix \ref{dr_estimates}). The effect is +22.6 percent without
covariates and +15.3 percent with them --- smaller, but positive, large, and significant. 

\FloatBarrier
\section{Fertility, Migration and Occupations} \label{further_res}

\subsection{Occupations}

Occupations are coded using HISCO \citep{leeuwen2002hisco} via
OccCANINE \citep{dahl2024breaking}, where the first digit denotes the
major category. The estimation framework from
Section~\ref{emp_strat} is applied to each of these major groups. Because many
parishes have zero workers in a given occupation, four outcome
transformations are used to separate extensive from intensive margin
effects \citep{roth2023loglike}:
\begin{enumerate}
    \item Extensive: $f_1(y_{it}) = 1[y_{it}>0]$
    \item Intensive: $f_2(y_{it}) = \log(y_{it})$ (parishes with $y_{it}>0$ in any pre-treatment year)
    \item Combined (log-transform): $f_3(y_{it}) = \log(y_{it}+1)$
    \item Combined (arcsinh-transform): $f_4(y_{it}) = \operatorname{arcsinh}(y_{it})$
\end{enumerate}

\begin{figure}
\begin{center}
  \caption{Impact of the Agger Channel on Occupational Structure in 1901} \label{fig:mech_occ}
  \includegraphics[width=0.8\textwidth]{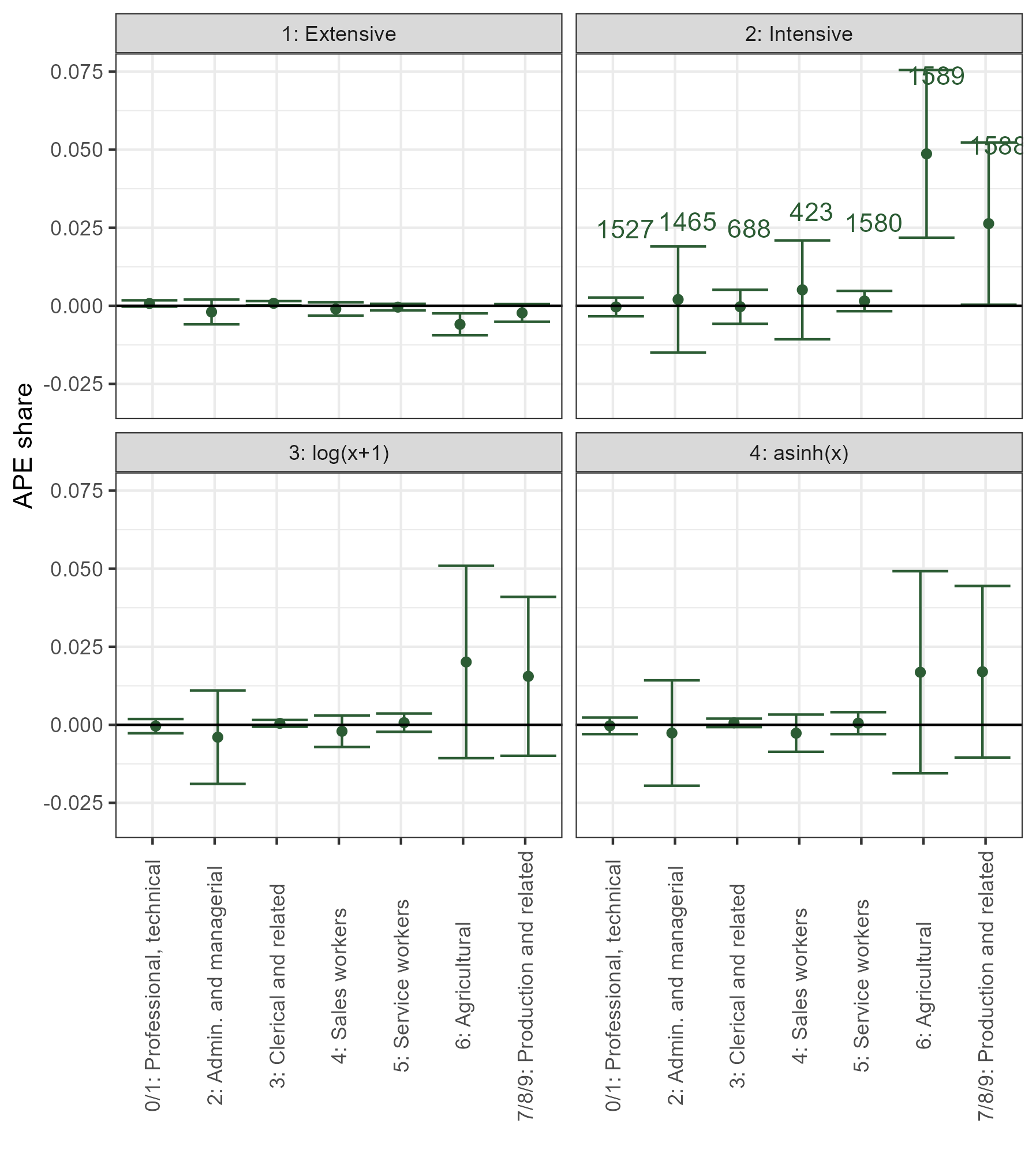}
  \parbox{1\textwidth}{
  \caption*{\footnotesize \textit{Notes:} This figure illustrates the average partial effect (as a share of parish size) on the occupational structure in 1901. The first panel shows the effects on the extensive margin. The second panel shows the effect on the intensive margin (with included number of parishes shown as well). Instead of choosing between intensive and extensive margins, the third panel uses the $log(x+1)$ transformation and the fourth panel uses the inverse hyperbolic sine transformation. The results are based on the Dummy definition of being affected by the channel. Appendix D.1 shows a table of all results including results based on market access. The error bars represent 95 percent confidence intervals corrected for multiple testing using the Bonferroni correction. This is based on standard errors, which are clustered at the parish level.}
}
\end{center}
\end{figure}

I convert raw coefficients to an \textit{Average Partial Effect
share} ($APE\,share$) to scale effects relative to parish size ---
a 50 percent rise in the number of priests from two to three is not a
structural shift. $APE\,share_j$ is the average number of individuals
gained in occupation $j$, expressed as a share of average West
Limfjord parish population in 1901:

\begin{equation}
    \label{eq:ape_share}
    APE\,share_j = \frac{\overline{Occ_{j, 1901}}}{\overline{Pop_{1901}}} \hat{\beta_j},
\end{equation}

\noindent
where $\overline{Occ_{j, 1901}}$ is the mean count in occupation $j$,
$\overline{Pop_{1901}}$ is the mean West Limfjord parish population in
1901, and $\hat{\beta_j}$ is the estimated coefficient. The formula is
applied to all four specifications ($f_1$--$f_4$); Figure~\ref{fig:mech_occ}
presents results for all four. The interpretation is cleanest under
the intensive log specification ($f_2$), where $\hat{\beta}_j$
approximates a proportional change and $\hat{\beta}_j \times
\overline{Occ_{j,1901}}$ directly gives the absolute worker gain. For
$f_3$ and $f_4$ the approximation is similar at typical count values.
For the extensive margin ($f_1$), $\hat{\beta}_j$ is a probability
effect, so the formula gives the implied gain in expected workers from
the extensive margin rather than a direct head count. Testing 504
parameters across categories, years, and specifications creates a
multiple-testing problem; results are reported for 1901 only, with
standard errors adjusted using the Bonferroni correction.

Figure~\ref{fig:mech_occ} shows the Bonferroni-corrected APE shares
for 1901. Two categories show significant effects on the intensive
margin: Agriculture (HISCO~6, coefficient 0.197, $p<0.01$) and
Manufacturing (HISCO~7/8/9, coefficient 0.221, $p<0.05$). No other
category is significant. For Agriculture, the extensive margin is
significantly \emph{negative} (HISCO~6, Dummy: $-0.024^{***}$),
meaning the breach reduced the number of parishes with agricultural
workers while increasing employment in those that retained them ---
a consolidation pattern. For Manufacturing the intensive margin
dominates; the extensive margin is negative but not significant at
conventional levels after correction. Across both sectors, growth
operated through parishes that already had workers, not by drawing
entirely new parishes into those activities. Full results for both the dummy and market
access approaches are in Appendix~\ref{all_occ_results}.

Figure~\ref{fig:mech_occ2} breaks down Agriculture and Manufacturing
into subcategories. Confidence intervals were not estimated for this
figure given the vast number of comparisons involved making multiple 
comparisons problem blow up; it is descriptive and shows
direction, not the basis for inference. Within Manufacturing, the
growth is concentrated in textiles and generic factory work ---
occupations that signal early industrialization. Within Agriculture,
the largest point estimate is for fishing, consistent with improved
access to North Sea fish stocks \citep{Poulsen2007}.\footnote{See event-study 
plots for fishing and spinning are in Appendix~\ref{fishing_spinning}.}

\begin{figure}
\begin{center}
  \caption{Effects on Detailed Occupational Structure} \label{fig:mech_occ2}
  \includegraphics[width=1\textwidth]{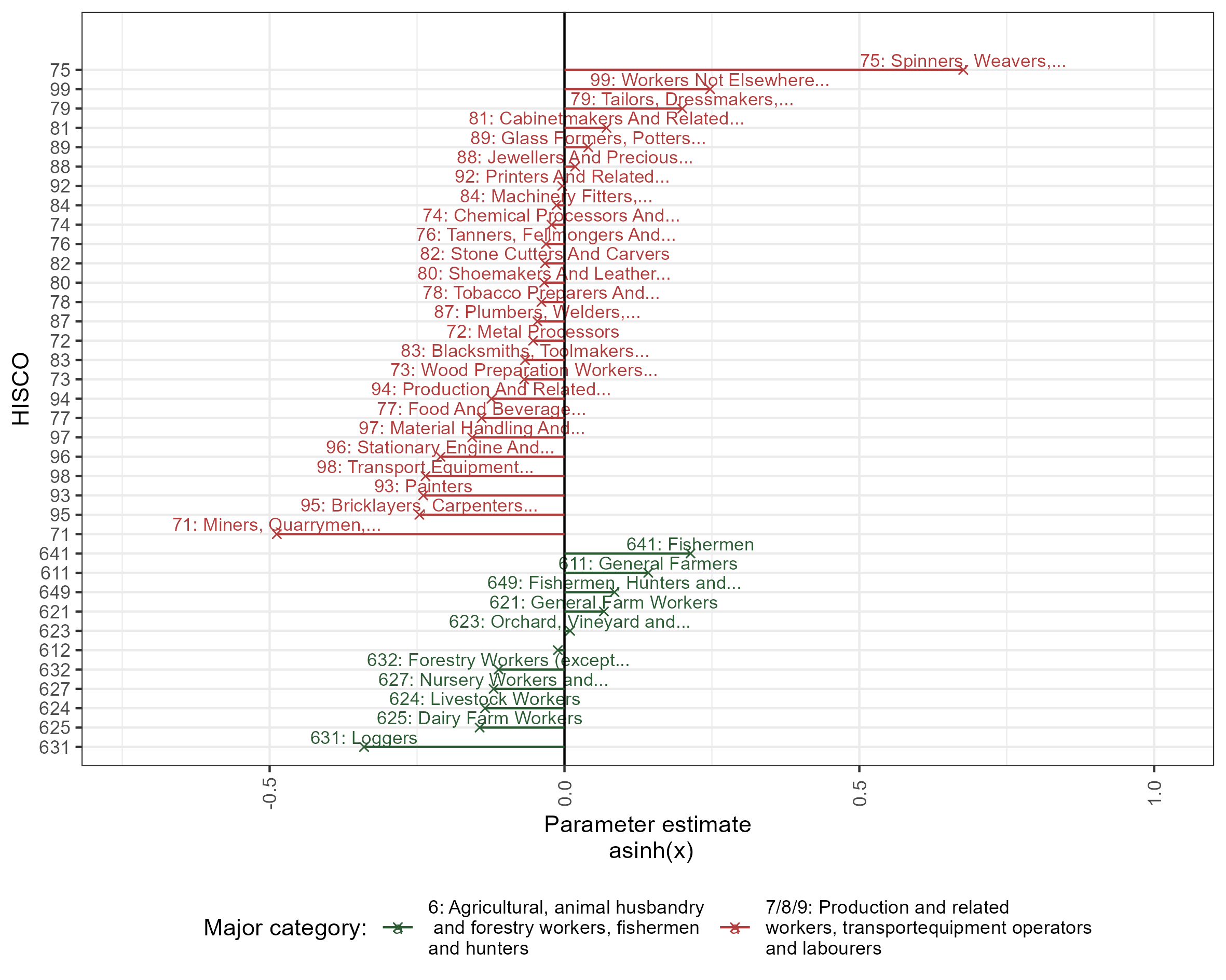}
  \parbox{1\textwidth}{
  \caption*{\footnotesize \textit{Notes:} This figure depicts the relative changes in subcategories of agricultural and manufacturing occupations (HISCO codes starting with 6 and 7/8/9). The plot presents the results using the arcsinh transformation and the dummy-definition of being affected by the channel. Qualitatively similar results using alternative approaches are available in the online repository.}
}
\end{center}
\end{figure}

\subsection{Fertility and Migration}

\begin{figure}
    \centering
    \caption{Effects on fertility and internal migration}
    \begin{subfigure}[b]{0.45\textwidth}
        \centering
        \caption{Fertility (MA approach)} \label{fig:fert_ma}
        \includegraphics[width=\textwidth]{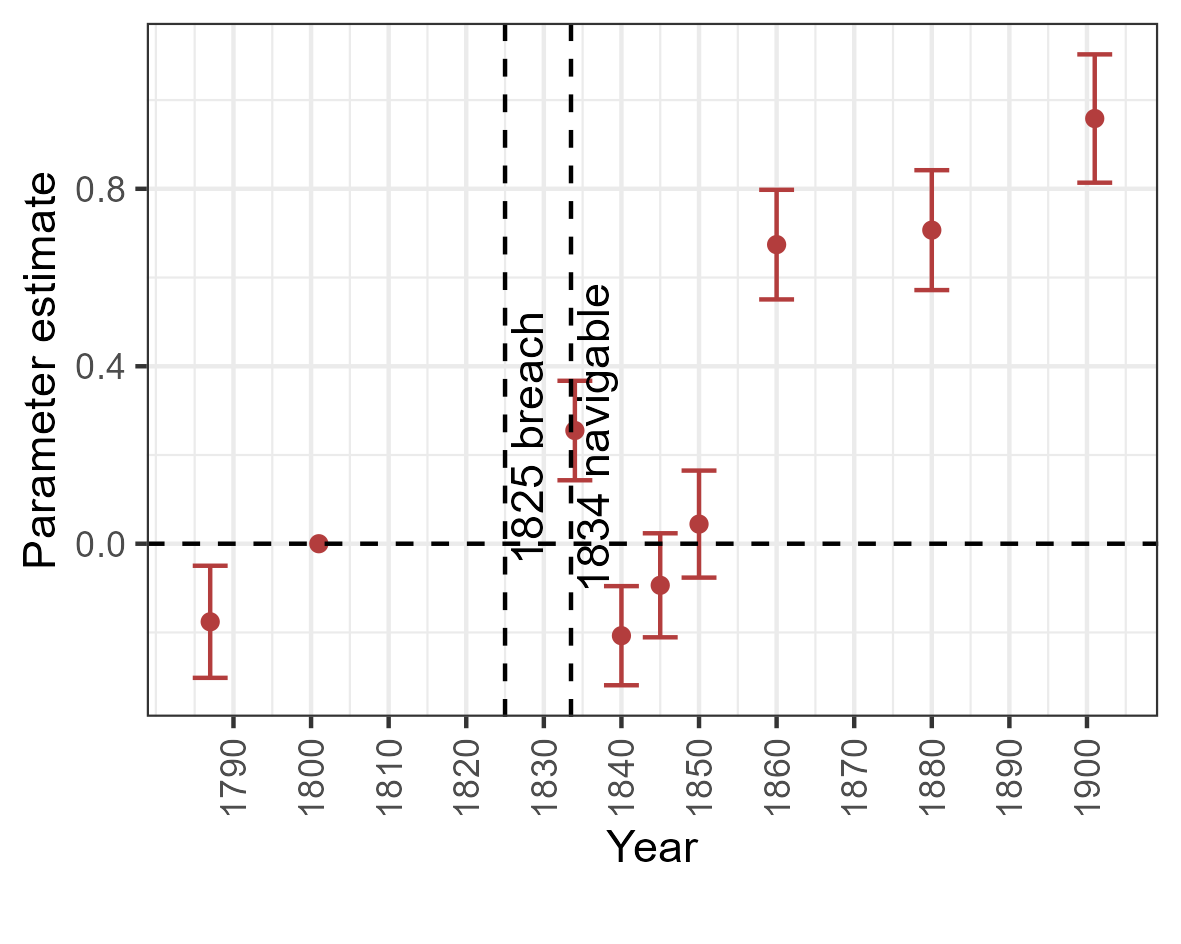}
    \end{subfigure}
    \hfill
    \begin{subfigure}[b]{0.45\textwidth}
        \centering
        \caption{Fertility (dummy approach)} \label{fig:fert_dummy}
        \includegraphics[width=\textwidth]{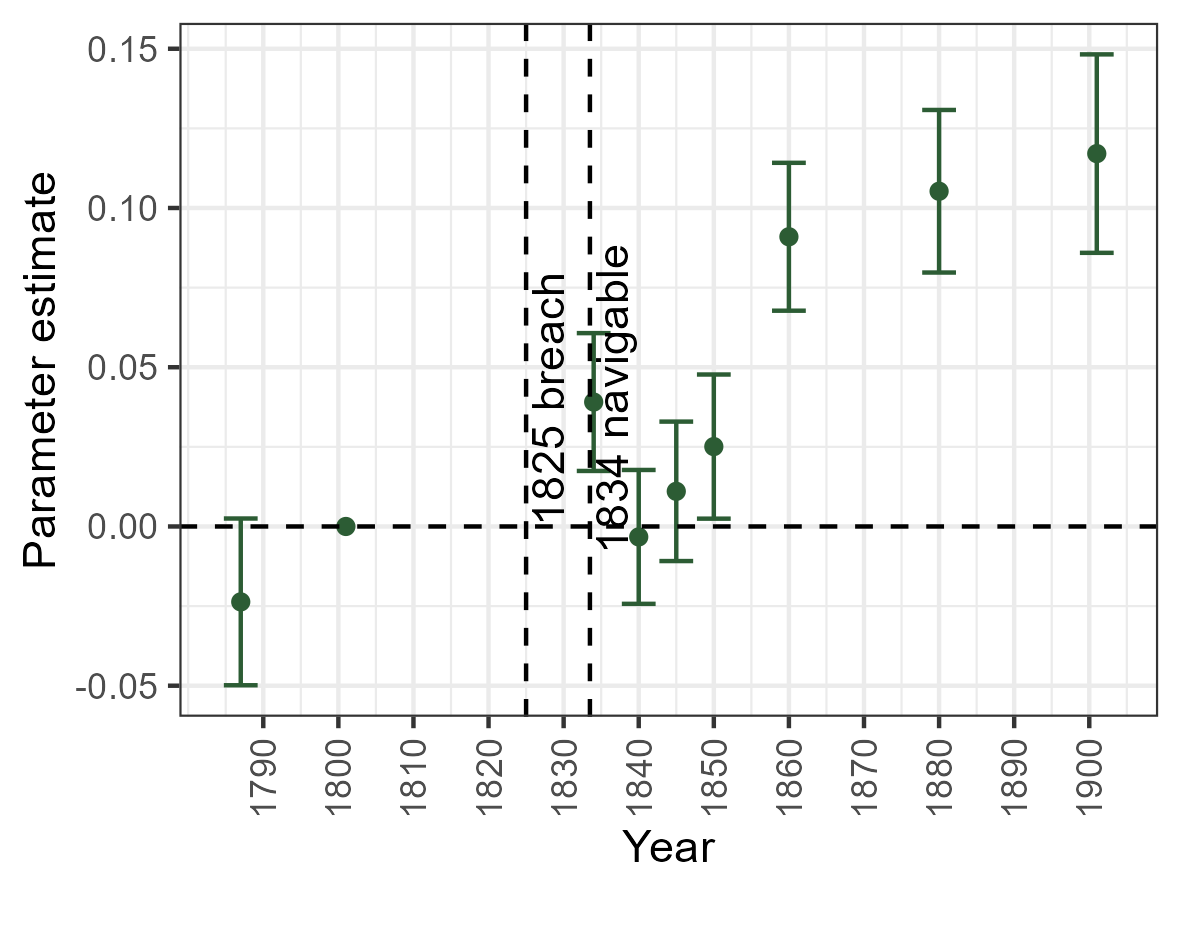}
    \end{subfigure}
    \vspace{0.45cm}
    \begin{subfigure}[b]{0.45\textwidth}
        \centering
        \caption{Internal migration (MA approach)} \label{fig:migr_ma}
        \includegraphics[width=\textwidth]{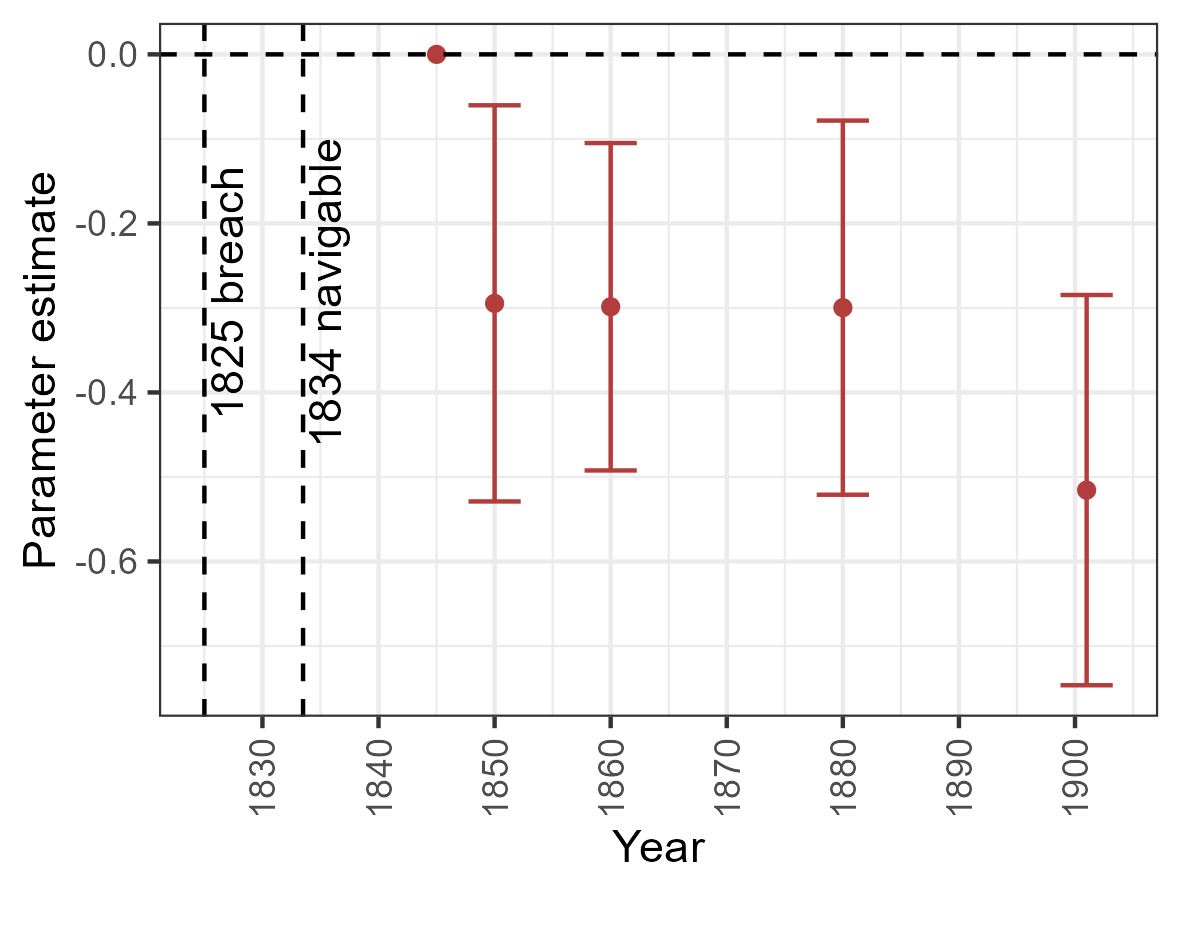}
    \end{subfigure}
    \hfill
    \begin{subfigure}[b]{0.45\textwidth}
        \centering
        \caption{Internal migration (dummy approach)} \label{fig:migr_dummy}
        \includegraphics[width=\textwidth]{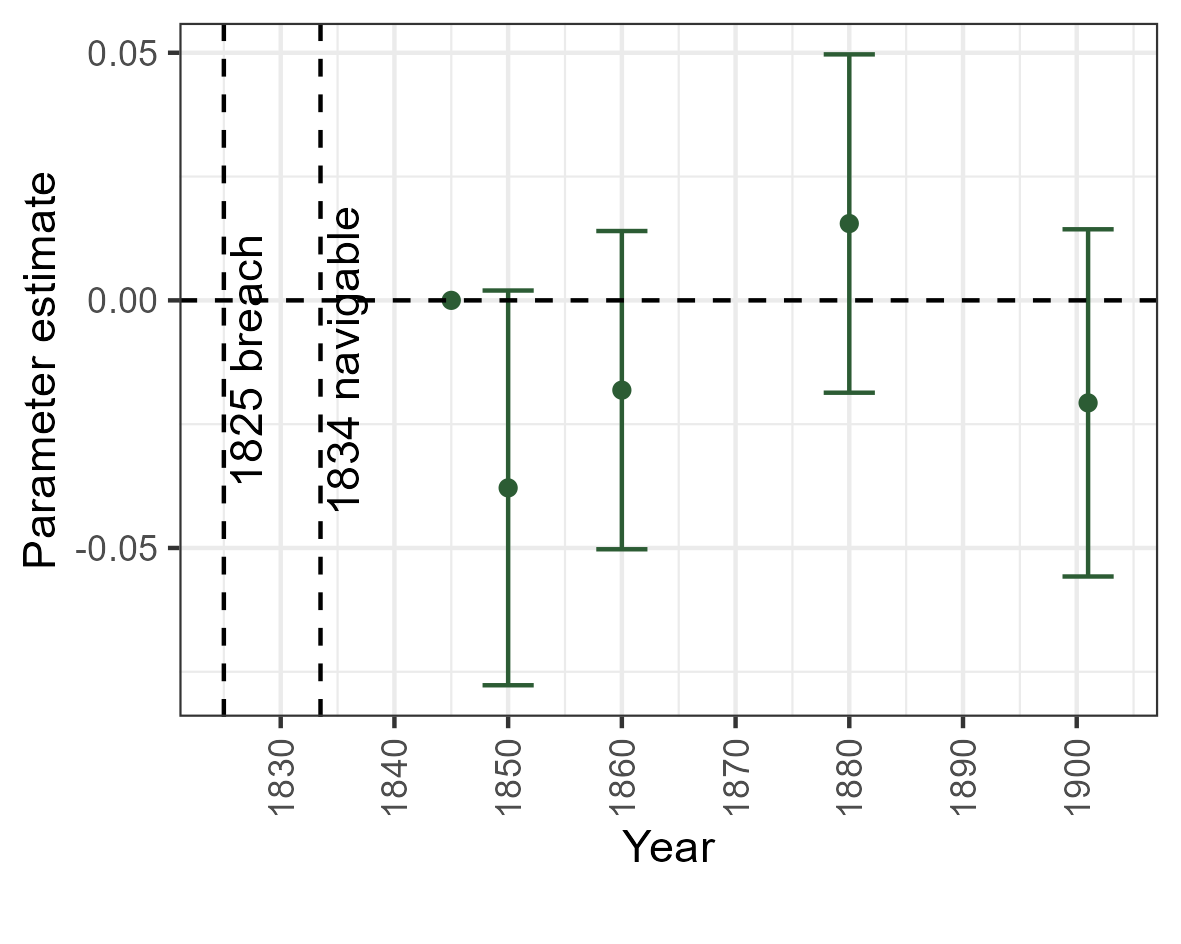}
    \end{subfigure}
    \parbox{0.9\textwidth}{
    \caption*{\footnotesize \textit{Notes:} This shows the effect on indicators of fertility and migration. Panel (a) and panel (b) show the effect on the child-women ratio. Panels (c) and (d) show the effect on the number of people born in a different county, than where they usually live as a share of the total population in that parish. The results of panels (a) and (b) indicate that the channel caused fertility to increase. The results of panels (c) and (d) indicate, that migration either contributed negatively to the increase in population or did not contribute at all.  \\ \textit{Source: Danish census data}}
}
    \label{fig:migr_fert}
\end{figure}

Post-Malthusian dynamics predict that an intrinsic improvement in
living standards raises fertility \citep{Jensen2022, Klemp2016}; mere
reallocation of prosperity from elsewhere would raise migration
instead. The channel raised fertility by 11.6 percent by 1901 ---
a market access elasticity of 0.96 (Figure~\ref{fig:migr_fert},
panels a--b). Migration either declined or was unchanged (panels
c--d). Population growth was intrinsic, not drawn from elsewhere.

The fertility estimate is smaller than the population 
effect\footnote{For comparison the estimate on the market access 
approach is 1.591 and dummy approach: 0.239 in 1901. Reported in
Appendix~\ref{extra_pop_res}, Table~\ref{pop_reg_table}.} 
because fertility is a rate and population is a stock: a higher
birth rate compounding across generations produces a larger
proportional gain in the stock than in the rate. As a robustness 
exercise, I do a back-of-the envelope conversation of the fertility 
result into an implied population path under that given fertility: 
cumulating the effect, I find 1.593 under the market access approach and 
0.238 under the dummy approach by 1901, matching the observed 
population effects with high precision (Appendix~\ref{boe_fertility}).

The timing of the fertility effect mirrors the three-stage mechanism
in Section~\ref{hist}: a spike as early as 1834 during the fish
windfall, a collapse as catches fell, then a sustained rise from the
1860s as trade expanded. The population grew younger: age-group
results in Appendix~\ref{effects_by_age_group} show more children
born and surviving beyond infancy.

\FloatBarrier

\section{The Reverse Natural Experiment} \label{arch}

\subsection{Method and data}
The same channel closed sometime between 1086 and 1208, in a society
that shares the geography but nothing else. Medieval Denmark
differed from nineteenth-century Denmark in culture, religion,
technology, and institutions in every way. If the 
effect of the channel replicates in reverse in that society,
confounders specific to the 19th-century context are not to blame. 

Economic activity in the medieval period is measured using
archaeological finds as a proxy. Settlement Scaling Theory predicts
that larger, more economically active settlements generate more
archaeological material \citep{Ortman2020}, with recent application
in economics
\citep{Davis2002, Bakker2021Phonecians, Allen2023, Barjamovic2019}.
The data come from the Danish national registry \textit{Fund og
Fortidsminder}, which geo-references all sites documented by Danish
museums and the National Agency of Culture and Palaces. The full
registry contains 290,524 findings spanning the Paleolithic to the
modern period. I restrict to 750--1500~CE and to the two types most
informative about economic life: coins (evidence of trade) and
buildings (evidence of settlement density). This yields 3,411 coin
findings and 4,396 building findings, matched to parishes using the
same borders as the census data.

Each finding carries a date range rather than a precise year.
Finding~ID~338 is a coin from the parish of Pedersker on Bornholm,
dated 1300--1535. A naive approach records a ‘1’ for every year in
that range, but this overrepresents findings with wide date ranges
relative to those dated precisely, introducing a systematic bias. Instead, the date
range is treated as a probability distribution.\footnote{A uniform
distribution is the basis of the results shown here; results under a
normal distribution are in Appendix~\ref{norm_arch}.} Bayes’ rule
gives the inverse probability $P(\text{finding}|t)$ for each
parish-year; Appendix~\ref{math_note} derives this analytically. In
practice a Monte Carlo procedure samples from these distributions to
build a panel of estimated parish-level activity, which enters the
empirical framework from Section~\ref{emp_strat}. Standard errors are
constructed by resampling the Monte Carlo draws in a clustered
bootstrap, so dating uncertainty is propagated into
inference.\footnote{Each bootstrap iteration resamples clusterwise
from the Monte Carlo draws used in panel construction.}

\begin{figure}
    \centering
    \caption{Soil type propensity scores}
    \begin{subfigure}[b]{0.45\textwidth}
        \centering
        \caption{Propensity score before matching} \label{fig:prop1}
        \includegraphics[width=\textwidth]{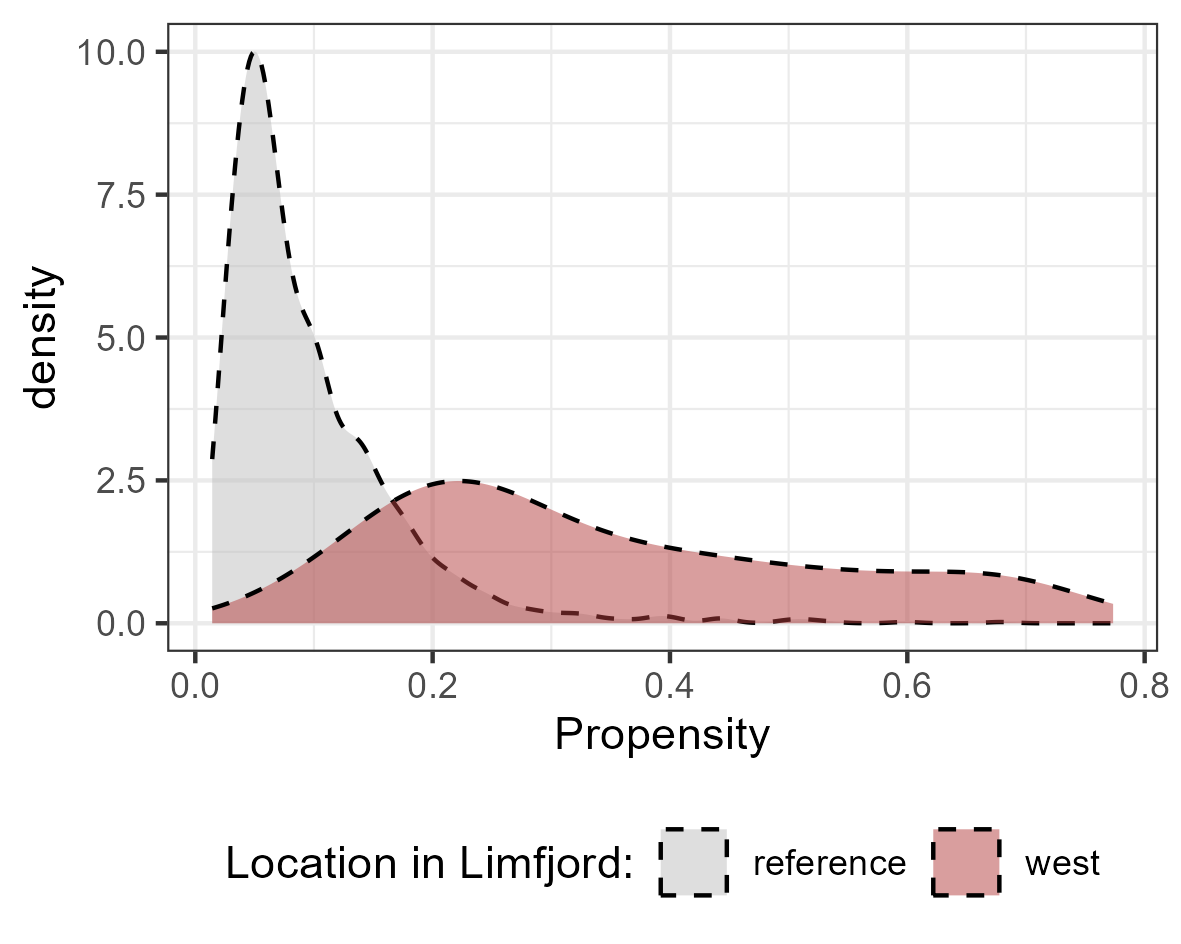}
    \end{subfigure}
    \hfill
    \begin{subfigure}[b]{0.45\textwidth}
        \centering
        \caption{Propensity score after matching} \label{fig:prop2}
        \includegraphics[width=\textwidth]{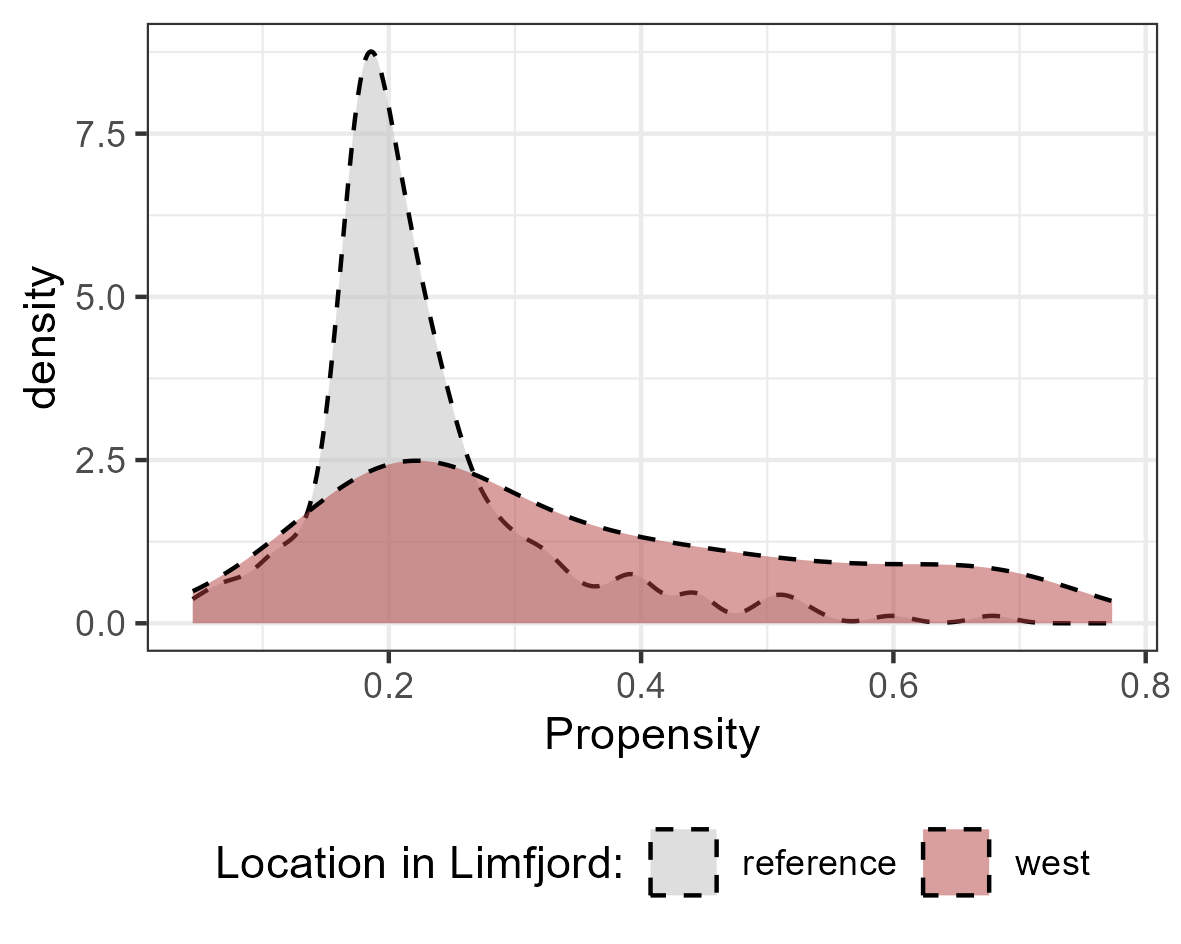}
    \end{subfigure}
    \parbox{0.9\textwidth}{
    \caption*{\footnotesize \textit{Notes:} Propensity scores between the West Limfjord and the reference group before and after matching. Propensity scores are estimated using extreme gradient boost. Matching is done with a \textit{greedy} matching procedure in random order.}
}
    \label{fig:prop_score}
\end{figure}

Soil type is a potential confounder: fertile soils in the western
Limfjord may have shaped development independently
\citep{HeavyPlough2016, WinnersAndLosers2022}, and soil conditions
also affect the preservation of archaeological material. I address
this with propensity score matching, using XGBoost
\citep{chen2016xgboost} on soil type data \citep{Pedersen2019} to
match each West Limfjord parish to the most similar unaffected parish.
Only soil types present in at least 10 percent of parishes are used to
avoid overfitting on rare categories. Each parish is matched once, via
a greedy without-replacement algorithm. Figure~\ref{fig:prop_score}
shows the propensity score distributions before and after matching.

\begin{figure}
    \centering
    \caption{Rate of coin findings}
    \label{fig:arch_desc}
    \includegraphics[width=\textwidth]{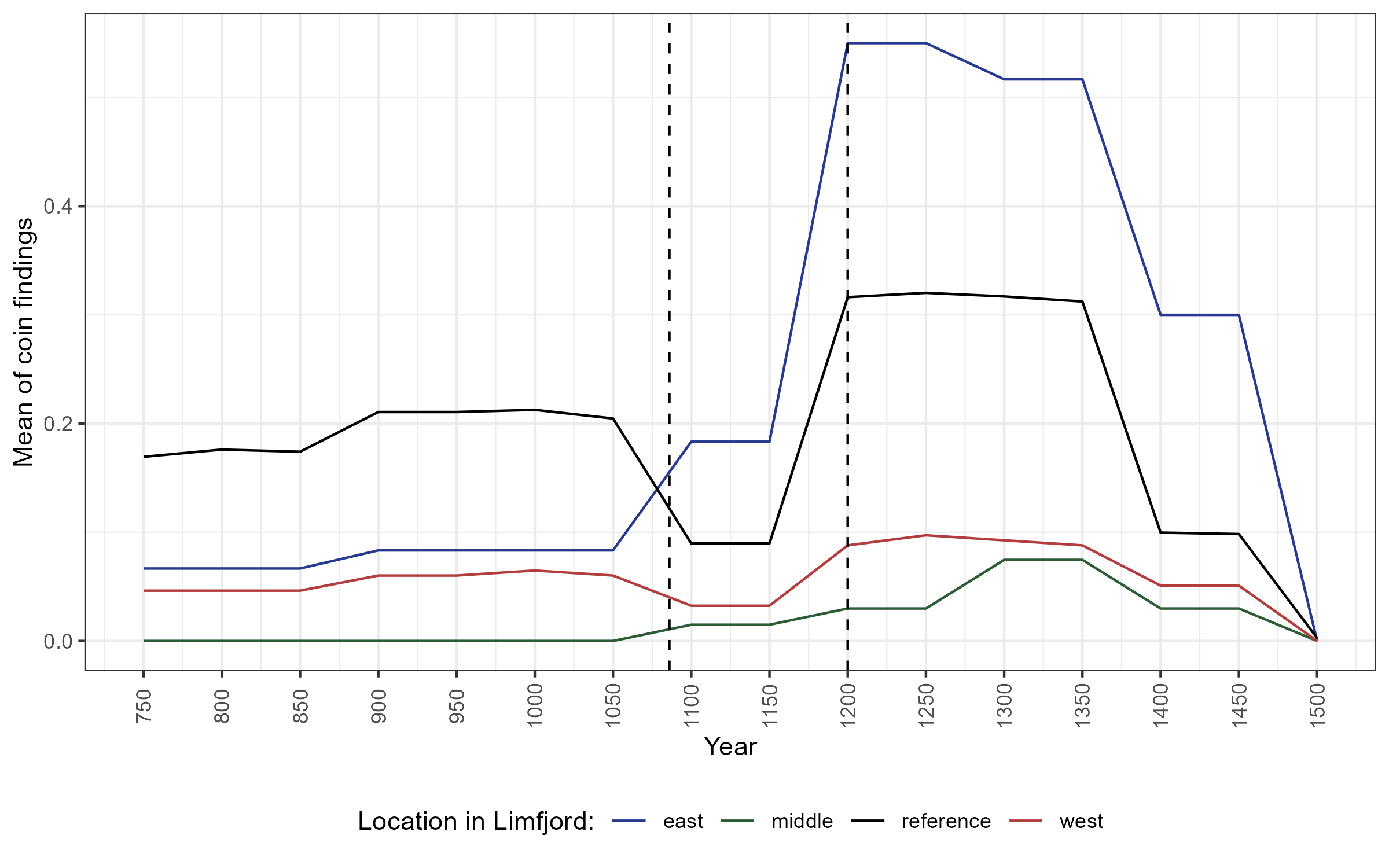}
    \parbox{0.9\textwidth}{
        \caption*{\footnotesize \textit{Notes:} This plot shows the average number of coin findings per year, counted without adjusting for dating range uncertainty, for the West, Middle, and East Limfjord regions, as well as for the rest of Denmark. \\ \textit{Source: Danish registry of archaeological findings}}
    }
\end{figure}

Figure~\ref{fig:arch_desc} shows coin findings by year across the
four regions, counted without adjusting for dating uncertainty. After
the channel closed (1086--1208), findings in the East Limfjord and the
rest of Denmark grew steadily; the West and Middle Limfjord stagnated.
The divergence is visible in raw counts before any adjustment.

\FloatBarrier
\subsection{Results}
\begin{figure}
    \centering
    \caption{Archaeological results}
    \begin{subfigure}[b]{0.45\textwidth}
        \centering
        \caption{Coins: Market access approach} \label{fig:arch1a}
        \includegraphics[width=\textwidth]{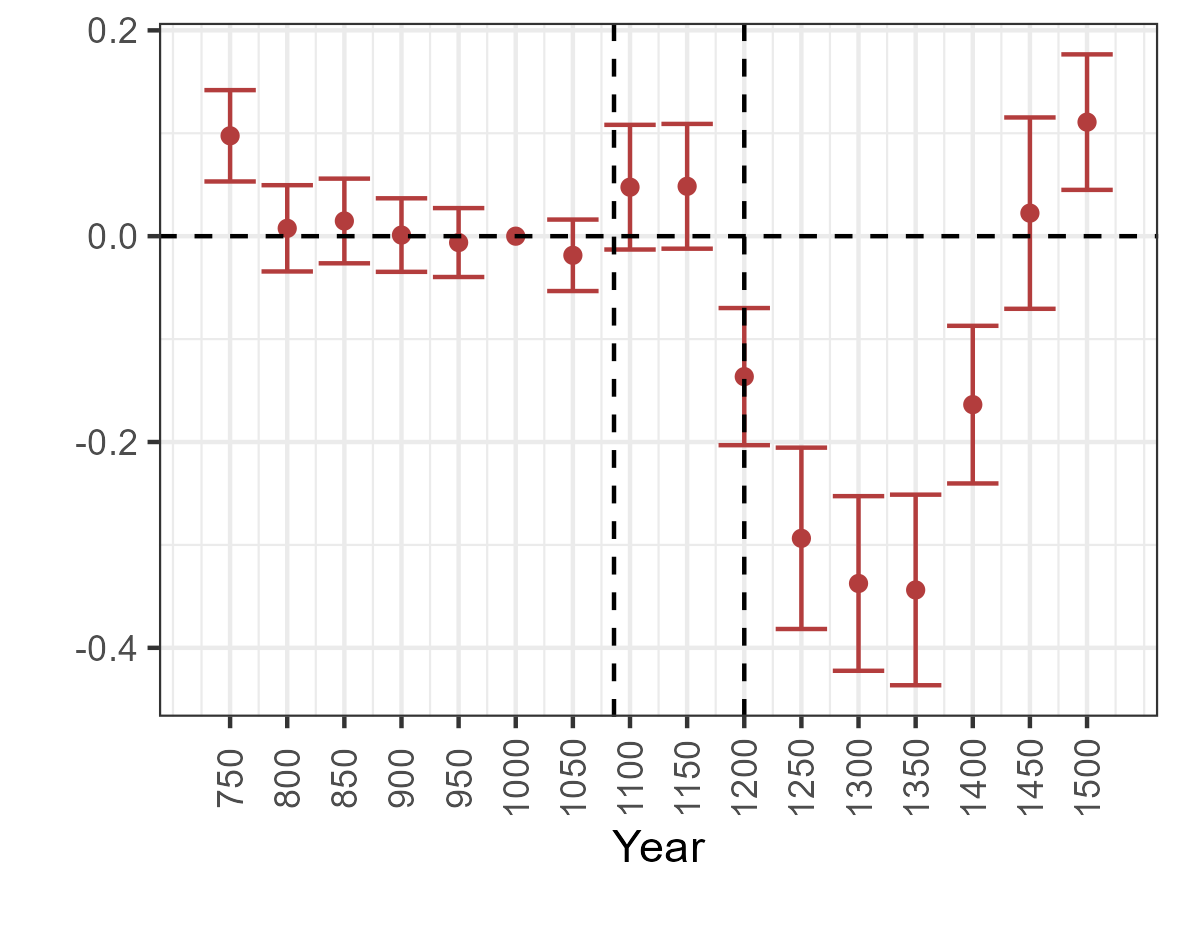}
    \end{subfigure}
    \hfill
    \begin{subfigure}[b]{0.45\textwidth}
        \centering
        \caption{Coins: Dummy approach} \label{fig:arch1b}
        \includegraphics[width=\textwidth]{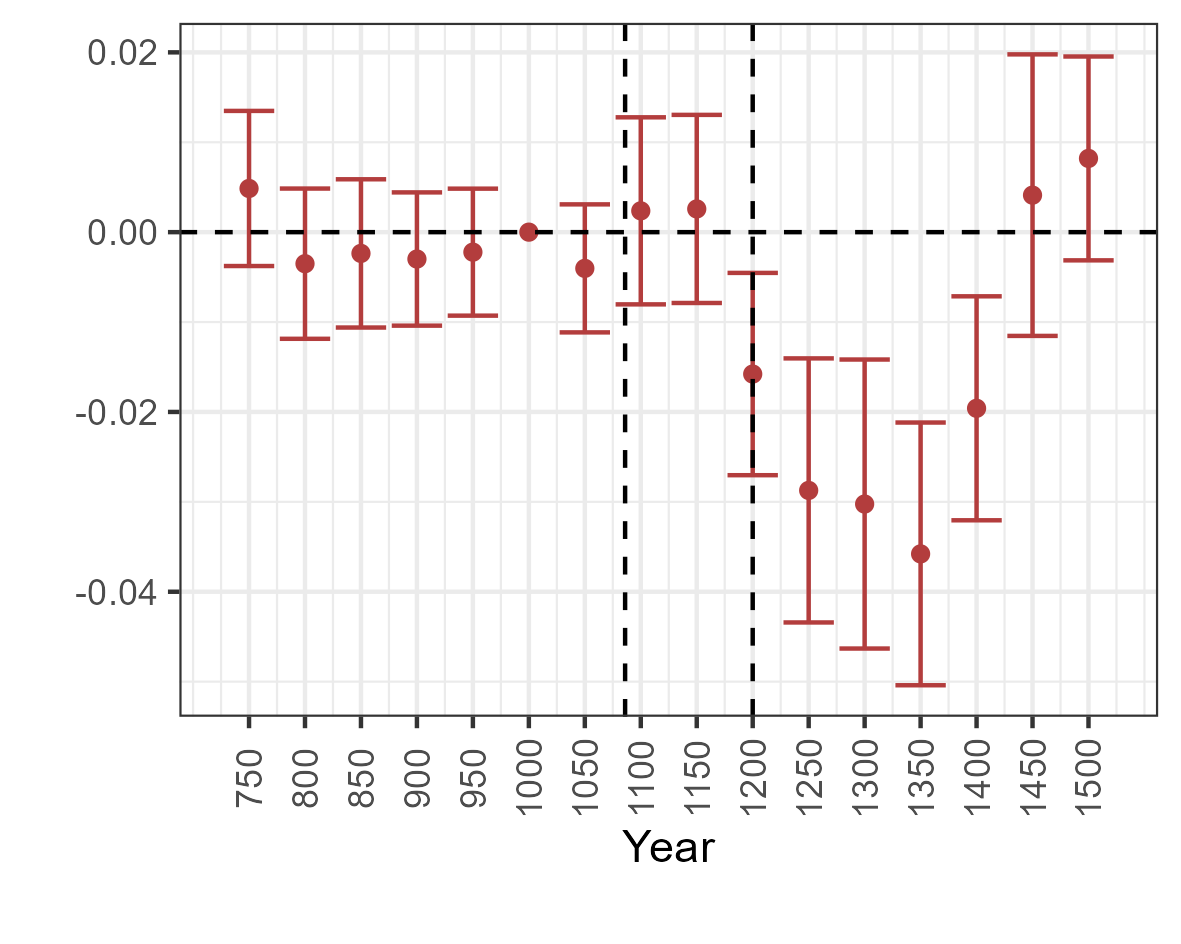}
    \end{subfigure}
    \vspace{0.45cm}
    \begin{subfigure}[b]{0.45\textwidth}
        \centering
        \caption{Buildings: Market access approach} \label{fig:arch1c}
        \includegraphics[width=\textwidth]{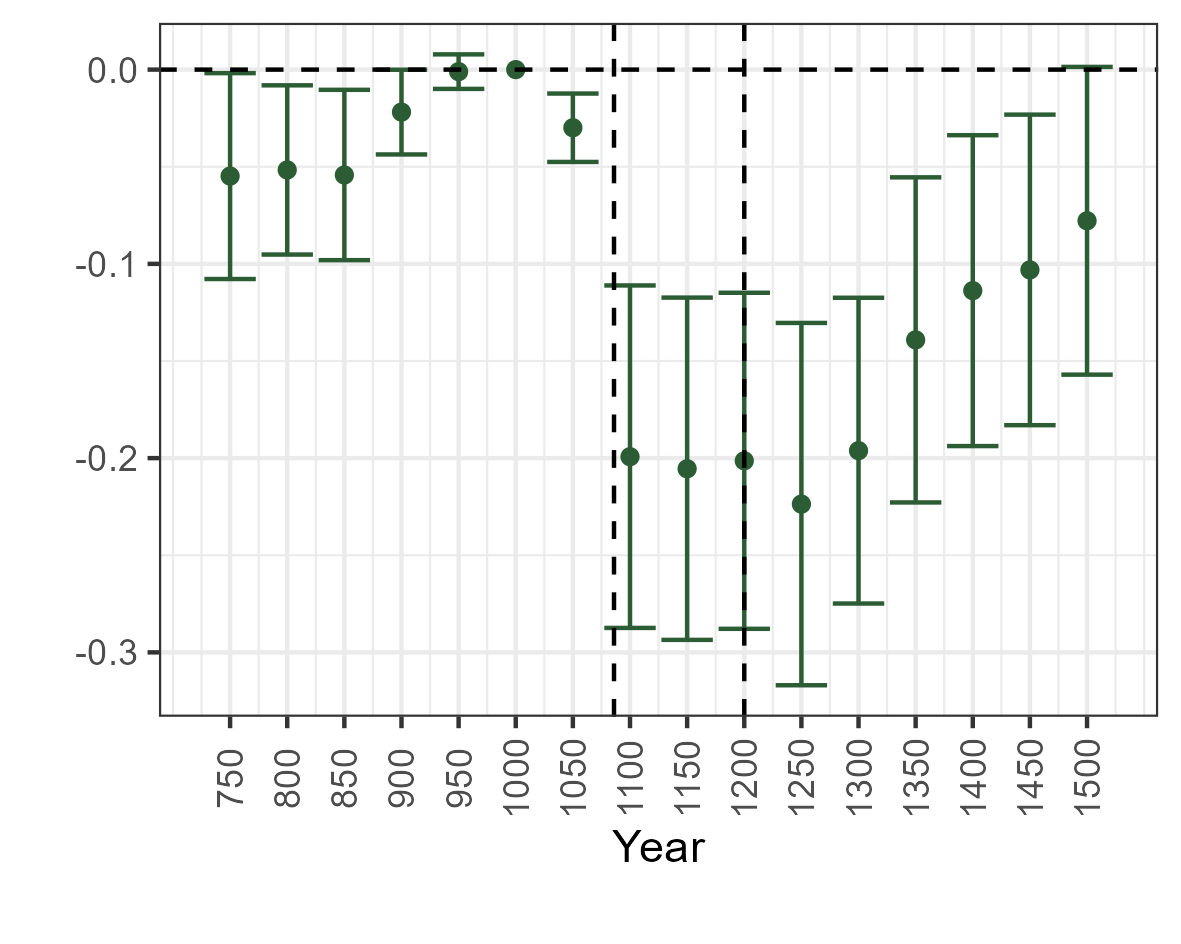}
    \end{subfigure}
    \hfill
    \begin{subfigure}[b]{0.45\textwidth}
        \centering
        \caption{Buildings: Dummy approach} \label{fig:arch1d}
        \includegraphics[width=\textwidth]{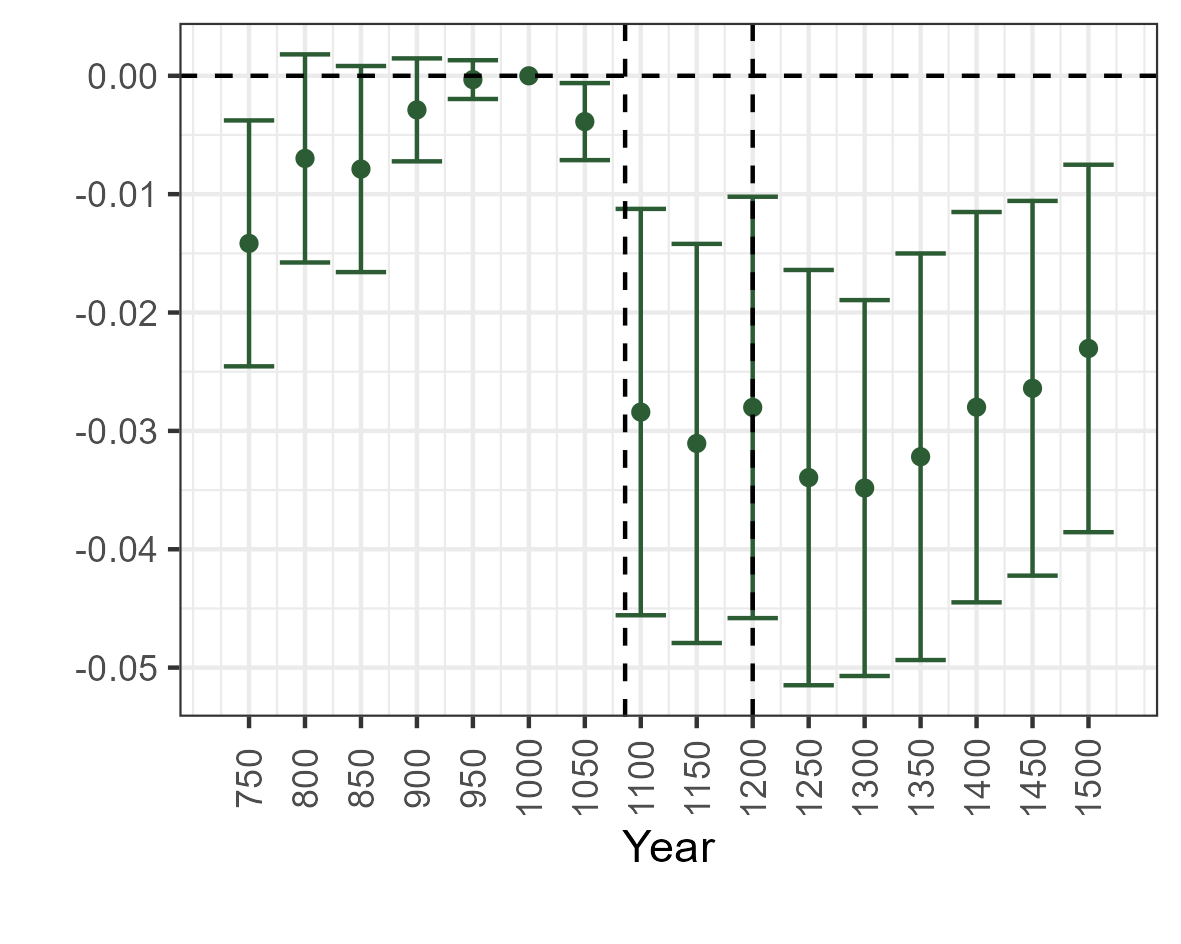}
    \end{subfigure}
    \parbox{1\textwidth}{
\caption*{\footnotesize \textit{Notes:} This shows the archaeological regression results. Standard errors in parentheses are clustered at the parish level, bootstrapped from Monte Carlo draws. Columns 1-4 show results using the full sample of all of Denmark. Columns 5-8 show results for a matched sample. All the even columns show results using the dummy definition of being affected. All the uneven columns show results using the change in market access approach. The outcome is the probability that a given finding type was generated in the area covered by that parish within $\pm$25 years of the reported year. A parameter estimate for all years 750, 800, ..., 1500 can be found in the Appendix \ref{all_arch_param}. \\ \textit{Source: Danish registry of archaeological findings}}
}
    \label{fig:arch_reg_boot}
\end{figure}

\begin{figure}
    \centering
    \caption{Distribution of parameter estimates in 1350}
    \begin{subfigure}[b]{0.45\textwidth}
        \centering
        \caption{Coins: Market access approach} \label{fig:distri_a}
        \includegraphics[width=\textwidth]{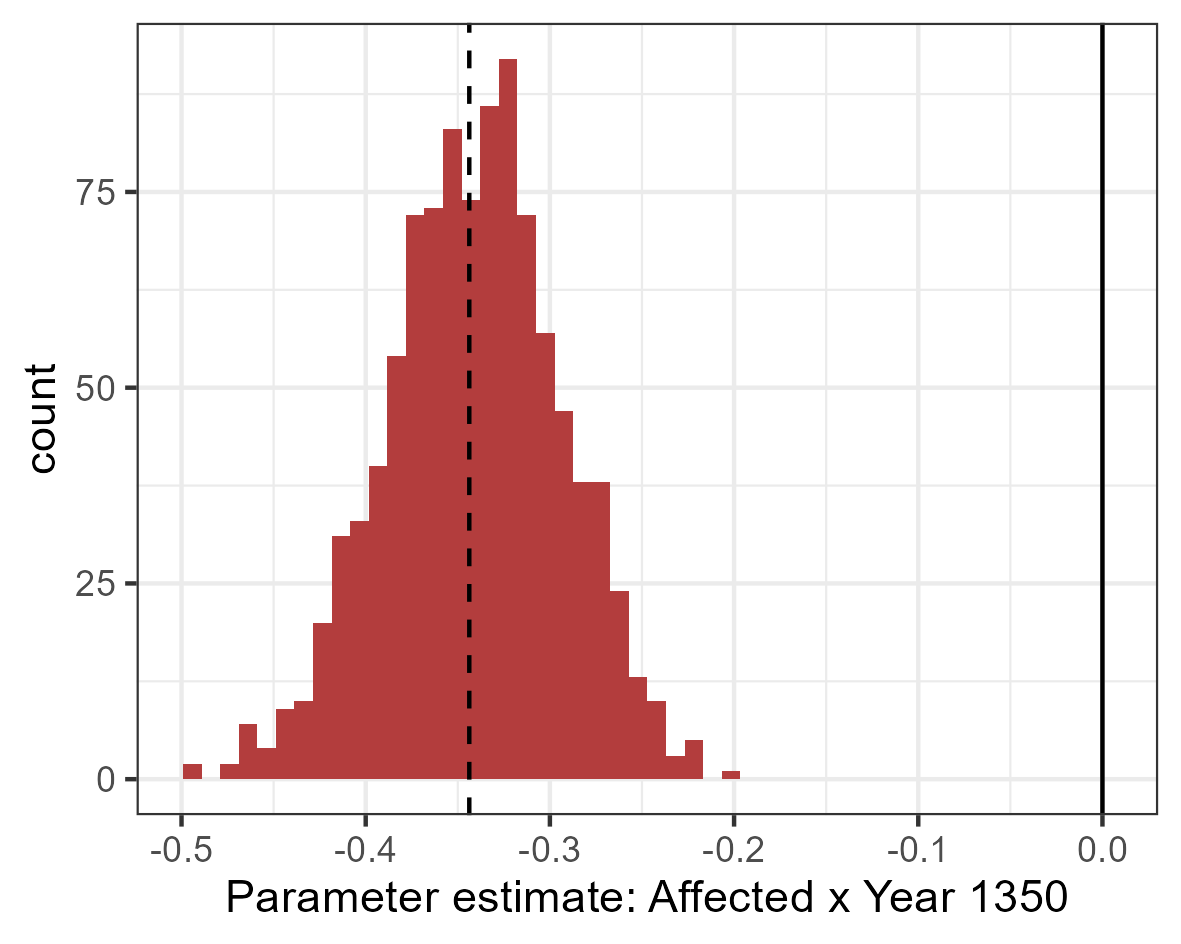}
    \end{subfigure}
    \hfill
    \begin{subfigure}[b]{0.45\textwidth}
        \centering
        \caption{Coins: Dummy approach} \label{fig:distri_b}
        \includegraphics[width=\textwidth]{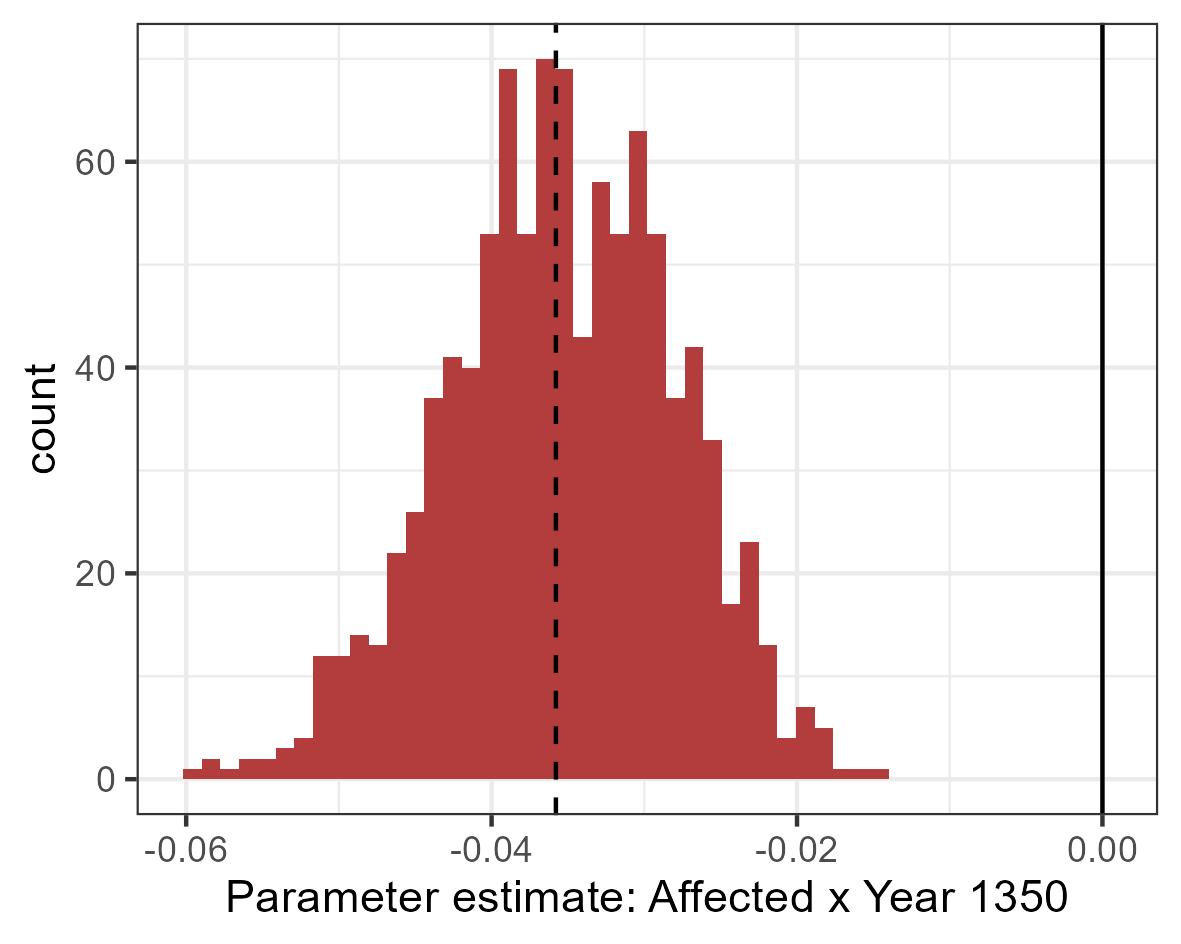}
    \end{subfigure}
    \vspace{0.45cm}
    \begin{subfigure}[b]{0.45\textwidth}
        \centering
        \caption{Buildings: Market access approach} \label{fig:distri_c}
        \includegraphics[width=\textwidth]{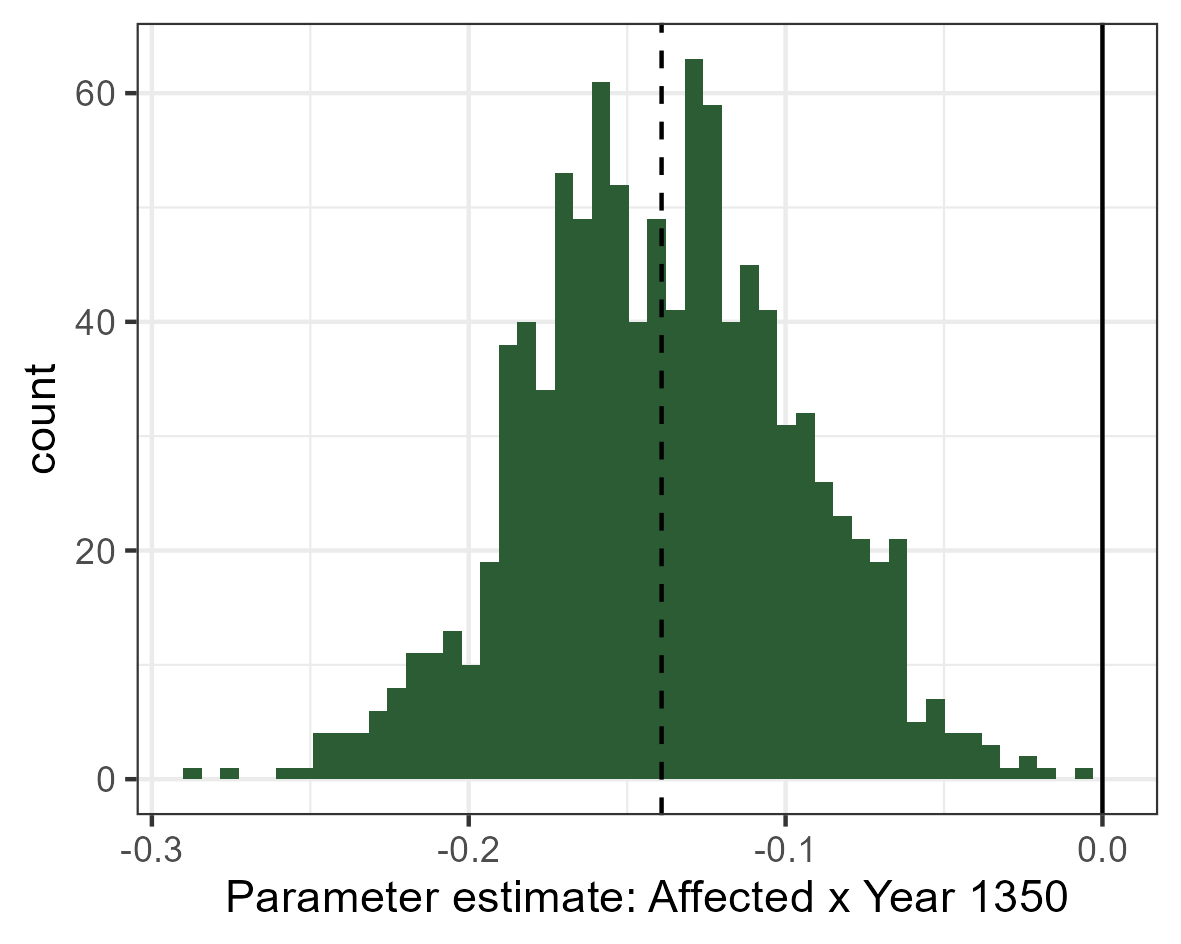}
    \end{subfigure}
    \hfill
    \begin{subfigure}[b]{0.45\textwidth}
        \centering
        \caption{Buildings: Dummy approach} \label{fig:distri_d}
        \includegraphics[width=\textwidth]{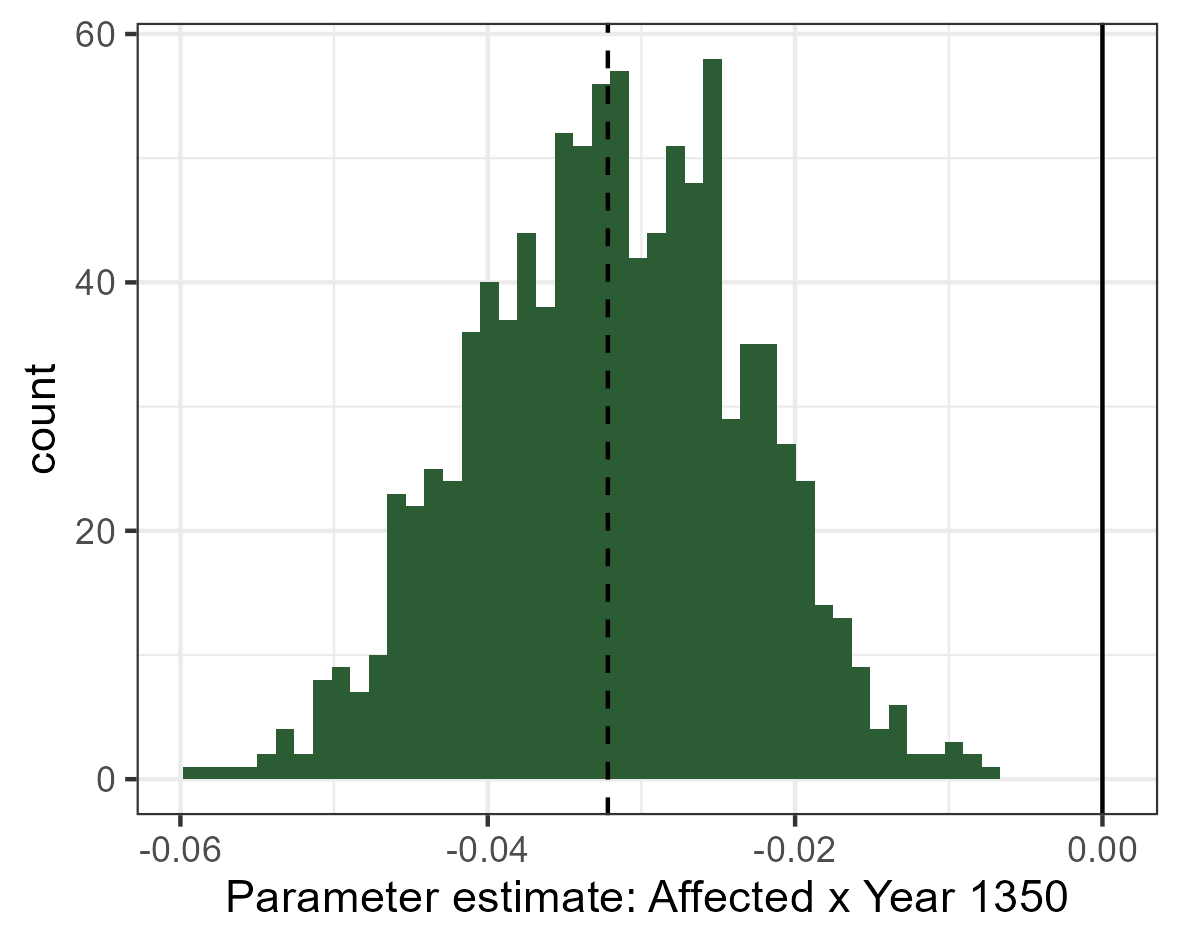}
    \end{subfigure}

    \parbox{0.9\textwidth}{
    
    \caption*{\footnotesize \textit{Notes:} This shows 1000 draws from the bootstrap procedure, which takes classical (clustered) statistical uncertainty as well as dating uncertainty into account. Panel (a) shows the distribution of the effect on coin using the market access approach. Panel (b) shows the effect on coins using the dummy approach. Panel (c) shows the effect on buildings using the market access approach. Finally panel (d) shows the effect on buildings using the dummy approach. The dotted line is the point estimate.  \\ \textit{Source: Danish registry of archaeological findings}}
}    \label{fig:arch_reg}
\end{figure}

The closure of the Limfjord channel reduced the probability of
archaeological activity in West Limfjord parishes for centuries after.
Figure~\ref{fig:arch_reg} and Table~\ref{tab:arch1} report these
effects; Figure~\ref{fig:arch_reg_boot} shows the parameter
distribution at 1350 from the bootstrap procedure
(Appendix~\ref{math_note}). The outcome is the probability that a coin
finding or building was generated in a parish within $\pm 25$ years of
the reported year.

The market access approach gives percentage-point changes per
log-point change in market access. A one log-point decrease in market
access reduced the probability of any coin finding in 1350 by 0.34 or
0.23 percentage points against a pre-shock mean of 2.0 percent
(columns 1 and 5), and the probability of any building finding by 0.14
or 0.11 percentage points against a mean of 4.2 percent (columns 3
and 7). The dummy approach gives the change relative to other
parishes: by 1350, the probability of coin findings had fallen by 3.6
or 3.0 percentage points (columns 2 and 6), and the probability of
building findings by 3.3 or 2.6 percentage points (columns 4 and 8).

Effects are close to zero before 1086--1208 and decline sharply
thereafter (Appendix~\ref{all_arch_param} tabulates all years). A
natural concern is that the Monte Carlo procedure inflates standard
errors artificially, but the bootstrap resamples the Monte Carlo draws
clusterwise, so dating uncertainty is fully propagated into inference.
Figure~\ref{fig:arch_reg_boot} shows the resulting parameter
distribution at 1350.

The channel closure caused decline in the West
Limfjord for centuries. A direct comparison to the 1825 census
estimates is not possible, but the parameters are of similar sign and
magnitude for two channels in the same location across seven
centuries. First-nature geography is the cause of the location of
prosperity across two very different societies.

\begin{landscape}
\begin{table}
\centering
\caption{Archaeological regression results} \label{tab:arch1}
\footnotesize
\begin{tabular}{lcccccccc}
      \tabularnewline \midrule \midrule
                                & \multicolumn{4}{c}{Full sample} & \multicolumn{4}{c}{Matched sample} \\
   \cmidrule(lr){2-5} \cmidrule(lr){6-9}
   Outcome:                     & \multicolumn{2}{c}{Coin findings} & \multicolumn{2}{c}{Buildings} & \multicolumn{2}{c}{Coin findings} & \multicolumn{2}{c}{Buildings} \\
   \midrule
   \emph{Variables}\\
   Year 950 $\times$ Affected   & -0.0058         & -0.0024         & -0.0046         & -0.0010         & -0.0011         & -0.0001         & -0.0070  & -0.0011\\   
                                & (0.0172)        & (0.0036)        & (0.0047)        & (0.0009)        & (0.0349)        & (0.0050)        & (0.0075) & (0.0012)\\   
   Year 1050 $\times$ Affected  & -0.0143         & -0.0035         & -0.0378$^{***}$ & -0.0050$^{***}$ & -0.0125         & -0.0018         & -0.0098  & -0.0021\\   
                                & (0.0184)        & (0.0037)        & (0.0089)        & (0.0017)        & (0.0445)        & (0.0063)        & (0.0142) & (0.0021)\\   
   Year 1150 $\times$ Affected  & 0.0509          & 0.0027          & -0.2123$^{***}$ & -0.0319$^{***}$ & 0.1104$^{**}$   & 0.0160$^{**}$   & -0.0494  & -0.0157\\   
                                & (0.0318)        & (0.0054)        & (0.0444)        & (0.0084)        & (0.0505)        & (0.0075)        & (0.0741) & (0.0107)\\   
   Year 1250 $\times$ Affected  & -0.2905$^{***}$ & -0.0287$^{***}$ & -0.2283$^{***}$ & -0.0346$^{***}$ & -0.2566$^{***}$ & -0.0268$^{**}$  & -0.1187  & -0.0308$^{**}$\\   
                                & (0.0447)        & (0.0076)        & (0.0480)        & (0.0090)        & (0.0753)        & (0.0112)        & (0.0869) & (0.0124)\\   
   Year 1350 $\times$ Affected  & -0.3412$^{***}$ & -0.0355$^{***}$ & -0.1446$^{***}$ & -0.0333$^{***}$ & -0.2320$^{***}$ & -0.0302$^{***}$ & -0.1112  & -0.0260$^{**}$\\
                                & (0.0490)        & (0.0079)        & (0.0416)        & (0.0084)        & (0.0787)        & (0.0117)        & (0.0779) & (0.0121)\\
   \midrule
   Observations                 & 29,568          & 29,568          & 29,568          & 29,568          & 6,912           & 6,912           & 6,912    & 6,912\\  
   Mean outcome (pre-shock)     & 0.0203          & 0.0203          & 0.0419          & 0.0419          & 0.0203          & 0.0203          & 0.0419   & 0.0419\\
   \midrule \midrule
\end{tabular}

\parbox{1.3\textwidth}{
\caption*{\footnotesize \textit{Notes:} Results from regressions on the probability that a given finding type was generated within $\pm$25 years in a parish. Standard errors (in parentheses) are clustered at the parish level and derived via bootstrapping of Monte Carlo samples. Columns 1-4 use the full sample and columns 5-8 use a matched sample; even columns use the dummy measure and odd columns the change in market access approach. Estimates for all years (750, 800, …, 1500) are provided in Appendix \ref{all_arch_param}. *** $p<0.01$, ** $p<0.05$, * $p<0.10$. \\ \textit{Source: Danish registry of archaeological findings}}
}
\end{table}
\end{landscape}

\FloatBarrier
\section{Conclusion} \label{conclusion}

A storm that lasted one night reshaped a region for the rest of the century. The 1825 
breach of Agger Isthmus gave the West Limfjord direct access to North Sea 
shipping lanes, and by 1901 affected parishes were 27.0 percent more populous 
than comparable Danish areas --- growth driven by higher fertility and a rising 
share of fishermen and manufacturing workers, not by people moving in from 
elsewhere. Geography seemingly shapes the location of prosperity.

The medieval mirror sharpens the conclusion beyond what a single
experiment can establish. When a channel in the same location closed
sometime between 1086 and 1208, coin findings in the West Limfjord
fell by 0.23--0.34 percentage points per log-point loss in market
access (Table~\ref{tab:arch1}, columns~1 and~5 at Year~1350; pre-shock
mean 2.0 percent), and building findings by 0.11--0.14 percentage
points (columns~3 and~7; mean 4.2 percent). The
geographic shock was the same in location, opposite in sign. Everything
else --- culture, religion, technology, institutions --- differed across
seven centuries. The result flipped with the geography. Confounders specific 
to the 19th-century context are not to blame. 

Geography works through market access, agglomeration, and structural change 
simultaneously --- which is why it is both harder to identify than soil quality 
or coal deposits, and more consequential than any single-channel mechanism. 
Where \citet{Ahlfeldt2015} showed that second-nature geography shapes the 
location of prosperity, this paper provides the analogous evidence for 
first-nature geography, filling the gap in the causal literature. The same 
approach could be applied wherever geography changed abruptly --- the silting 
of the medieval channel at Bruges \citep{Houtte1966, Charlier2011}, or the 
shifting fortunes of K\"{o}nigsberg/Kaliningrad as its Baltic access was 
alternately opened and closed \citep{Britannica2018}.

As climate change accelerates, so will geographic change. The Northwest Passage
is opening. Greenland is acquiring new coasts \citep{kavan2025new}.
Delta cities flood. We are not free of geomorphology --- we remain,
as much as any medieval Viking or nineteenth-century Limfjord
fisherman, its prisoners, and occasionally --- by a perfect storm ---
its beneficiaries.

\newpage
\bibliographystyle{apacite}
\bibliography{main}

@book{milkandbutter,
author = {Lampe, Markus and Sharp, Paul},
isbn = {9780226549644},
publisher = {University of Chicago Press},
series = {University of Chicago Press Economics Books},
title = {{A Land of Milk and Butter}},
url = {https://ideas.repec.org/b/ucp/bkecon/9780226549507.html},
year = {2018}
}

@misc{EtymFjord,
author = {Harper, D},
booktitle = {Online Etymology Dictionary},
title = {{Etymology of fjord}},
url = {https://www.etymonline.com/word/fiord},
urldate = {2022-05-29},
year = {2022}
}

@article{ORourkeCoal2021,
author = {Fernihough, Alan and O'Rourke, Kevin Hjortsh{\o}j},
doi = {10.1093/ej/ueaa117},
issn = {0013-0133},
journal = {The Economic Journal},
number = {635},
pages = {1135--1149},
title = {{Coal and the European Industrial Revolution}},
url = {https://doi.org/10.1093/ej/ueaa117},
volume = {131},
year = {2020}
}

@article{Jensen2022,
author = {Jensen, Peter Sandholt and Pedersen, Maja Uhre and Radu, Cristina Victoria and Sharp, Paul Richard},
doi = {https://doi.org/10.1016/j.eeh.2021.101437},
issn = {0014-4983},
journal = {Explorations in Economic History},
pages = {101437},
title = {{Arresting the Sword of Damocles: The transition to the post-Malthusian era in Denmark}},
url = {https://www.sciencedirect.com/science/article/pii/S0014498321000644},
volume = {84},
year = {2022}
}

@article{Berger2019a,
author = {Berger, Thor},
doi = {10.1016/j.eeh.2019.06.002},
issn = {10902457},
journal = {Explorations in Economic History},
number = {October 2018},
pages = {101277},
publisher = {Elsevier Inc.},
title = {{Railroads and Rural Industrialization: evidence from a Historical Policy Experiment}},
url = {https://doi.org/10.1016/j.eeh.2019.06.002},
volume = {74},
year = {2019}
}

@misc{ThistedAmtsavis1834,
author = {Amtsavis, Thisted},
booktitle = {Thisted Amtsavis},
title = {{Thisted Amtsavis}},
url = {http://hdl.handle.net/109.3.1/uuid:bef130f1-8d10-4151-a8cf-fb6d80caa0e6},
year = {1834}
}

@article{Redding2008,
author = {Redding, Stephen J. and Sturm, Daniel M},
journal = {American Economic Review},
number = {5},
pages = {1766--1797},
title = {{The Costs of Remoteness: Evidence from German Division and Reunification}},
url = {https://www.jstor.org/stable/29730152},
volume = {98},
year = {2008}
}

@article{Petersen1877,
author = {Petersen, J. Chr.},
journal = {Geografisk Tidskrift},
title = {{Om Aggertangen f{\o}r og nu}},
volume = {1},
year = {1877}
}

@article{Atack2010,
author = {Atack, Jeremy and Bateman, Fred and Haines, Michael and Margo, Robert A},
issn = {01455532, 15278034},
journal = {Social Science History},
number = {2},
pages = {171--197},
publisher = {Cambridge University Press},
title = {{Did Railroads Induce or Follow Economic Growth? Urbanization and Population Growth in the American Midwest, 1850-1860}},
url = {http://www.jstor.org/stable/40587344},
volume = {34},
year = {2010}
}

@book{Mortensen2018,
address = {Aarhus},
author = {Mortensen, Lars Boje},
isbn = {9788791844320},
pages = {100},
publisher = {Aarhus Universitetsforlag},
title = {{Saxo}},
year = {2018}
}

@article{Svalgaard1977,
author = {Svalgaard, Robert},
journal = {{\AA}rbog for Thy, Mors og Vester Hanherred},
pages = {35--57},
title = {{Toldvagtskibe ved Agger Kanal samt toldvagtskibet "Thybor{\o}n"}},
url = {https://www.arkivthy.dk/images/Aarbog/1977/Svalgaard, Robert   Toldvagtskibet ved Agger kanal samt told.pdf https://www.arkivthy.dk/},
year = {1977}
}

@article{chen2016xgboost,
author = {Chen, Tianqi and Guestrin, Carlos},
journal = {CoRR},
title = {{XGBoost: {A} Scalable Tree Boosting System}},
url = {http://arxiv.org/abs/1603.02754},
volume = {abs/1603.0},
year = {2016}
}

@book{Schovelin1891,
author = {Schovelin, Julius},
pages = {1--172},
title = {{Blade af den Danske Dampskibsfartshistorie}},
url = {https://slaegtsbibliotek.dk/905231.pdf},
year = {1891}
}

@article{Berger2017,
author = {Berger, Thor and Enflo, Kerstin},
doi = {10.1016/j.jue.2015.09.001},
issn = {0094-1190},
journal = {Journal of Urban Economics},
pages = {124--138},
publisher = {Elsevier Inc.},
title = {{Locomotives of local growth : The short- and long-term impact of railroads in Sweden R}},
url = {http://dx.doi.org/10.1016/j.jue.2015.09.001},
volume = {98},
year = {2017}
}

@article{Poulsen2022,
author = {Poulsen, Bo},
issn = {2506-6730},
journal = {Journal for the History of Environment and Society},
number = {1},
pages = {129--158},
publisher = {Brepols Publishers},
title = {{Between Adaptation and Mitigation. The Nineteenth-century North Sea Storm Surges and the Entangled Socio-Natural Transformation of the Limfjord Region, Denmark}},
volume = {6},
year = {2022}
}

@article{Donaldson2016,
author = {Donaldson, Dave and Hornbeck, Richard},
doi = {10.1093/qje/qjw002.Advance},
journal = {The Quarterly Journal of Economics},
number = {2},
pages = {799--858},
title = {{Railroad and American Economic Growth: A 'Market Access' Approach}},
volume = {131},
year = {2016}
}

@misc{mathiesen2022linklives,
address = {Denmark},
author = {Mathiesen, N. and Robinson, O. and Thomsen, A. and Revuelta-Eugercios, B.},
publisher = {Danish National Archives/University of Copenhagen},
title = {{Link-Lives Data v.1.2.1}},
url = {https://www.rigsarkivet.dk/udforsk/link-lives-data/},
year = {2022}
}

@article{Spejlborg2012,
author = {Spejlborg, Marie B{\o}nl{\o}kke},
journal = {Danmarkshistorien.dk},
title = {{Knud den Store ca. 995-1035}},
url = {https://danmarkshistorien.dk/vis/materiale/knud-den-store-ca-995-1035/},
year = {2012}
}

@article{Callaway2021did,
author = {Callaway, Brantly and Sant'Anna, Pedro H C},
doi = {https://doi.org/10.1016/j.jeconom.2020.12.001},
issn = {0304-4076},
journal = {Journal of Econometrics},
number = {2},
pages = {200--230},
title = {{Difference-in-Differences with multiple time periods}},
url = {https://www.sciencedirect.com/science/article/pii/S0304407620303948},
volume = {225},
year = {2021}
}

@book{Trap3,
address = {Copenhagen},
author = {Trap, J. P. and Falbe-Hansen, V. and Westergaard, H and Weitemeyer, H.},
edition = {3rd},
publisher = {Universitetsboghandler G. E. C. Gad},
title = {{Kongeriget Danmark (Trap Danmark 3rd edition)}},
url = {http://runeberg.org/trap/},
year = {1906}
}

@article{Bogart2019,
author = {Bogart, Dan and Lefors, Michael and Satchell, A E M},
doi = {https://doi.org/10.1016/j.eeh.2018.08.005},
issn = {0014-4983},
journal = {Explorations in Economic History},
pages = {1--24},
title = {{Canal carriers and creative destruction in English transport}},
url = {https://www.sciencedirect.com/science/article/pii/S0014498318300263},
volume = {71},
year = {2019}
}

@article{Houtte1966,
author = {Houtte, J A Van},
issn = {00130117, 14680289},
journal = {The Economic History Review},
number = {1},
pages = {29--47},
publisher = {[Economic History Society, Wiley]},
title = {{The Rise and Decline of the Market of Bruges}},
url = {http://www.jstor.org/stable/2592791},
volume = {19},
year = {1966}
}

@article{Poulsen2007,
author = {Poulsen, Bo and Holm, Poul and MacKenzie, Brian R.},
doi = {10.1016/j.fishres.2007.07.014},
issn = {01657836},
journal = {Fisheries Research},
number = {2-3},
pages = {181--195},
title = {{A long-term (1667-1860) perspective on impacts of fishing and environmental variability on fisheries for herring, eel, and whitefish in the Limfjord, Denmark}},
volume = {87},
year = {2007}
}

@book{ThistedLokalhistorie1974,
address = {Thisted},
author = {Balle, Torsten and S{\o}ndberg, Bjarne and Bjerregaard, Mogens and Svalgaard, Robert and Miltersen, J{\o}rgen and J{\o}rgensen, Marius},
publisher = {Historisk Samfund for Thy og Mors, Thisted Bogtrykkeri},
title = {{Thisted k{\o}bstads historie - Historisk {\AA}rbog for Thy og Mors og Vester Hanherred}},
url = {https://slaegtsbibliotek.dk/911456.pdf},
year = {1974}
}

@article{Christensen2004,
author = {Christensen, J T and Cedhagen, T and Hylleberg, J},
doi = {10.1080/00364820410002640},
issn = {00364827},
journal = {Sarsia},
number = {6},
pages = {379--387},
title = {{Late-Holocene salinity changes in Limfjorden, Denmark}},
volume = {89},
year = {2004}
}

@article{WinnersAndLosers2022,
author = {Boberg-Fazli{\'{c}}, Nina and Lampe, Markus and {Martinelli Lasheras}, Pablo and Sharp, Paul},
doi = {https://doi.org/10.1016/j.jdeveco.2021.102813},
issn = {0304-3878},
journal = {Journal of Development Economics},
pages = {102813},
title = {{Winners and losers from agrarian reform: Evidence from Danish land inequality 1682–1895}},
url = {https://www.sciencedirect.com/science/article/pii/S0304387821001656},
volume = {155},
year = {2022}
}

@book{Aagard1802,
address = {Viborg},
author = {Aagard, Knud},
publisher = {Author's own publisher},
title = {{Physisk, oeconomisk og topographisk Beskrivelse over Thye}},
url = {https://bibliotek.slaegt.dk/cgi-bin/koha/opac-detail.pl?biblionumber=28303},
year = {1802}
}

@article{Hornbeck2019,
author = {Hornbeck, Richard and Rotemberg, Martin},
journal = {NBER Working Paper},
number = {26594},
title = {{Railroads, Reallocation, and the Rise of American Manufacturing}},
url = {https://www.nber.org/papers/w26594},
year = {2019}
}

@article{Pajung2012,
author = {Pajung, Stefan},
journal = {danmarkshistorien.dk},
title = {{Knud den Hellige, ca. 1042-1086}},
url = {https://danmarkshistorien.dk/vis/materiale/knud-den-hellige-ca-1042-1086/},
year = {2012}
}

@article{HeavyPlough2016,
author = {Andersen, Thomas Barnebeck and Jensen, Peter Sandholt and Skovsgaard, Christian Volmar},
doi = {10.1016/j.jdeveco.2015.08.006},
issn = {0304-3878},
journal = {Journal of Development Economics},
pages = {133--149},
publisher = {Elsevier B.V.},
title = {{The heavy plow and the agricultural revolution in Medieval Europe}},
url = {http://dx.doi.org/10.1016/j.jdeveco.2015.08.006},
volume = {118},
year = {2016}
}

@article{Pedersen2019,
author = {Pedersen, S. and Hermansen, B. and Nathan, C. and Tougaard, L.},
journal = {GEUS},
title = {{Surface geology map of Denmark 1:200 000, version 2}},
url = {https://eng.geus.dk/products-services-facilities/data-and-maps/maps-of-denmark},
year = {2019}
}

@article{Degn1989,
author = {Degn, Ole},
journal = {Historie/Jyske Samlinger},
title = {{Byer, byhierarkier og byudvikling}},
url = {https://tidsskrift.dk/historiejyskesamling/article/view/40197},
year = {1989}
}

@article{Rasmussen1966,
author = {Rasmussen, Holger},
journal = {Skalk},
number = {5},
pages = {18--27},
title = {{Vel kaldes det tilsammen Limfjorden}},
year = {1966}
}

@article{Charlier2011,
author = {Charlier, Roger H.},
doi = {10.2112/10A-00003.1},
issn = {07490208},
journal = {Journal of Coastal Research},
number = {4},
pages = {746--756},
title = {{The Zwin: From golden inlet to nature reserve}},
volume = {27},
year = {2011}
}

@article{Barjamovic2019,
author = {Barjamovic, Gojko and Chaney, Thomas and Coşar, Kerem and Horta{\c{c}}su, Ali},
doi = {10.1093/qje/qjz009},
issn = {0033-5533},
journal = {The Quarterly Journal of Economics},
number = {3},
pages = {1455--1503},
title = {{Trade, Merchants, and the Lost Cities of the Bronze Age}},
url = {https://doi.org/10.1093/qje/qjz009},
volume = {134},
year = {2019}
}

@article{Marczinek2022,
author = {Marczinek, Maximilian and Maurer, Stephan E and Rauch, Ferdinand},
journal = {CEPR Discussion paper},
number = {DP16957},
title = {{Trade Persistence and Trader Identity - Evidence from the Demise of the Hanseatic League}},
url = {https://cepr.org/active/publications/discussion_papers/dp.php?dpno=16957},
year = {2022}
}

@article{SantosSilva2022,
author = {{Santos Silva}, J M C and Tenreyro, Silvana},
doi = {10.1007/s10258-021-00203-w},
issn = {1617-9838},
journal = {Portuguese Economic Journal},
title = {{The Log of Gravity at 15}},
url = {https://doi.org/10.1007/s10258-021-00203-w},
year = {2022}
}

@article{Lampe2015DanesUK,
author = {Lampe, Markus and Sharp, Paul},
doi = {10.1093/ereh/hev013},
issn = {1361-4916},
journal = {European Review of Economic History},
number = {4},
pages = {432--453},
title = {{How the Danes discovered Britain: the international integration of the Danish dairy industry before 1880}},
url = {https://doi.org/10.1093/ereh/hev013},
volume = {19},
year = {2015}
}

@book{pedersen2014,
author = {Pedersen, Anne and Wilson, David M},
isbn = {9788788415872},
publisher = {National Museum of Denmark, Jutland Archaeological Society},
title = {{Aggersborg - The Viking-Age settlement and fortress}},
year = {2014}
}

@article{Bogart2022,
author = {Bogart, Dan and You, Xuesheng and Alvarez-Palau, Eduard J and Satchell, Max and Shaw-Taylor, Leigh},
doi = {https://doi.org/10.1016/j.jue.2021.103390},
issn = {0094-1190},
journal = {Journal of Urban Economics},
pages = {103390},
title = {{Railways, divergence, and structural change in 19th century England and Wales}},
url = {https://www.sciencedirect.com/science/article/pii/S0094119021000723},
volume = {128},
year = {2022}
}

@book{Diamond1997,
address = {New York, London},
author = {Diamond, Jared},
edition = {1st},
isbn = {0-393-31755-2},
publisher = {W. W. Norton \& Company},
title = {{Guns, Germs and Steel - The Fates of Human Societies}},
year = {1997}
}

@article{Silva2006,
author = {Silva, J.M.C. Santos and Tenreyro, Silvana},
journal = {The Review of Economics and Statistics},
number = {4},
pages = {641--658},
title = {{The Log of Gravity}},
url = {https://www.jstor.org/stable/40043025},
volume = {88},
year = {2006}
}

@misc{ViborgStift1852,
author = {{Viborg Stiftstidende}},
booktitle = {Viborg Stiftstidende},
title = {{Viborg Stiftstidende}},
url = {http://hdl.handle.net/109.3.1/uuid:2980a160-6418-42a7-a227-df25503f6134},
year = {1852}
}

@article{Klem1967,
author = {Klem, Knud},
journal = {M/S Museet for S{\o}farts {\aa}rbog},
pages = {125--152},
title = {{Hjuldamperen IRIS og Limfjordsfarten}},
url = {https://tidsskrift.dk/mfs_aarbog/article/view/96199},
volume = {26},
year = {1967}
}

@misc{roth2023loglike,
archivePrefix = {arXiv},
arxivId = {econ.EM/2212.06080},
author = {Chen, Jiafeng and Roth, Jonathan},
eprint = {2212.06080},
primaryClass = {econ.EM},
title = {{Log-like? Identified ATEs defined with zero-valued outcomes are (arbitrarily) scale-dependent}},
url = {http://www.jonathandroth.com},
year = {2023}
}

@article{Santanna2020DRDID,
author = {Sant'Anna, Pedro H C and Zhao, Jun},
doi = {https://doi.org/10.1016/j.jeconom.2020.06.003},
issn = {0304-4076},
journal = {Journal of Econometrics},
number = {1},
pages = {101--122},
title = {{Doubly robust difference-in-differences estimators}},
url = {https://www.sciencedirect.com/science/article/pii/S0304407620301901},
volume = {219},
year = {2020}
}

@book{sildeboom2022,
author = {{\O}rnbjerg, Jakob},
editor = {Oldrup, Thomas},
isbn = {978 87 7219 680 0},
pages = {100},
publisher = {Aarhus Universitetsforlag},
title = {{Det vilde sildeboom 1703}},
url = {https://unipress.dk/udgivelser/d/det-vilde-sildeboom/},
year = {2022}
}

@article{Bakker2021Phonecians,
author = {Bakker, Jan David and Maurer, Stephan and Pischke, J{\"{o}}rn-Steffen and Rauch, Ferdinand},
doi = {10.1162/rest_a_00902},
issn = {0034-6535},
journal = {The Review of Economics and Statistics},
number = {October},
pages = {1--14},
title = {{Of Mice and Merchants: Connectedness and the Location of Economic Activity in the Iron Age}},
volume = {103},
year = {2021}
}

@article{Dalgaard2020,
  author    = {Dalgaard, Carl-Johan and Knudsen, Anne Sofie B. and Selaya, Pablo},
  title     = {{The bounty of the sea and long-run development}},
  journal   = {Journal of Economic Growth},
  volume    = {25},
  number    = {3},
  pages     = {259--295},
  year      = {2020},
  doi       = {10.1007/s10887-020-09181-8},
  url       = {https://doi.org/10.1007/s10887-020-09181-8},
  issn      = {1573-7020}
}

@misc{MinisterietforFodevarer2022,
author = {{Ministry of Food}, Agriculture and Fisheries},
booktitle = {Fishing statistics},
title = {{Fiskeristatistik - dynamisk landingstabel}},
url = {https://fiskeristyrelsen.dk/fiskeristatistik},
year = {2022}
}

@article{Lassen1883,
author = {Lassen, K C},
journal = {Tidsskrift for Land{\o}konomi},
number = {65},
title = {{Udviklingen af Dampskibstrafikken fra Danmark til}},
url = {https://tidsskrift.dk/tidsskriftlandoekonomi/issue/view/6869},
year = {1883}
}

@article{Davis2002,
author = {Davis, Donald R and Weinstein, David E},
doi = {10.1257/000282802762024502},
journal = {American Economic Review},
month = {dec},
number = {5},
pages = {1269--1289},
title = {{Bones, Bombs, and Break Points: The Geography of Economic Activity}},
url = {https://www.aeaweb.org/articles?id=10.1257/000282802762024502},
volume = {92},
year = {2002}
}

@misc{RoskildeAmt1836,
author = {{Roskilde Amts og Advertissementstidende}},
booktitle = {Roskilde Amts og Advertissementstidende},
title = {{Roskilde Amts og Advertissementstidende}},
url = {http://hdl.handle.net/109.3.1/uuid:ae3932f7-2868-4c8e-9524-d3115a891af3},
year = {1836}
}

@book{Bergsoee1844,
address = {Copenhagen},
author = {Bergs{\o}e, Adolph Frederik},
publisher = {Author's own press},
title = {{Den danske Stats Statistik, vol I}},
year = {1844}
}

@article{Harris1954,
author = {Harris, Chauncy D},
journal = {Annals of the Association of American Geographers},
number = {4},
pages = {315--348},
title = {{The Market as a Factor in the Localization of Industry in the United States}},
volume = {44},
year = {1954}
}

@article{Atack2008,
author = {Atack, Jeremy and Haines, Michael and Margo, Robert},
doi = {10.11126/stanford/9780804771856.003.0007},
isbn = {9780804771856},
journal = {National Bureau of Economic Research, Inc, NBER Working Papers},
title = {{Railroads and the Rise of the Factory: Evidence for the United States, 1850-70}},
year = {2008}
}

@book{saxo,
author = {Grammaticus, Saxo},
title = {{Gesta Danorum (Translation by Grundtvig 1818)}},
url = {http://runeberg.org/saxo/},
year = {1208}
}

@article{Ravn1993,
author = {Ravn, Thomas Bloch},
journal = {Skalk},
number = {3},
pages = {18--24},
title = {{Gennembrud}},
year = {1993}
}

@article{rauch2022a,
author = {Rauch, F and Maurer, S},
doi = {https://doi.org/10.1093/oep/gpac009},
journal = {Oxford Economic Papers},
publisher = {Oxford University Press},
title = {{Economic geography aspects of the Panama Canal}},
year = {2022}
}

@article{Bosker2022,
author = {Bosker, Maarten},
doi = {https://doi.org/10.1016/j.regsciurbeco.2021.103677},
issn = {0166-0462},
journal = {Regional Science and Urban Economics},
pages = {103677},
title = {{City origins}},
url = {https://www.sciencedirect.com/science/article/pii/S0166046221000375},
volume = {94},
year = {2022}
}

@book{Poulsen2019,
address = {Aarhus},
author = {Poulsen, Bo},
edition = {1st},
editor = {Bejder, Peter},
isbn = {978 87 7184 823 6},
pages = {100},
publisher = {Aarhus Universitetsforlag},
title = {{Stormflod}},
url = {https://unipress.dk/bogserier/100-danmarkshistorier/},
year = {2019}
}

@article{Henderson2018,
author = {Henderson, J. Vernon and Squires, Tim and Storeygard, Adam and Weil, David},
doi = {10.1093/qje/qjx030.Advance},
journal = {Quarterly journal of economics},
number = {November},
pages = {1--50},
title = {{THE GLOBAL DISTRIBUTION OF ECONOMIC ACTIVITY: NATURE, HISTORY, AND THE ROLE OF TRADE}},
url = {http://www.lse.ac.uk/geography-and-environment/assets/Documents/THE-GLOBAL-DISTRIBUTION-OF-ECONOMIC-ACTIVITY.pdf},
year = {2018}
}

@article{Ortman2020,
author = {Ortman, Scott and Lobo, Jos{\'{e}}},
doi = {10.1126/sciadv.aba5694},
issn = {23752548},
journal = {Science Advances},
number = {25},
pmid = {32596462},
title = {{Smithian growth in a nonindustrial society}},
volume = {6},
year = {2020}
}

@article{Turnbull1987,
author = {Turnbull, Gerard},
issn = {00130117, 14680289},
journal = {The Economic History Review},
number = {4},
pages = {537--560},
publisher = {[Economic History Society, Wiley]},
title = {{Canals, Coal and Regional Growth during the Industrial Revolution}},
url = {http://www.jstor.org/stable/2596392},
volume = {40},
year = {1987}
}

@article{Dijkstra1959,
author = {Dijkstra, E W},
doi = {10.1007/BF01386390},
issn = {0945-3245},
journal = {Numerische Mathematik},
number = {1},
pages = {269--271},
title = {{A note on two problems in connexion with graphs}},
url = {https://doi.org/10.1007/BF01386390},
volume = {1},
year = {1959}
}

@article{Roesdahl2009,
author = {Roesdahl, Else},
journal = {danmarkshistorien.dk},
title = {{De sidste {\aa}rtier af vikingetiden}},
url = {https://danmarkshistorien.dk/perioder/vikingetiden-ca-800-1050/de-sidste-aartier-af-vikingetiden/},
year = {2009}
}

@article{gobel2010oresundstolden,
author = {G{\o}bel, Erik},
journal = {M/S Museet for S{\o}farts {\aa}rbog},
pages = {41--72},
title = {Oeresundstolden og dens regnskaber 1497-1857},
volume = {69},
year = {2010}
}

@misc{Britannica2018,
author = {Britannica},
booktitle = {Encyclopedia Britannica},
title = {{Kaliningrad}},
url = {https://www.britannica.com/place/Kaliningrad},
year = {2018}
}

@article{Christensen1735,
author = {Christensen, Christian},
journal = {Historie/Jyske Samlinger},
title = {{Tilstanden i Thy og paa Mors ved Aar 1735}},
url = {http://img.kb.dk/tidsskriftdk/pdf/ho/ho_4rk_0003-PDF/ho_4rk_0003_75584.pdf},
volume = {4. r{\ae}kke,},
year = {1735}
}

@book{Dioerup1842Thisted,
author = {Di{\o}rup, C},
publisher = {Tr.},
series = {Bidrag til Kundskab om de danske Provindsers naervaerende Tilstand i oekonomisk Henseende},
title = {{Thisted Amt}},
url = {https://books.google.co.uk/books?id=FOdRAAAAcAAJ},
year = {1842}
}

@article{Feyrer2021,
author = {Feyrer, James},
doi = {10.1016/j.jdeveco.2021.102708},
issn = {03043878},
journal = {Journal of Development Economics},
number = {May},
pages = {102708},
publisher = {Elsevier B.V.},
title = {{Distance, trade, and income — The 1967 to 1975 closing of the Suez canal as a natural experiment}},
url = {https://doi.org/10.1016/j.jdeveco.2021.102708},
volume = {153},
year = {2021}
}

@article{VanEtten2017,
author = {van Etten, Jacob},
doi = {10.18637/jss.v076.i13},
journal = {Journal of Statistical Software},
number = {13},
pages = {1--21},
title = {{R Package gdistance: Distances and Routes on Geographical Grids}},
url = {https://www.jstatsoft.org/index.php/jss/article/view/v076i13},
volume = {76},
year = {2017}
}

@misc{Feldbaek2015,
author = {Feldb{\ae}k, Ole},
booktitle = {Den Store Danske www.lex.dk},
title = {{Engl{\ae}nderkrigene ('The English Wars')}},
url = {https://denstoredanske.lex.dk/Engl{\ae}nderkrigene},
year = {2015}
}

@article{soundtoll_data,
author = {Veluwenkamp, Jan and Woude, Siem},
journal = {Database},
title = {{SoundToll Registers Online}},
url = {www.soundtoll.nl/index.php/en/welkom},
year = {2009}
}

@article{Klemp2016,
author = {Klemp, Marc and M{\o}ller, Niels Framroze},
issn = {03470520, 14679442},
journal = {The Scandinavian Journal of Economics},
number = {4},
pages = {841--867},
publisher = {[Wiley, The Scandinavian Journal of Economics]},
title = {{Post-Malthusian Dynamics in Pre-Industrial Scandinavia}},
url = {http://www.jstor.org/stable/45097668},
volume = {118},
year = {2016}
}

@misc{dahl2024breaking,
      title={Breaking the HISCO Barrier: Automatic Occupational Standardization with OccCANINE}, 
      author={Christian Møller Dahl and Torben Johansen and Christian Vedel},
      year={2024},
      eprint={2402.13604},
      archivePrefix={arXiv},
      primaryClass={cs.CL},
      url={https://arxiv.org/abs/2402.13604}, 
}

@techreport{Seror2020Random,
address = {Montr\'{e}al},
author = {Marlon Seror},
copyright = {http://www.econstor.eu/dspace/Nutzungsbedingungen},
language = {eng},
number = {2020-17},
publisher = {Universit\'{e} du Qu\'{e}bec \`{a} Montr\'{e}al, \'{E}cole des sciences de la gestion (ESG UQAM), D\'{e}partement des sciences \'{e}conomiques},
title = {Random river: Trade and rent extraction in imperial China},
type = {Document de travail},
url = {http://hdl.handle.net/10419/234817},
year = {2020}
}

@article{Allen2023,
Author = {Allen, Robert C. and Bertazzini, Mattia C. and Heldring, Leander},
Title = {{The Economic Origins of Government}},
Journal = {American Economic Review},
Volume = {113},
Number = {10},
Year = {2023},
Month = {October},
Pages = {2507-45},
DOI = {10.1257/aer.20201919},
URL = {https://www.aeaweb.org/articles?id=10.1257/aer.20201919}
}

@article{Hornung2015,
 ISSN = {15424766, 15424774},
 URL = {http://www.jstor.org/stable/24539267},
 author = {Erik Hornung},
 journal = {Journal of the European Economic Association},
 number = {4},
 pages = {699--736},
 publisher = {Oxford University Press},
 title = {RAILROADS AND GROWTH IN PRUSSIA},
 urldate = {2024-03-17},
 volume = {13},
 year = {2015}
}

@article{Ahlfeldt2015,
author = {Ahlfeldt, Gabriel M. and Redding, Stephen J. and Sturm, Daniel M. and Wolf, Nikolaus},
title = {{The Economics of Density: Evidence From the Berlin Wall}},
journal = {Econometrica},
volume = {83},
number = {6},
pages = {2127-2189},
doi = {https://doi.org/10.3982/ECTA10876},
url = {https://onlinelibrary.wiley.com/doi/abs/10.3982/ECTA10876},
eprint = {https://onlinelibrary.wiley.com/doi/pdf/10.3982/ECTA10876},
year = {2015}
}

@article{Matranga2024,
    author = {Matranga, Andrea},
    title = "{The Ant and the Grasshopper: Seasonality and the Invention of Agriculture}",
    journal = {The Quarterly Journal of Economics},
    pages = {qjae012},
    year = {2024},
    issn = {0033-5533},
    doi = {10.1093/qje/qjae012},
    url = {https://doi.org/10.1093/qje/qjae012},
    eprint = {https://academic.oup.com/qje/advance-article-pdf/doi/10.1093/qje/qjae012/57352214/qjae012.pdf},
}

@article{dk_hisco_data,
author = {Vedel, Christian and Dahl, Christian Møller and Johansen, Torben S. D.},
publisher = {Harvard Dataverse},
journal = {Harvard Dataverse},
title = {{HISCO codes for Danish Census data}},
year = {2024},
version = {V3},
doi = {10.7910/DVN/WZILNI}
}

@book{leeuwen2002hisco,
  title={{HISCO: Historical International Standard Classification of Occupations}},
  author={Leeuwen, M.H.D. and Maas, I. and Miles, A.},
  isbn={9789058671967},
  lccn={2002494856},
  url={https://books.google.dk/books?id=EMPtAAAAIAAJ},
  year={2002},
  publisher={Leuven University Press}
}

@article{boehm2024,
  author = {Boehm, J. and Chaney, T.},
  title = {{Trade and the End of Antiquity}},
  year = {2024},
  institution = {CEPR Press},
  number = {No. 19459},
  address = {Paris \& London},
  journal = {CEPR discussion paper},
  url = {https://cepr.org/publications/dp19459}
}

@article{roth2023trendingDiD,
title = {What’s trending in difference-in-differences? A synthesis of the recent econometrics literature},
journal = {Journal of Econometrics},
volume = {235},
number = {2},
pages = {2218-2244},
year = {2023},
issn = {0304-4076},
doi = {https://doi.org/10.1016/j.jeconom.2023.03.008},
url = {https://www.sciencedirect.com/science/article/pii/S0304407623001318},
author = {Jonathan Roth and Pedro H.C. Sant’Anna and Alyssa Bilinski and John Poe}
}

@article{kavan2025new,
  author    = {Kavan, J. and Szczypińska, M. and Kochtitzky, W. and others},
  title     = {New coasts emerging from the retreat of Northern Hemisphere marine-terminating glaciers in the twenty-first century},
  journal   = {Nature Climate Change},
  volume    = {15},
  pages     = {528--537},
  year      = {2025},
  doi       = {10.1038/s41558-025-02282-5},
}

@misc{MirakletPaaHeden2025,
  author    = {Kristine Holm-Jensen},
  title     = {Miraklet på heden – Den midtjyske tekstilindustri},
  year      = {2025},
  url       = {https://danmarkshistorien.lex.dk/Miraklet_p%C3%A5_heden_%E2%80%93_Den_midtjyske_tekstilindustri},
  note      = {Accessed: 2025-05-17},
  publisher = {Danmarkshistorien – Lex.dk}
}

@misc{gorges2025tracksmodernity,
      title={{Tracks to Modernity: Railroads, Growth, and Social Movements in Denmark}}, 
      author={Tom Görges and Magnus Ørberg Rove and Paul Sharp and Christian Vedel},
      year={2025},
      eprint={2502.21141},
      archivePrefix={arXiv},
      primaryClass={econ.GN},
      url={https://arxiv.org/abs/2502.21141}, 
}

@article{gibbons2024,
title = {The spatial impacts of a massive rail disinvestment program: The Beeching Axe},
journal = {Journal of Urban Economics},
volume = {143},
pages = {103691},
year = {2024},
issn = {0094-1190},
doi = {https://doi.org/10.1016/j.jue.2024.103691},
url = {https://www.sciencedirect.com/science/article/pii/S0094119024000615},
author = {Stephen Gibbons and Stephan Heblich and Edward W. Pinchbeck}
}

@book{Schade1811Mors,
  title     = {Beskrivelse over {\O}en Mors. Med Kobbere},
  author    = {Schade, C.},
  year      = {1811},
  publisher = {trykt \ldots{} hos Albert Borch},
  url       = {https://books.google.dk/books?id=OuIXXij--DYC}
}

\normalsize
\newpage
\setcounter{table}{0}
\setcounter{figure}{0}
\setcounter{section}{0}
\renewcommand*{\thesection}{\Alph{section}}
\renewcommand{\thefigure}{A\arabic{figure}}
\renewcommand{\thetable}{A\arabic{table}}
\pagenumbering{roman}

\addcontentsline {toc}{part}{APPENDIX}

\begin{title}

    \begin{center}
        
        \LARGE
        \textbf{Online Appendix} \\
        \Large
        A Perfect Storm: First-Nature Geography and Economic Development \\
        
        \vspace{0.5cm}
        \large
        Christian Vedel, University of Southern Denmark,\\
        \small
        \vspace{0.25cm}
        christian-vs@sam.sdu.dk; 
        
    \end{center}

    \localtableofcontents 
        
    \vfill
    
\end{title}

\section{Extended historical background} \label{app:hist}

This appendix provides additional detail on the setting, the closure, the pre-1825 economic baseline, and the infrastructure response to the breach.

The Limfjord sits at the heart of Scandinavia, positioned between modern-day Norway and Sweden, and during the Viking Age its dual openings made it a safe shortcut for ships and Viking expeditions.\footnote{Other examples include Gothenburg, Trondheim, Stavanger, Odense, and Roskilde.} At its center, Aggersborg grew prominent as the site of one of the largest ring castles in Scandinavia, built in the 10th century and symbolizing the consolidation of power by the emerging Danish state \citep{pedersen2014}.

The last historical evidence of a western opening dates to 1085, when King Canute IV assembled the last Viking fleet in the Limfjord to press his claim to the English throne --- following the precedent of his great-uncle Canute II (``the great''), who once ruled a North Sea empire encompassing England, Denmark, Norway, and parts of Sweden \citep{Spejlborg2012}. His soldiers rebelled, the fleet never sailed, and Canute IV was killed in Saint Alban's Church in Odense \citep{Pajung2012}. Danish Vikings never again attempted to conquer England \citep{Roesdahl2009}. The channel closed sometime between 1086 and 1208, a range established by cross-referencing Saxo \citep{saxo} [book XIII, section 5], \citet{Mortensen2018}, and geological evidence from \citet{Christensen2004}. The potential of a western reopening was not lost on contemporaries: as early as 1811, a local observer speculated in print about what would happen if the sea naturally broke through the isthmus again \citep[pp.~76--78]{Schade1811Mors}. That possibility became reality fourteen years later.

By 1672, Aalborg's dominance was complete: it was the largest Danish market town after Copenhagen, while the west lagged far behind \citep{Degn1989}. The institutional apparatus that enforced this geography is visible in the court record. Western market towns repeatedly sought the right to trade independently of Aalborg merchants across the 16th and 17th centuries; they lost every time \citep[pp.~78--89]{ThistedLokalhistorie1974}. One argument advanced in these cases was that Lübeck did not constitute a foreign country --- and that direct trade with it should therefore be permitted without Aalborg's intermediation, suggesting how embedded the salt-for-herring relationship had become. In practice, western towns had to route their trade eastward via the difficult Løgstør shallows, paying local boatmen to transfer cargo, before reaching Aalborg (see Figure~\ref{fig:main_map}). Only limited trade occurred along the western routes \citep{Poulsen2019}. In 1800, Thisted shipped 6,993 barrels of barley and 6,832 barrels of oats: 31 and 47 percent respectively came out directly via the open coast beaches --- a hazardous shortcut --- while the rest went via Aalborg \citep[p.~30, p.~234]{Aagard1802, Christensen1735}. Most of this trade went to Norway in exchange for construction timber. The export growth documented in the main paper is measured against these 1800 figures.

The sparsely populated Agger parish, with only 388 inhabitants in the 1801 census, bore the breach directly: residents were offered money to relocate to safer areas, and many settled further into the Limfjord \citep{Poulsen2019, Poulsen2022}. Newspaper reports from the period document the subsequent arrival of ships from England, marking what contemporaries called a golden age of trade.\footnote{Example newspaper reports can be found in \citet{ThistedAmtsavis1834, RoskildeAmt1836, ViborgStift1852}.}

The breach set off a sustained infrastructure response. Fearing the Agger channel would silt up, authorities attempted to deepen the Løgstør passage in 1843 \citep[p.~311, p.~4]{Bergsoee1844}; the effort failed. They commissioned the Frederik VII canal instead, completed in 1861 \citep{Petersen1877}. Steamships arrived in 1842, inaugurating the first Danish steamship route to England and connecting Aalborg, Copenhagen, and other Limfjord towns \citep{Klem1967, Schovelin1891, Lampe2015DanesUK}, though the channel's shallow depth capped vessel size \citep{Lassen1883}. As navigational challenges persisted and the Agger channel eventually silted up, Esbjerg emerged as the key port for Danish--British trade \citep{Lampe2015DanesUK}. A second storm in 1862 opened the Thyborøn channel at approximately the same location; it was navigable by 1867 and remains in use today \citep{Petersen1877, Ravn1993}. Coastal groins followed in 1875 \citep{Trap3}; Thyborøn, at the channel's mouth, now accounts for 25 percent of the Danish fishing catch \citep{MinisterietforFodevarer2022}. Throughout the 20th century a relatively strong textile industry also prospered in the region \citep{MirakletPaaHeden2025}.\footnote{Individual port expansions: Thisted received a new port in 1841 \citep[pp.~384--386]{Dioerup1842Thisted}; Struer expanded in 1856 and 1864 \citep[vol.~V, p.~467]{Trap3}; Lemvig built a port in 1857 \citep[vol.~V, p.~474]{Trap3}; Nykøbing Mors purchased its port in 1843 \citep[vol.~IV, p.~214]{Trap3}; Løgstør expanded in 1852 \citep[vol.~IV, pp.~399--400]{Trap3}.}

\section{Details of market access computation} \label{details_ma}

Market access is computed as: 

\begin{equation}
\label{eq:MA2_a}
\textit{MA}_p = \sum_{h \in \mathscr{H}} [CostDist(p, h; \alpha) + 1]^\theta.
\end{equation}

Most of this is defined in the main paper, but here follows details on $\theta$ and CostDist(): $\theta$ determines the spread of the market potential function. It reflects the elasticity to distance. A large absolute value of $\theta$ corresponds to a very localized effect of a change to market potential. The standard $\theta = -1$ is used in the main specification.  This is the original value suggested by \cite{Harris1954}, but is also used in \cite{rauch2022a} and is very close to what is empirically estimated by \cite{Redding2008}. However, it is plausible that other values are more appropriate e.g. $\theta = -8$ as estimated by \cite{Donaldson2016}. Robustness checks with $\theta \in (-1, -2, -4, -8, -16)$ can be found in this appendix. It makes no qualitative difference in the conclusions.  

The change in market access before and after the breach comes from changing the definition of $\mathscr{H}$. We can think of $\textit{MA}_p$ as a function for each parish, $p$, that maps a set of ports to a measure of market access:

\begin{equation}
\label{eq:MA99}
\textit{MA}_p = f_p(\mathscr{H}) = \sum_{h \in \mathscr{H}} [CostDist(p, h; \alpha) + 1]^\theta.
\end{equation}

At first $\textit{MA}_p$ is computed with $\mathscr{H}_{\textit{before}}$ defined as the set of effective ports before 1825. Then $\textit{MA}_p$ is recomputed with $\mathscr{H}_{\textit{after}}$ defined as all pre-1825 ports but now with the West and Middle Limfjord ports added. A port is simply a location observed to have at least one ship leaving or arriving in the relevant period in the Sound Toll Register. As such the relevant variable, used in the regression is computed as

\begin{equation}
\label{eq:MA100}
\Delta \log(\textit{MA}_p) = \log(\textit{MA}_{p,\textit{after}}) - \log(\textit{MA}_{p,\textit{before}}),
\end{equation}

\noindent
where

\begin{equation}
\begin{split}
\label{eq:MA101}
\textit{MA}_{p,\textit{after}} &= f_p(\mathscr{H}_{\textit{after}})\\ 
\textit{MA}_{p,\textit{before}} &= f_p(\mathscr{H}_{\textit{before}}).
\end{split}
\end{equation}

The function $CostDist()$ is the result of the following optimization:

\begin{equation}
\label{eq:MA4}
CostDist(x, y):=\min_{r\in \mathscr{R}}\left[Dist^{water}_r(x, y) + \alpha\, Dist^{land}_r(x, y)\right]
\end{equation}

$CostDist(x,y)$ represents the cost of the shortest route $r^*$ between $x$ and $y$ in the set of all possible routes $\mathscr{R}$ given that land travel is $\alpha$ times more expensive than ocean travel. $\alpha = 10$ is used following \cite{Marczinek2022} and \cite{rauch2022a}. However robustness checks are carried out with $\alpha \in (1, 5, 10, 20, 50)$. 

Computing this optimized distance is a computationally hard problem, which was solved via the 'gdistance' R package \citep{VanEtten2017}. This implements Dijkstra's algorithm \citep{Dijkstra1959} which is the standard method for calculating cost distances. The algorithm takes a series of nodes with a given cost between them and finds the least cost path. The nodes in this case are a grid representing Denmark with two node types 'land' and 'water' and the corresponding relative cost to traverse it is $\alpha$ of equation \ref{eq:MA4}. From each grid cell, it is possible to travel to each of the 8 surrounding nodes (neighbouring grid cells) with the mean cost of the two nodes. The grid has a resolution of 500x500 m. This is the highest resolution which was computationally feasible. 

Around Løgstør (see historical background) shallow water forced traders to reload goods onto prams and pay the locals for transport. This area is encoded as having the same cost as land transportation. This is an upper bound on the market powers of the locals. Principally they could charge up to this cost before it would be more profitable to transport goods via land instead.

\section{Population results} \label{extra_pop_res}

\subsection{Full regression table}  \label{pop_reg_table}
\FloatBarrier
\begin{table}[H]
\centering
\caption{Regression results for population size} \label{tab:pop1}
\footnotesize
\begin{tabular}{lcc}
   \tabularnewline \midrule \midrule
   & \multicolumn{2}{c}{log(Pop)}\\
                          & Dummy          & Market Access\\  
                          & (1)            & (2)\\  
   \midrule
   Year 1787 $\times$ Affected  & 0.0113         & 0.0379\\   
                                & (0.0112)       & (0.0626)\\   
   Year 1834 $\times$ Affected  & 0.0178         & -0.0146\\   
                                & (0.0132)       & (0.0748)\\   
   Year 1840 $\times$ Affected  & -0.0059        & -0.2651$^{***}$\\   
                                & (0.0138)       & (0.0759)\\   
   Year 1845 $\times$ Affected  & -0.0012        & -0.2531$^{***}$\\   
                                & (0.0150)       & (0.0775)\\   
   Year 1850 $\times$ Affected  & 0.0040         & -0.2466$^{***}$\\   
                                & (0.0159)       & (0.0900)\\   
   Year 1860 $\times$ Affected  & 0.0366$^{**}$  & 0.0183\\   
                                & (0.0184)       & (0.0972)\\   
   Year 1880 $\times$ Affected  & 0.1411$^{***}$ & 0.8349$^{***}$\\   
                                & (0.0214)       & (0.1155)\\   
   Year 1901 $\times$ Affected  & 0.2392$^{***}$ & 1.591$^{***}$\\   
                                & (0.0282)       & (0.1607)\\   
   \midrule
   Observations                 & 14,301         & 14,301\\  
   \midrule \midrule
\end{tabular}
\parbox{0.9\textwidth}{
\caption*{\textit{Notes:} Cluster-robust standard errors in the parenthesis. Clustered at the parish level. Affected is either a dummy for being in the West Limfjord or improvement in market access, which is indicated by the headers of the results. *** $p< 0.01$ ** $p< 0.05$ * $p< 0.10$. \\ \textit{Source: Danish census data}}
}
\end{table}

\FloatBarrier
\subsection{Population multiverse} \label{pop_multiverse} 
\FloatBarrier

\begin{figure}[H]
    \centering
    \caption{Multiverse of the effect in different comparison groups and parameter choices, 1901}
    \begin{subfigure}[b]{0.45\textwidth}
        \centering
        \caption{Multiverse of control groups\\Dummy approach} \label{fig:mult1_1787}
        \includegraphics[width=\textwidth]{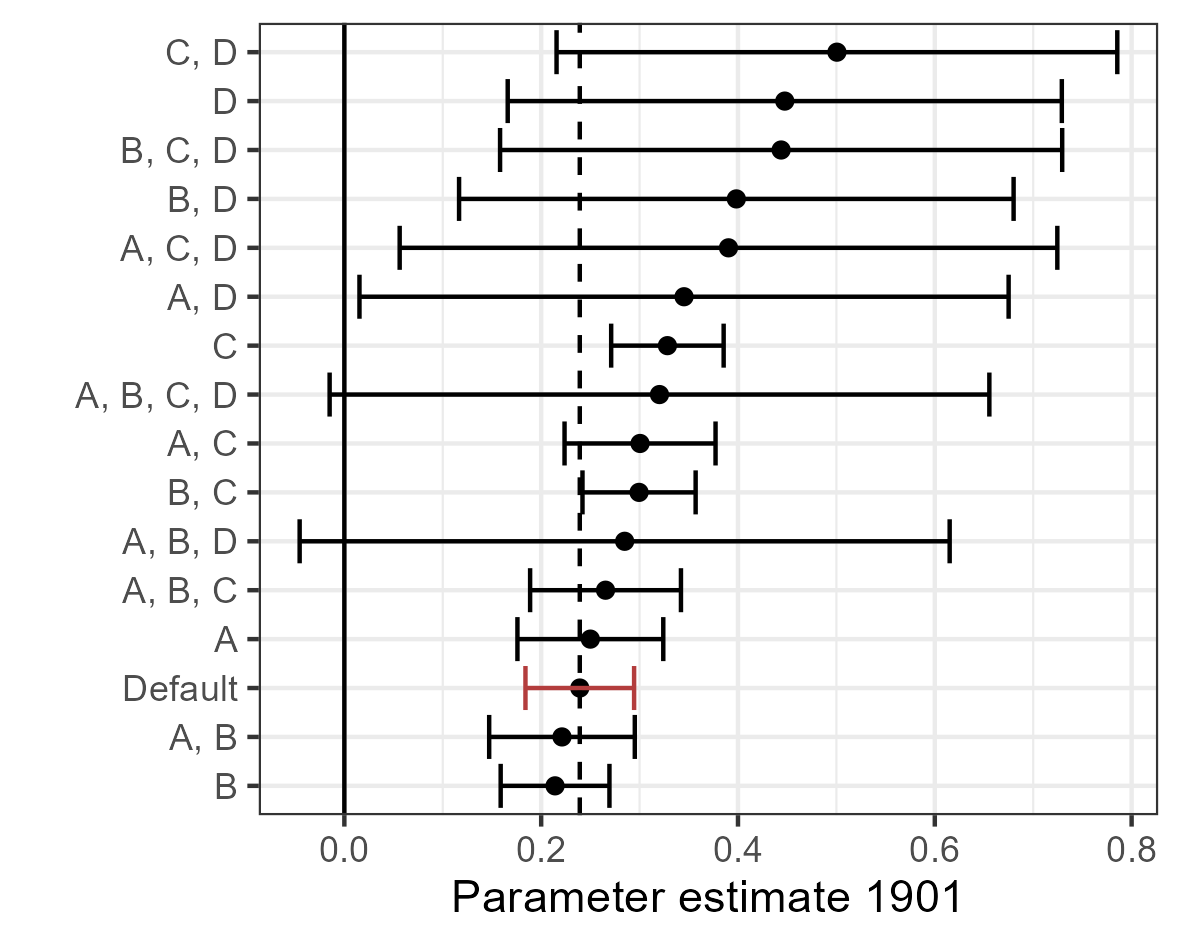}
    \end{subfigure}
    \hfill
    \begin{subfigure}[b]{0.45\textwidth}
        \centering
        \caption{Multiverse of control groups\\Market access approach} \label{fig:mult2_1787}
        \includegraphics[width=\textwidth]{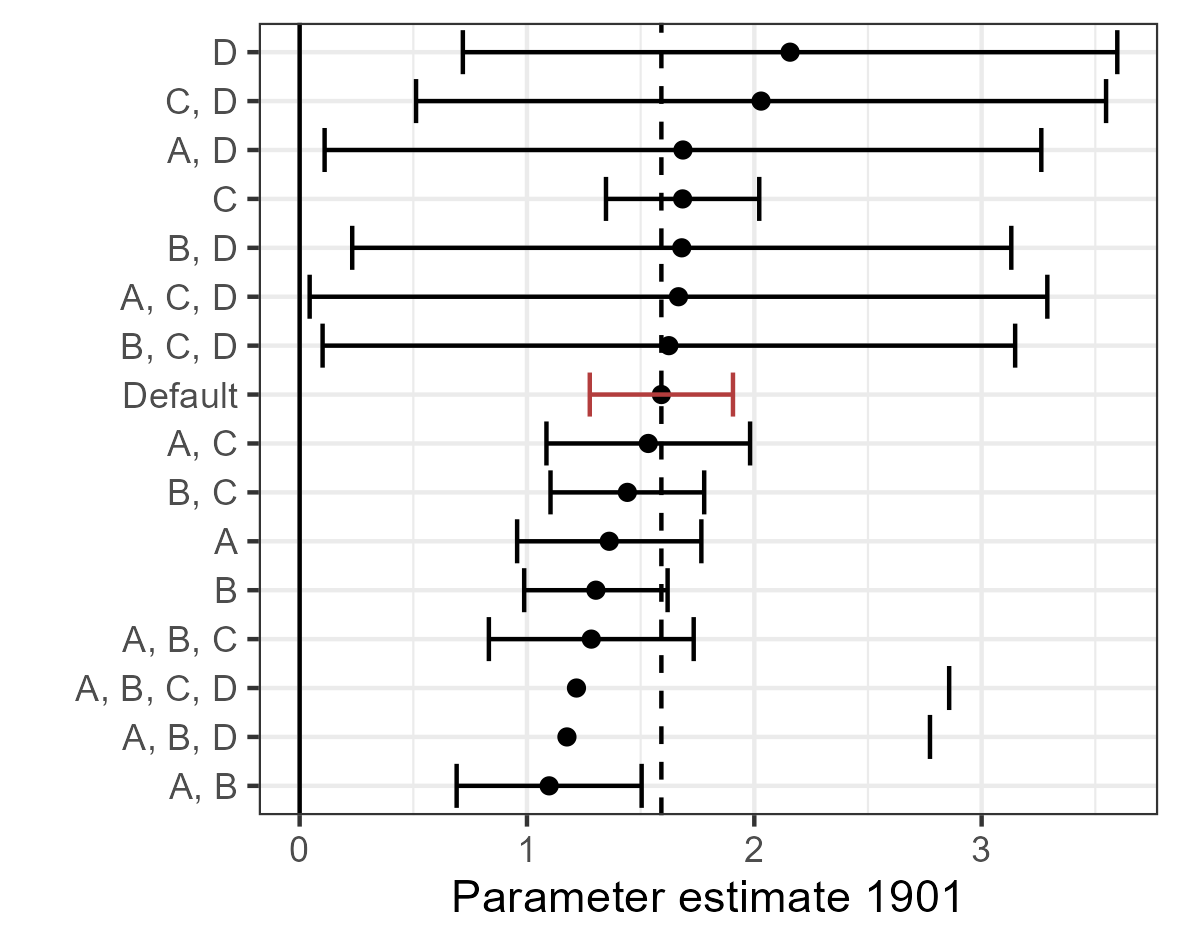}
    \end{subfigure}
    \vspace{0.45cm}
    \begin{subfigure}[b]{0.45\textwidth}
        \centering
        \caption{Multiverse of feasible parameters\\Market access approach} \label{fig:mult3_1787}
        \includegraphics[width=\textwidth]{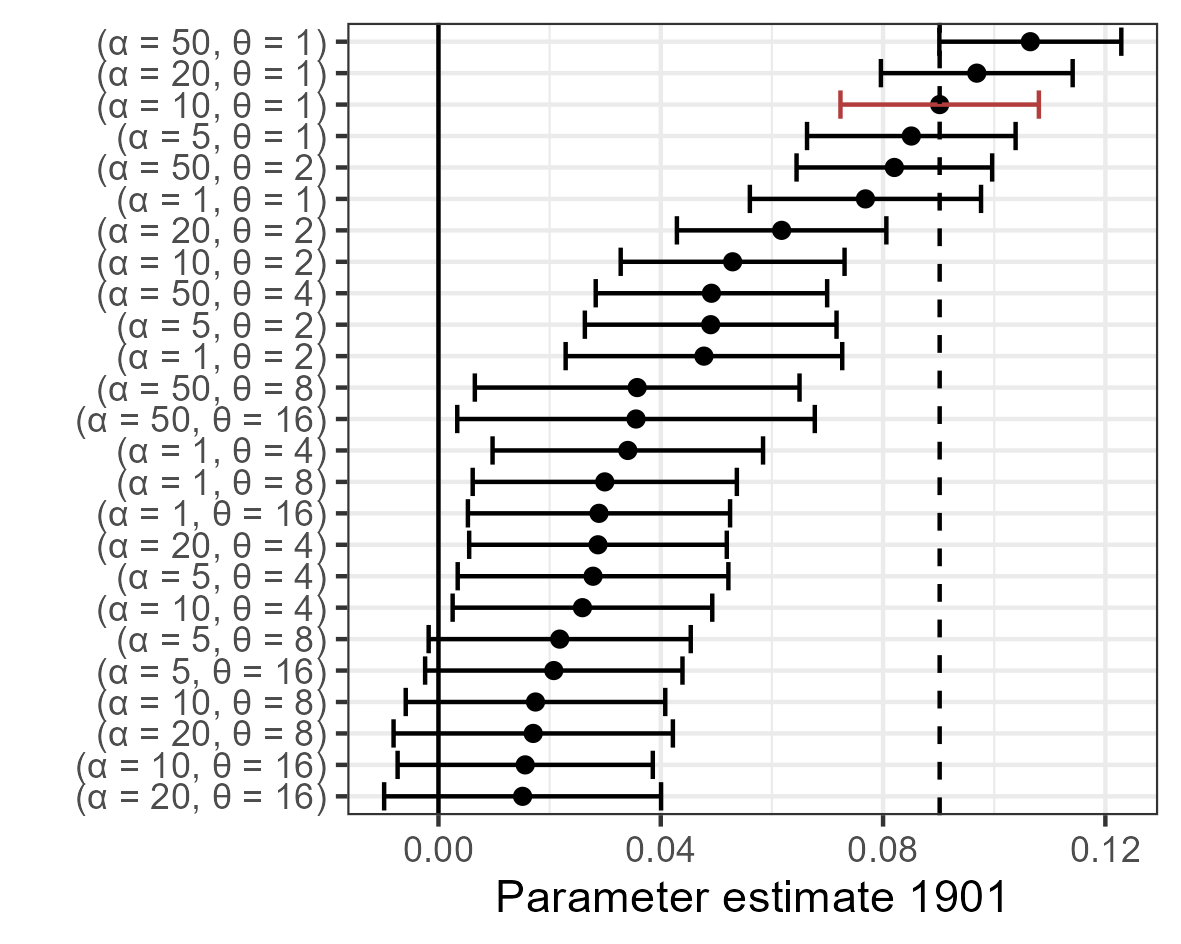}
    \end{subfigure}
    \parbox{0.9\textwidth}{
    \caption*{\footnotesize \textit{Notes:} This is the multiverse of parameter estimates of the effect in 1901 given different feasible choices that could have been made for how to run the analysis. In panel a and panel b, 'A', 'B', and 'C', represents subgroups of the data. 'A' is the result, when the regression is computed using only parishes with a centroid less than 5 km from the coast. 'B' is the subgroups, where all parishes in around Copenhagen are excluded, 'C' is the subgroup of parishes, where the control group does not contain any parishes within 100 km of the Limfjord. 'D' represents the result when using only parishes located within 5 km of a market town. Panel (c) represents the effect given different market access parameters. For enhanced comparability, the log change in market access is standardized to unit variance and zero mean. \\ \textit{Source: Danish census data}}
} \label{fig:pop2_1787}
\end{figure}

\begin{figure}[H]
    \centering
    \caption{Multiverse of the effect in different comparison groups and parameter choices, 1787}
    \begin{subfigure}[b]{0.45\textwidth}
        \centering
        \caption{Multiverse of control groups\\Dummy approach} \label{fig:mult1}
        \includegraphics[width=\textwidth]{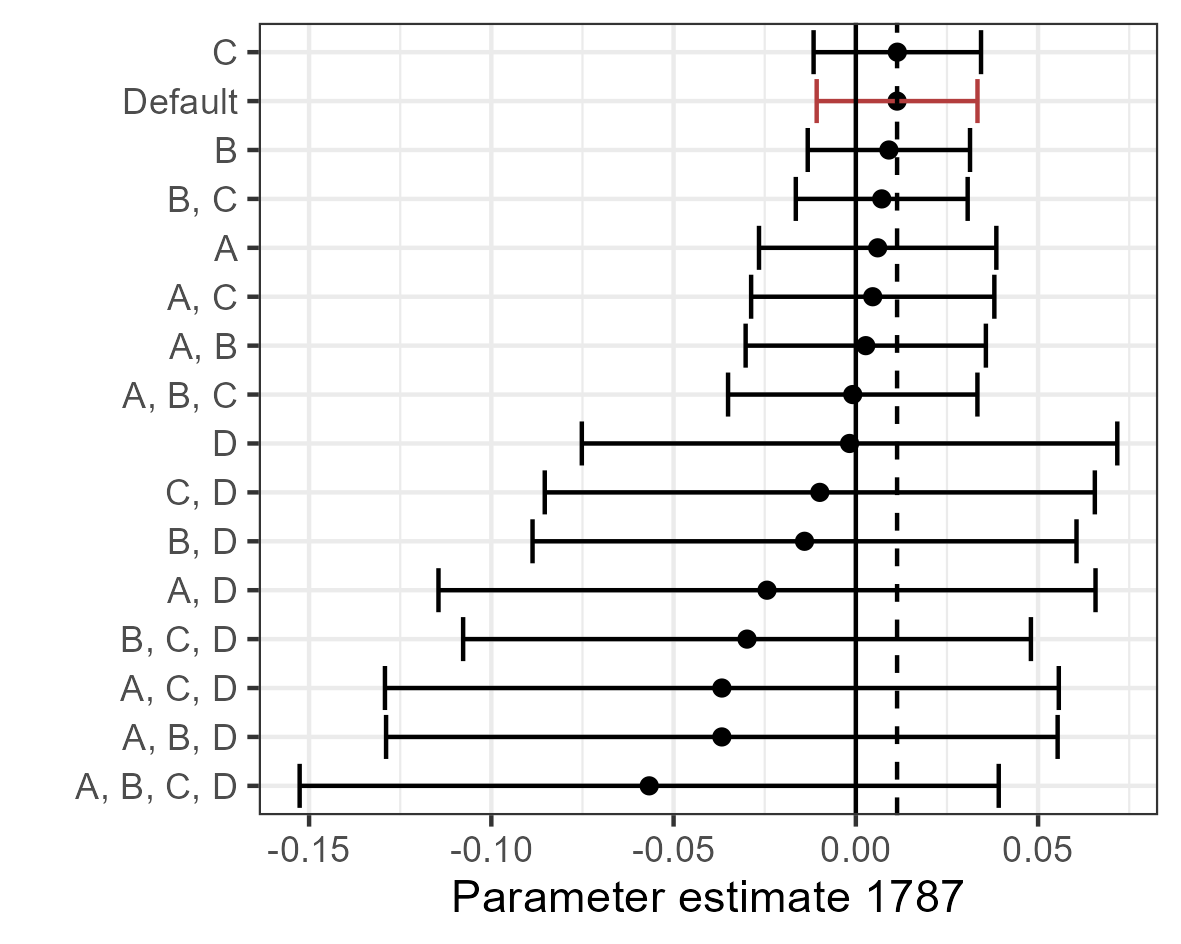}
    \end{subfigure}
    \hfill
    \begin{subfigure}[b]{0.45\textwidth}
        \centering
        \caption{Multiverse of control groups\\Market access approach} \label{fig:mult2}
        \includegraphics[width=\textwidth]{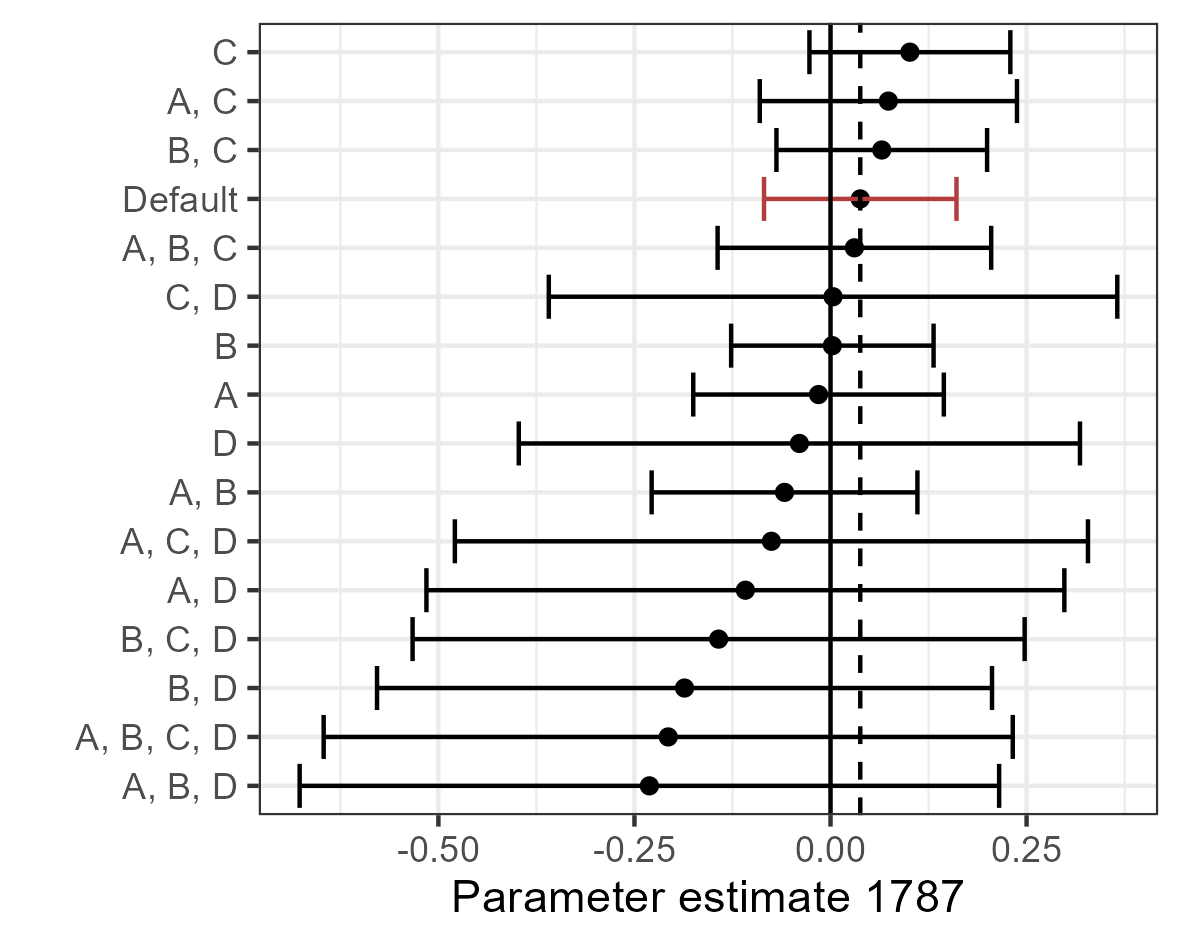}
    \end{subfigure}
    \vspace{0.45cm}
    \begin{subfigure}[b]{0.45\textwidth}
        \centering
        \caption{Multiverse of feasible parameters\\Market access approach} \label{fig:mult3}
        \includegraphics[width=\textwidth]{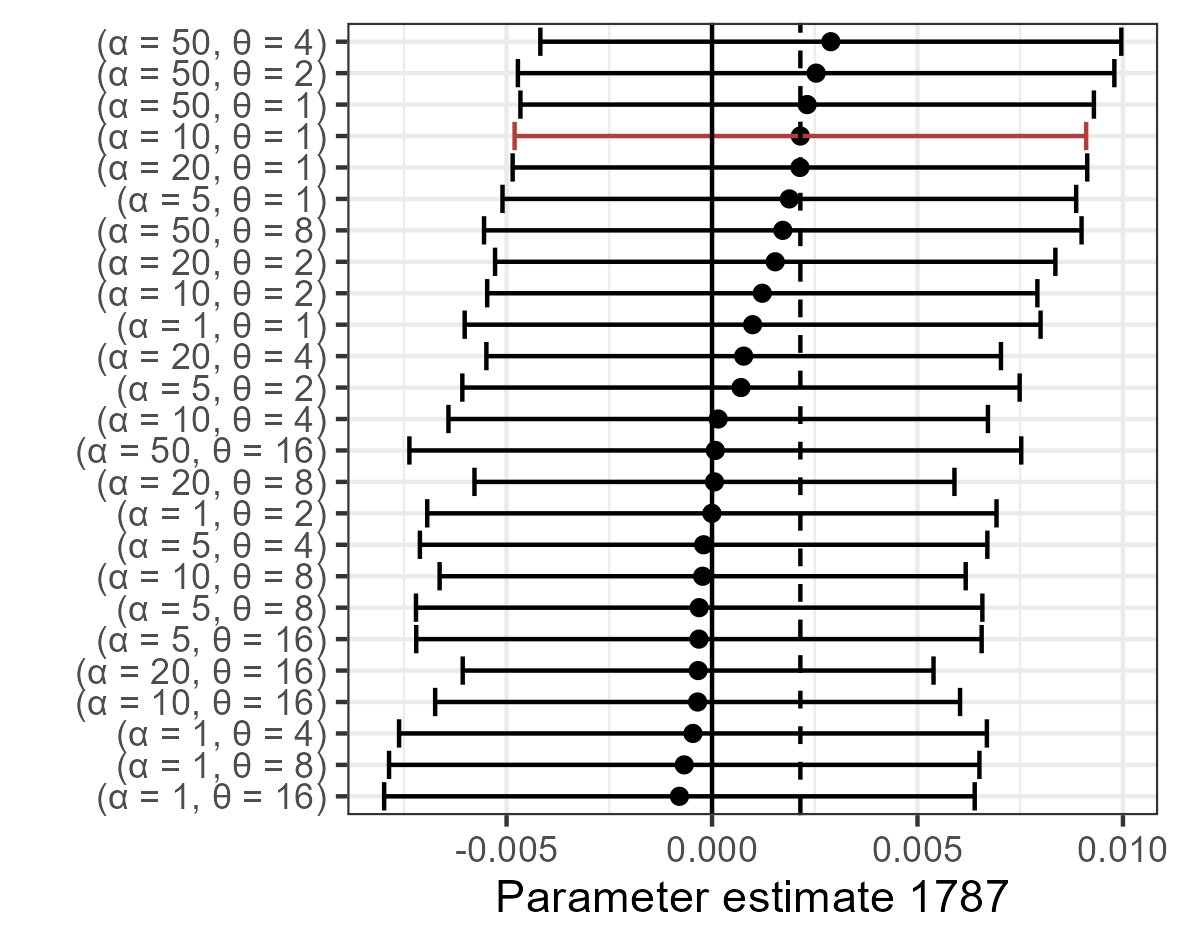}
    \end{subfigure}
    \caption*{Notes: Multiverse results for 1787. If any of these were different from zero it would indicate the existence of pre-trends. The dotted line indicate the default parameter estimate. The solid line is at zero.}
    \label{fig:pop2}
\end{figure}

\FloatBarrier
\subsection{Doubly robust estimates} \label{dr_estimates}
This section contains estimates based on the doubly-robust multi-period did estimator from \cite{Callaway2021did}. Table \ref{tab:cs_estimates} shows the results. As is default in their implementation, the first period is the reference. As consequence, the reference year here is 1787 rather than 1801 from the rest of my paper. To address concern over balance, covariates are included. The doubly robust method combines a propensity score and an outcome regression and is consistent if either of these are correctly specified \citep{Santanna2020DRDID}. Column 1 shows results without covariates. Column 2 shows results including pre-treatment covariates (occupation, young children per woman, and number of people in different age groups).

\begin{table}[H]
\centering
\caption{Callaway and Sant'Anna estimates} \label{tab:cs_estimates}
\footnotesize
\begin{tabular}{lcc}
   \tabularnewline \midrule \midrule
   Outcome: & \multicolumn{2}{c}{log(Population)}\\
            & (1)           & (2)\\  
   \midrule
    1801 & -0.0097 & -0.0215 \\
         & (0.0105) & (0.0137) \\
    1834 & 0.0219 & 0.0021 \\
         & (0.0134) & (0.0155) \\
    1840 & 0.0005 & -0.0113 \\
         & (0.0124) & (0.0162) \\
    1845 & 0.0050 & -0.0043 \\
         & (0.0151) & (0.0163) \\
    1850 & 0.0096 & -0.0056 \\
         & (0.0168) & (0.0184) \\
    1860 & 0.0392 & 0.0110 \\
         & (0.0168) & (0.0200) \\
    1880 & 0.1357* & 0.0844* \\
         & (0.0210) & (0.0241) \\
    1901 & 0.2267* & 0.1558* \\
         & (0.0292) & (0.0323) \\
   \midrule
   Observations & 14,301 & 14,058\\  
   \midrule \midrule
   \multicolumn{3}{l}{\emph{'*' confidence band (95 percent) does not cover 0}}\\
\end{tabular}
\parbox{0.6\textwidth}{
\caption*{\footnotesize \textit{Notes:} Effect using the estimator proposed by Callaway \& Sant’Anna (2021). Column (1) includes no covariates. Column (2) adjusts for demographic covariates, some of which are potentially bad controls, as they might be mediators.  \\ 
\textit{Source: Danish census data}}
}

\end{table}

\section{Mechanisms}

\FloatBarrier
\subsection{All occupational major categories estimates} \label{all_occ_results}

\begin{table}
    \centering
    \caption{Effect on occupation in 1901 (HISCO first digit 1 to 3)} \label{tab:occ1}
    \footnotesize
    \begin{tabular}{ccccc}
\toprule
hisco & Affected & Approach & Estimate & n\_parishes\\
\midrule
0/1 & MA & 2: Intensive & -0.658 (0.423) & 1527\\
0/1 & MA & 3: log(x+1) & -0.642 (0.337) & 1589\\
0/1 & MA & 4: asinh(x) & -0.595 (0.397) & 1589\\
0/1 & MA & 1: Extensive & 0.414 (0.139) & 1589\\
0/1 & Dummy & 1: Extensive & 0.064 (0.026) & 1589\\
0/1 & Dummy & 3: log(x+1) & -0.036 (0.059) & 1589\\
0/1 & Dummy & 2: Intensive & -0.031 (0.076) & 1527\\
0/1 & Dummy & 4: asinh(x) & -0.028 (0.069) & 1589\\
2 & MA & 3: log(x+1) & -0.871 (0.463) & 1589\\
2 & MA & 4: asinh(x) & -0.561 (0.521) & 1589\\
2 & MA & 2: Intensive & -0.22 (0.529) & 1465\\
2 & MA & 1: Extensive & 0.182 (0.148) & 1589\\
2 & Dummy & 3: log(x+1) & -0.078 (0.089) & 1589\\
2 & Dummy & 4: asinh(x) & -0.052 (0.1) & 1589\\
2 & Dummy & 2: Intensive & 0.04 (0.101) & 1465\\
2 & Dummy & 1: Extensive & -0.039 (0.024) & 1589\\
3 & MA & 1: Extensive & 1.128*** (0.259) & 1589\\
3 & MA & 4: asinh(x) & 0.697 (0.479) & 1589\\
3 & MA & 3: log(x+1) & 0.486 (0.381) & 1589\\
3 & Dummy & 1: Extensive & 0.178*** (0.045) & 1589\\
3 & Dummy & 4: asinh(x) & 0.137 (0.09) & 1589\\
3 & Dummy & 3: log(x+1) & 0.102 (0.072) & 1589\\
3 & Dummy & 2: Intensive & -0.047 (0.252) & 688\\
3 & MA & 2: Intensive & 0.043 (0.932) & 688\\
\bottomrule
\end{tabular}
\parbox{0.9\textwidth}{
\caption*{\footnotesize \textit{Notes:} Parameter estimate of the effect on occupational structure of the channel in 1901. Each row corresponds to a separate regression with all individuals with hisco codes starting with 0/1, 2 or 3 as outcome. The last column shows the number of parishes included in the regression, which is different from the full sample (1589) in the intensive margin estimates. As a rule of thumb, results with fewer than 100 observations should be entirely disregarded. *** $p< 0.01$ ** $p< 0.05$ * $p< 0.10$. Standard errors clustered on the parish level in parenthesis. All p-values are Bonferroni-corrected. \\ \textit{Source:} Danish census data.}
}
\end{table}

\begin{table}
    \centering
    \caption{Effect on occupation in 1901 (HISCO first digit 4 to 9)} \label{tab:occ2}
    \footnotesize
    \begin{tabular}{ccccc}
\toprule
hisco & Affected & Approach & Estimate & n\_parishes\\
\midrule
4 & MA & 4: asinh(x) & -1.342 (0.576) & 1589\\
4 & MA & 2: Intensive & 1.269 (1.129) & 423\\
4 & MA & 3: log(x+1) & -1.121 (0.489) & 1589\\
4 & Dummy & 2: Intensive & 0.224 (0.209) & 423\\
4 & Dummy & 4: asinh(x) & -0.159 (0.106) & 1589\\
4 & Dummy & 3: log(x+1) & -0.123 (0.09) & 1589\\
4 & MA & 1: Extensive & -0.123 (0.202) & 1589\\
4 & Dummy & 1: Extensive & -0.061 (0.038) & 1589\\
5 & MA & 4: asinh(x) & -0.943 (0.554) & 1589\\
5 & MA & 1: Extensive & -0.662*** (0.163) & 1589\\
5 & MA & 3: log(x+1) & -0.625 (0.462) & 1589\\
5 & MA & 2: Intensive & 0.512 (0.51) & 1580\\
5 & Dummy & 2: Intensive & 0.143 (0.092) & 1580\\
5 & Dummy & 3: log(x+1) & 0.065 (0.083) & 1589\\
5 & Dummy & 4: asinh(x) & 0.049 (0.099) & 1589\\
5 & Dummy & 1: Extensive & -0.041 (0.03) & 1589\\
6 & MA & 2: Intensive & 1.203*** (0.196) & 1589\\
6 & MA & 1: Extensive & -0.243*** (0.049) & 1589\\
6 & Dummy & 2: Intensive & 0.197*** (0.033) & 1589\\
6 & MA & 4: asinh(x) & -0.123 (0.322) & 1589\\
6 & Dummy & 3: log(x+1) & 0.081 (0.038) & 1589\\
6 & Dummy & 4: asinh(x) & 0.068 (0.039) & 1589\\
6 & MA & 3: log(x+1) & 0.025 (0.294) & 1589\\
6 & Dummy & 1: Extensive & -0.024*** (0.004) & 1589\\
7/8/9 & MA & 2: Intensive & 1.739*** (0.342) & 1588\\
7/8/9 & MA & 4: asinh(x) & 0.813 (0.39) & 1589\\
7/8/9 & MA & 3: log(x+1) & 0.709 (0.358) & 1589\\
7/8/9 & Dummy & 2: Intensive & 0.221** (0.066) & 1588\\
7/8/9 & MA & 1: Extensive & -0.218*** (0.05) & 1589\\
7/8/9 & Dummy & 4: asinh(x) & 0.143 (0.07) & 1589\\
7/8/9 & Dummy & 3: log(x+1) & 0.131 (0.064) & 1589\\
7/8/9 & Dummy & 1: Extensive & -0.019 (0.007) & 1589\\
\bottomrule
\end{tabular}
\parbox{0.9\textwidth}{
\caption*{\footnotesize \textit{Notes:} Parameter estimate of the effect on occupational structure of the channel in 1901. Each row corresponds to a separate regression with all individuals with hisco codes starting with 4, 5, 6, or 7/8/9 as outcome. The last column shows the number of parishes included in the regression, which is different from the full sample (1589) in the intensive margin estimates. As a rule of thumb, results with fewer than 100 observations should be entirely disregarded. *** $p< 0.01$ ** $p< 0.05$ * $p< 0.10$. Standard errors clustered on the parish level in parenthesis. All p-values are Bonferroni-corrected. \\ \textit{Source:} Danish census data.}
}
\end{table}

\FloatBarrier

\subsection{Event plot fishing and spinning} \label{fishing_spinning}
\begin{figure}
    \centering
    \caption{Fishermen and Spinners, Weavers, Knitters, Dyers And Related Workers}
    \begin{subfigure}[b]{0.45\textwidth}
        \centering
        \caption{Fishermen (MA approach)} \label{fig:fish_ma}
        \includegraphics[width=\textwidth]{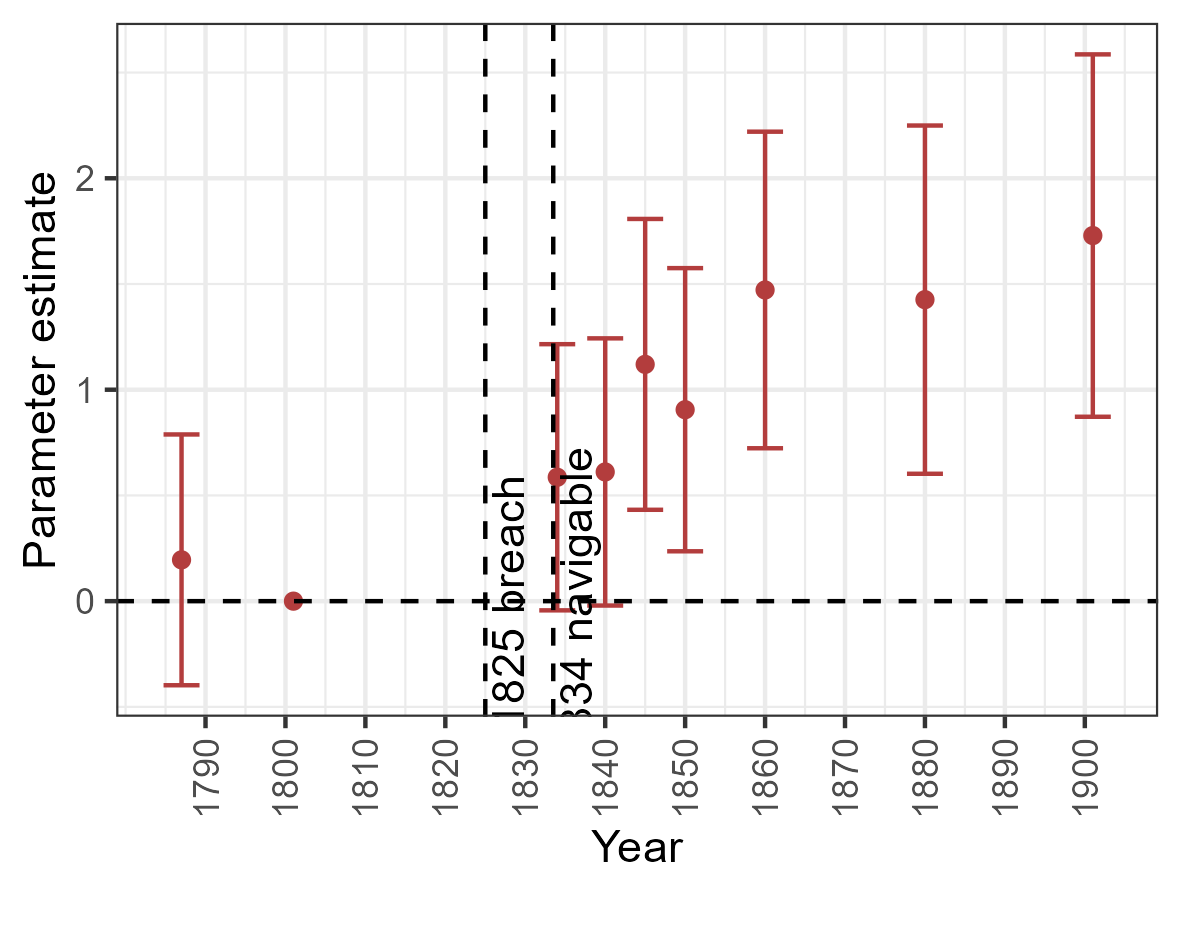}
    \end{subfigure}
    \hfill
    \begin{subfigure}[b]{0.45\textwidth}
        \centering
        \caption{Fishermen (Dummy approach)} \label{fig:fish_dummy}
        \includegraphics[width=\textwidth]{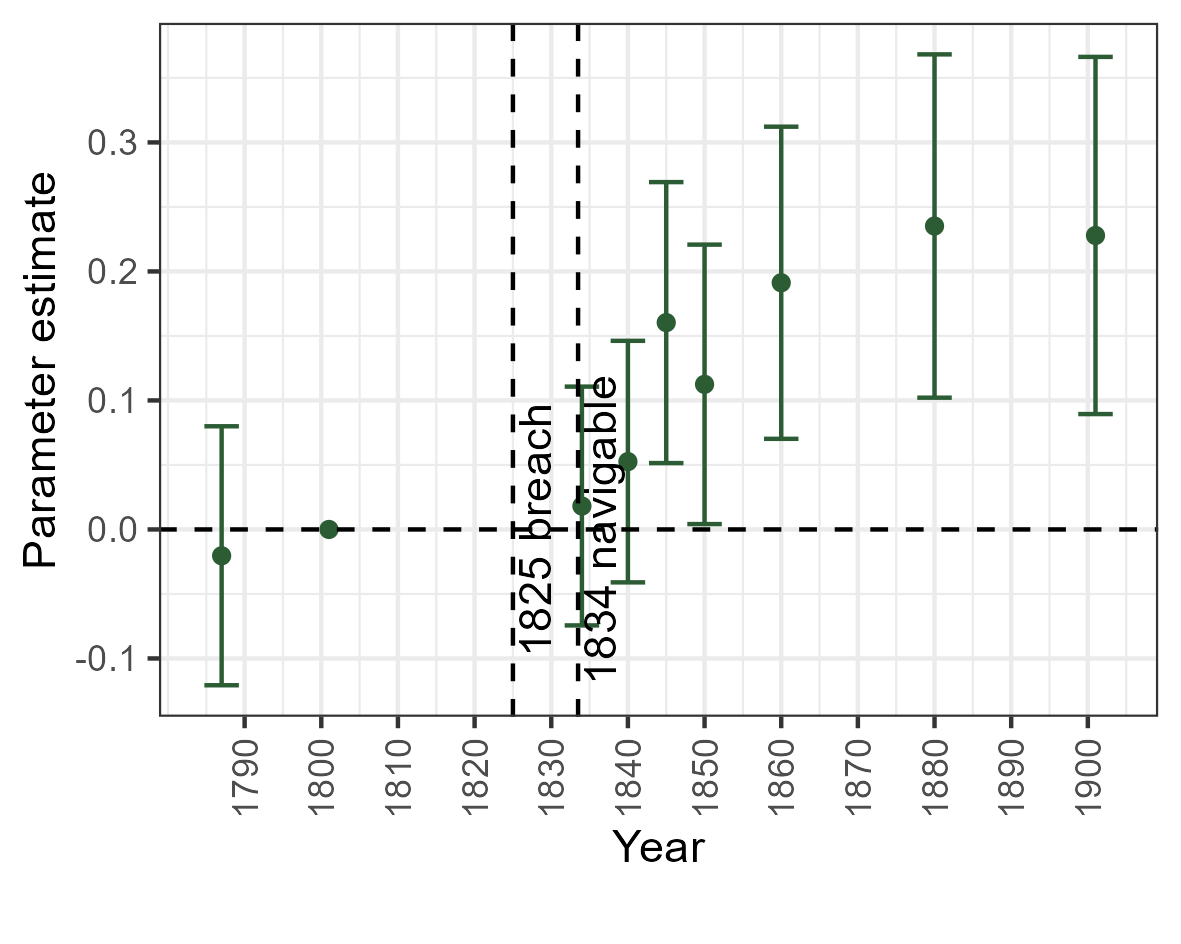}
    \end{subfigure}
    \vspace{0.45cm}
    \begin{subfigure}[b]{0.45\textwidth}
        \centering
        \caption{Spinners, weavers, knitters, dyers and related workers (MA approach)} \label{fig:spinners_ma}
        \includegraphics[width=\textwidth]{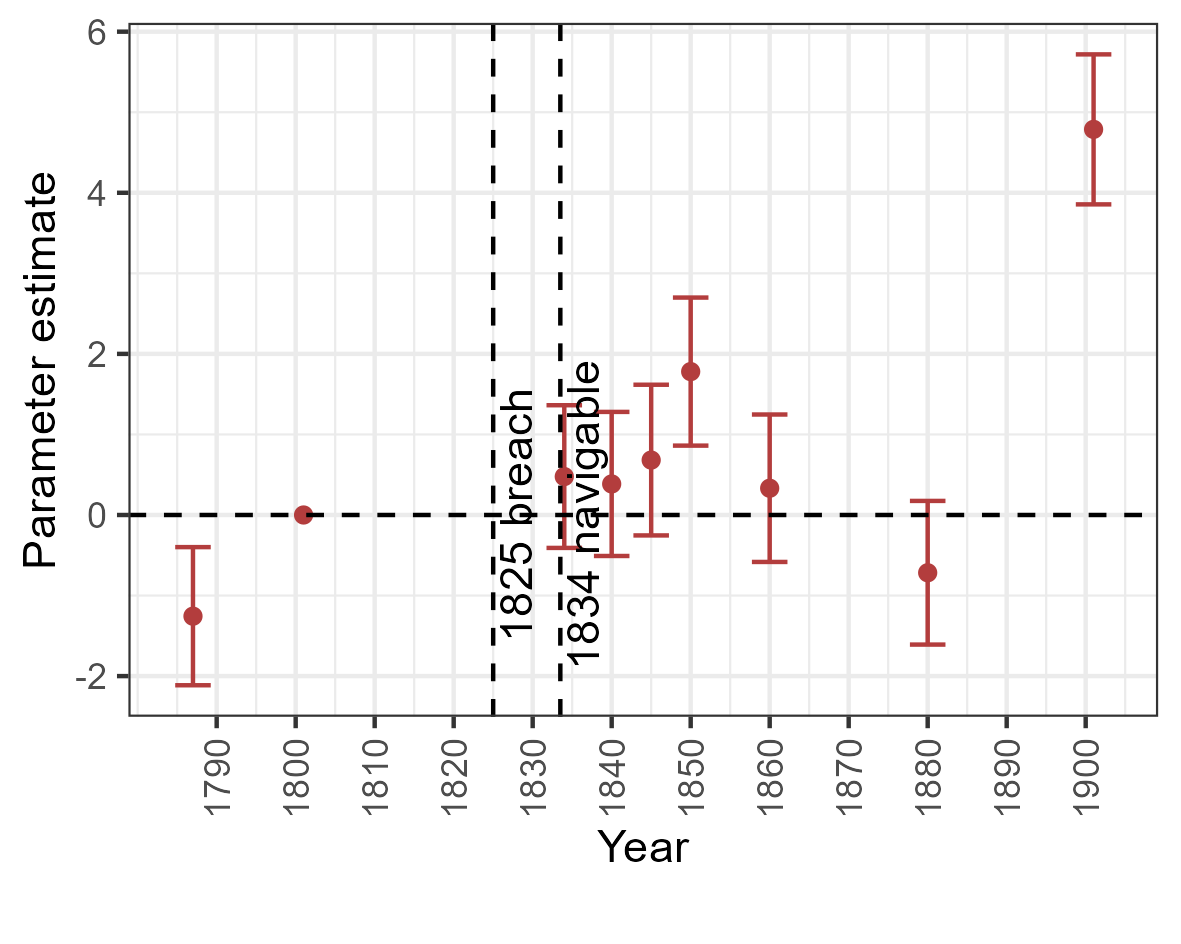}
    \end{subfigure}
    \hfill
    \begin{subfigure}[b]{0.45\textwidth}
        \centering
        \caption{Spinners, weavers, knitters, dyers and related workers (Dummy approach)} \label{fig:spinners_dummy}
        \includegraphics[width=\textwidth]{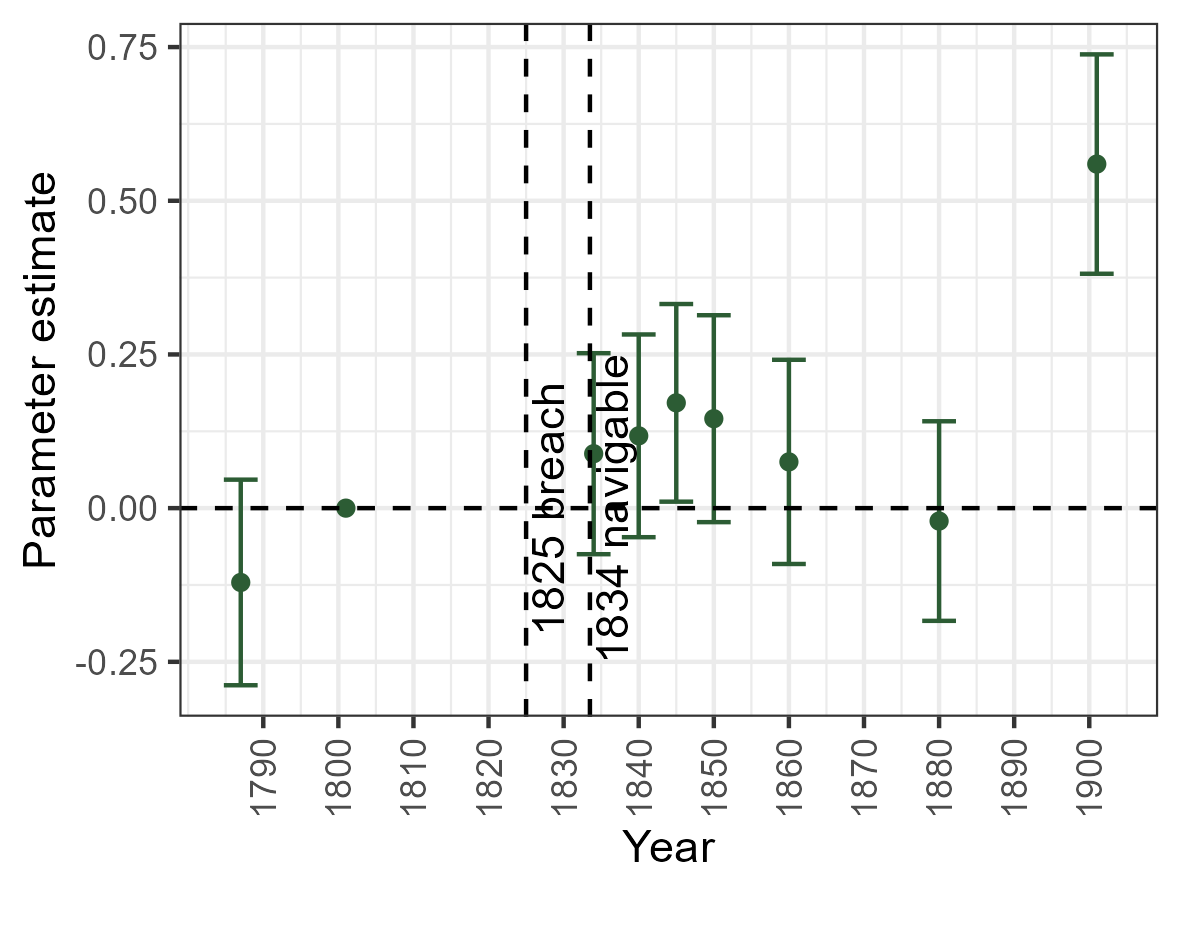}
    \end{subfigure}
    \parbox{0.9\textwidth}{
    \caption*{\footnotesize \textit{Notes:} Panel (a) and panel (b) shows event plots for the effect of the channel on the number of fishermen. Panel (c) and (d) shows the effect to the number of spinners, weavers, knitters, dyers and related workers (HISCO codes starting with 75).  \\ \textit{Source: Danish census data}}
}
    \label{fig:fishing_spinners}
\end{figure}

\FloatBarrier
\subsection{Effects by age group} \label{effects_by_age_group}

\begin{figure}
    \centering
    \caption{Age group composition}
    \begin{subfigure}[b]{0.45\textwidth}
        \centering
        \caption{Effect by age group (MA approach)} \label{fig:migr}
        \includegraphics[width=\textwidth]{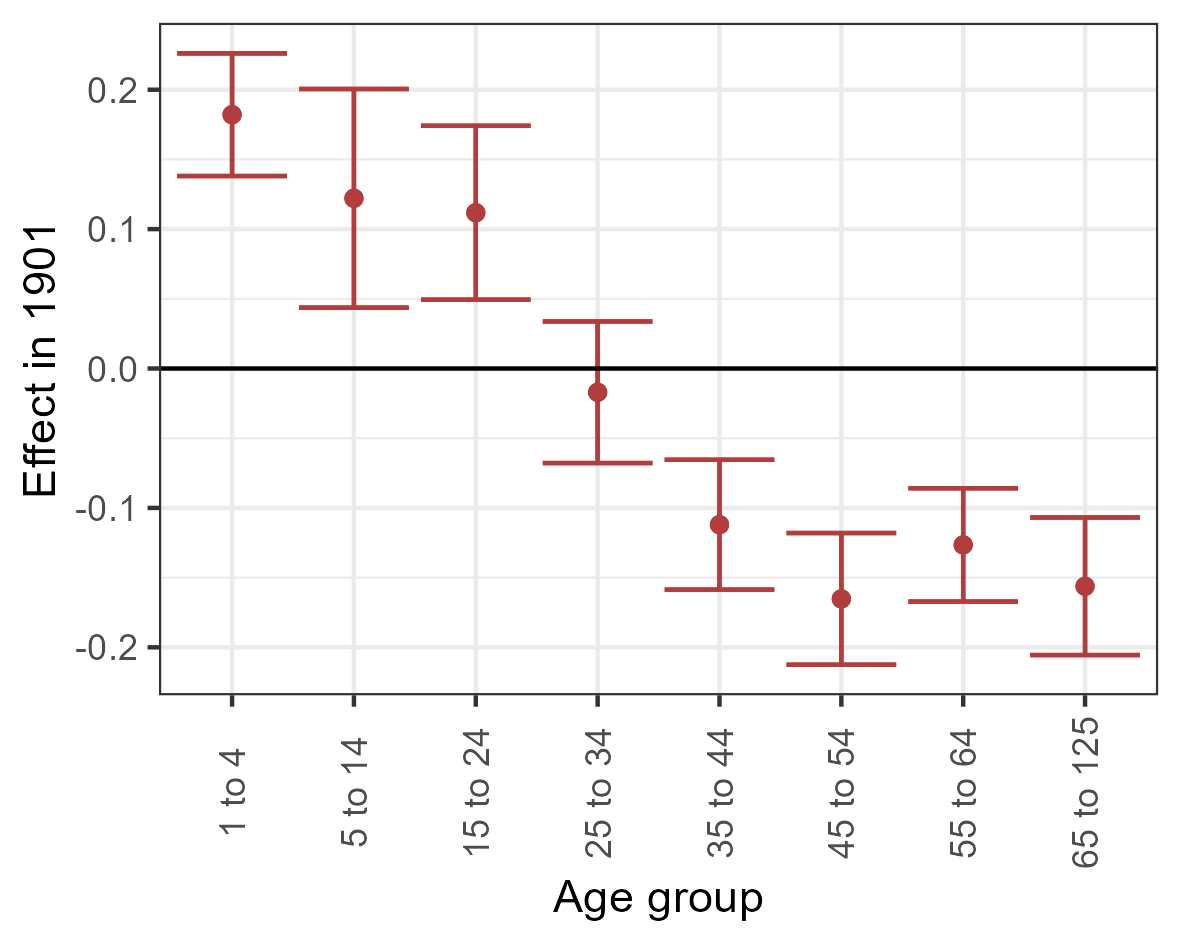}
    \end{subfigure}
    \hfill
    \begin{subfigure}[b]{0.45\textwidth}
        \centering
        \caption{Effect by age group (dummy approach)} \label{fig:fert}
        \includegraphics[width=\textwidth]{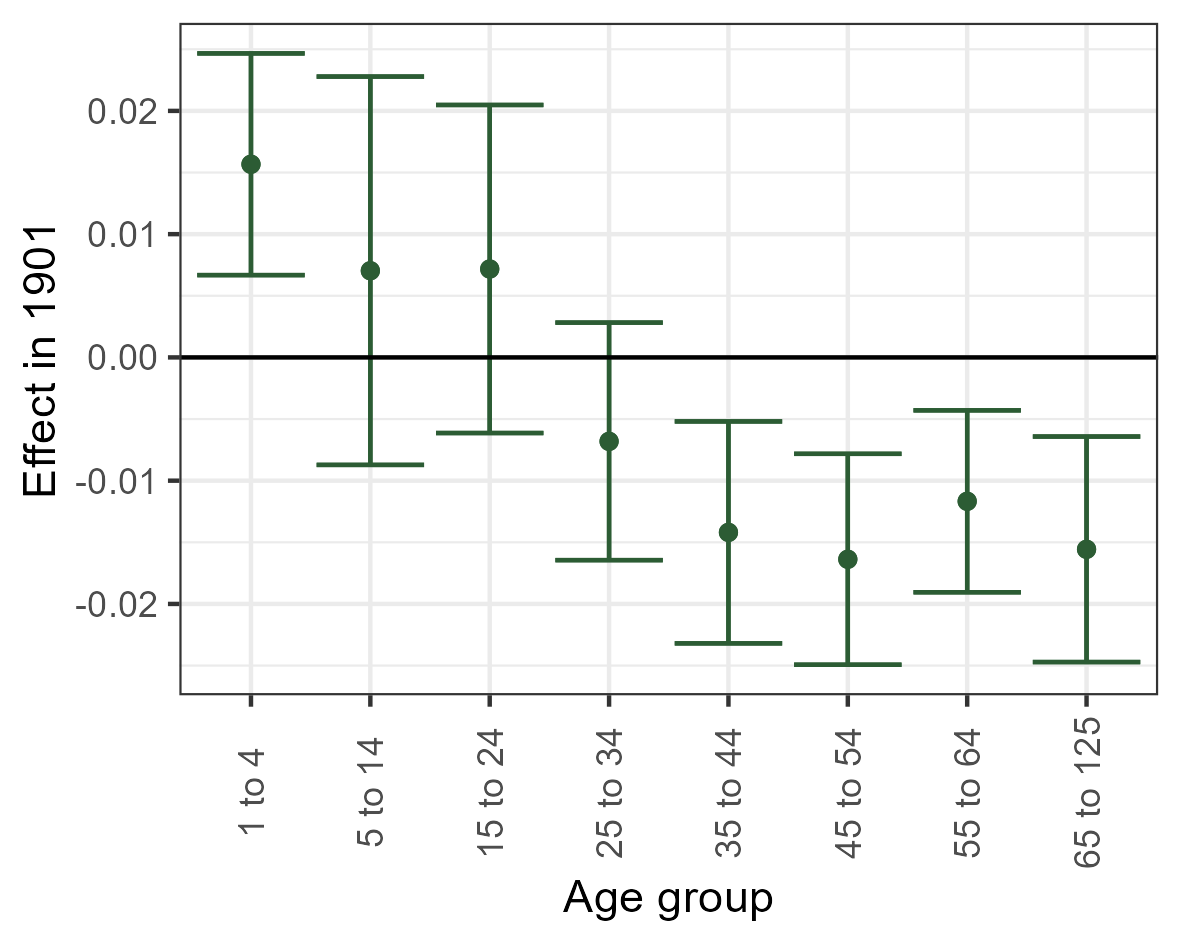}
    \end{subfigure}
    \parbox{0.9\textwidth}{
    \caption*{\footnotesize \textit{Notes:} Regression parameter in 1901 given the market access approach (panel a) and the dummy approach (panel b). The outcome of each regression is the size of the particular age group as a share of the total population. What this shows, is that the population in the affected parishes became comparatively younger.  \\ \textit{Source: Danish census data}}
}
    \label{fig:age_group}
\end{figure}

\FloatBarrier

\subsection{From fertility effect to population effect} \label{boe_fertility}

The fertility effect (0.96 under market access; 0.117 under the dummy approach) and the population effect (1.59; 0.239) are consistent once compounding is accounted for: one is a rate effect, the other a stock effect, and a persistent rise in the birth rate accumulates into a larger proportional gain in the stock than in the rate. This can be seen with a small back-of-the-envelope numerical integration as follows.

Let $\hat{\beta}_f(t)$ denote the event-study estimate of the effect of the channel on the child-women ratio in census year $t$, relative to 1801. Since the CWR counts children aged 0--5, dividing by five converts it to an annual birth rate effect. Each estimate is held constant over the inter-census interval centred on $t$, with 1825 as the left boundary for 1834. Figure~\ref{fig:boe_betaf} plots both effects as step functions; confidence bands are 95\% intervals from 500 parish cluster bootstrap draws.

Multiplying the annual effect by $w_f$ --- the 1801 share of women aged 15--44 in treated parishes --- gives the per-year fertility contribution to population growth (Figure~\ref{fig:boe_contrib}). Table~\ref{tab:boe_fertility} records the numbers for both the market access and dummy approaches. $w_f$ is 0.219 in 1801 and hovers around the same value for the entire period. 

Cumulating these contributions gives the total fertility contribution to the population stock (keeping mortality and migration constant), compared in Figure~\ref{fig:boe_cumul} to the observed population effect. Before 1860 the two series diverge: the salinity shock depressed population through channels beyond fertility. From 1860 they track closely, and by 1901 the cumulative contribution matches the observed population effect exactly under both approaches. This is indicative that the fertility channel very closely accounts for the observed population density outcome.

\begin{table}[h!]
    \centering
    \caption{Back-of-envelope fertility contribution to population effect}
    
\begin{tabular}{cccccccc}
\toprule
\multicolumn{2}{c}{ } & \multicolumn{3}{c}{Market Access} & \multicolumn{3}{c}{Dummy} \\
\cmidrule(l{3pt}r{3pt}){3-5} \cmidrule(l{3pt}r{3pt}){6-8}
Year & $\Delta t$ & $\hat{\beta}_f$ & BoE pop & Obs pop & $\hat{\beta}_f$ & BoE pop & Obs pop\\
\midrule
1834 & 12.0 & 0.255 & 0.134 & -0.015 & 0.039 & 0.020 & 0.018\\
1840 & 5.5 & -0.207 & 0.084 & -0.265 & -0.003 & 0.020 & -0.006\\
1845 & 5.0 & -0.094 & 0.064 & -0.253 & 0.011 & 0.022 & -0.001\\
1850 & 7.5 & 0.044 & 0.078 & -0.247 & 0.025 & 0.030 & 0.004\\
1860 & 15.0 & 0.674 & 0.520 & 0.018 & 0.091 & 0.090 & 0.037\\
1880 & 20.5 & 0.707 & 1.154 & 0.835 & 0.105 & 0.184 & 0.141\\
1901 & 10.5 & 0.958 & 1.593 & 1.591 & 0.117 & 0.238 & 0.239\\
\bottomrule
\end{tabular}

    \parbox{0.9\textwidth}{\caption*{\footnotesize \textit{Notes:} $\hat{\beta}_f$ --- event-study estimate of the effect of the channel on the child-women ratio in census year $t$, relative to 1801. $\Delta t$ --- inter-census interval length in years, centred on $t$ (1825 as left boundary for 1834). BoE pop $= (w_f/5)\sum_{s\leq t}\hat{\beta}_f(s)\,\Delta t_s$, where $w_f$ is the 1801 share of women aged 15--44 in treated parishes. Obs pop --- event-study estimate of the effect of the channel on log population. \textit{Source: Danish census data.}}}
    \label{tab:boe_fertility}
\end{table}

\begin{figure}[h!]
    \centering
    \caption{Fertility effect by census year (step function)}
    \includegraphics[width=\textwidth]{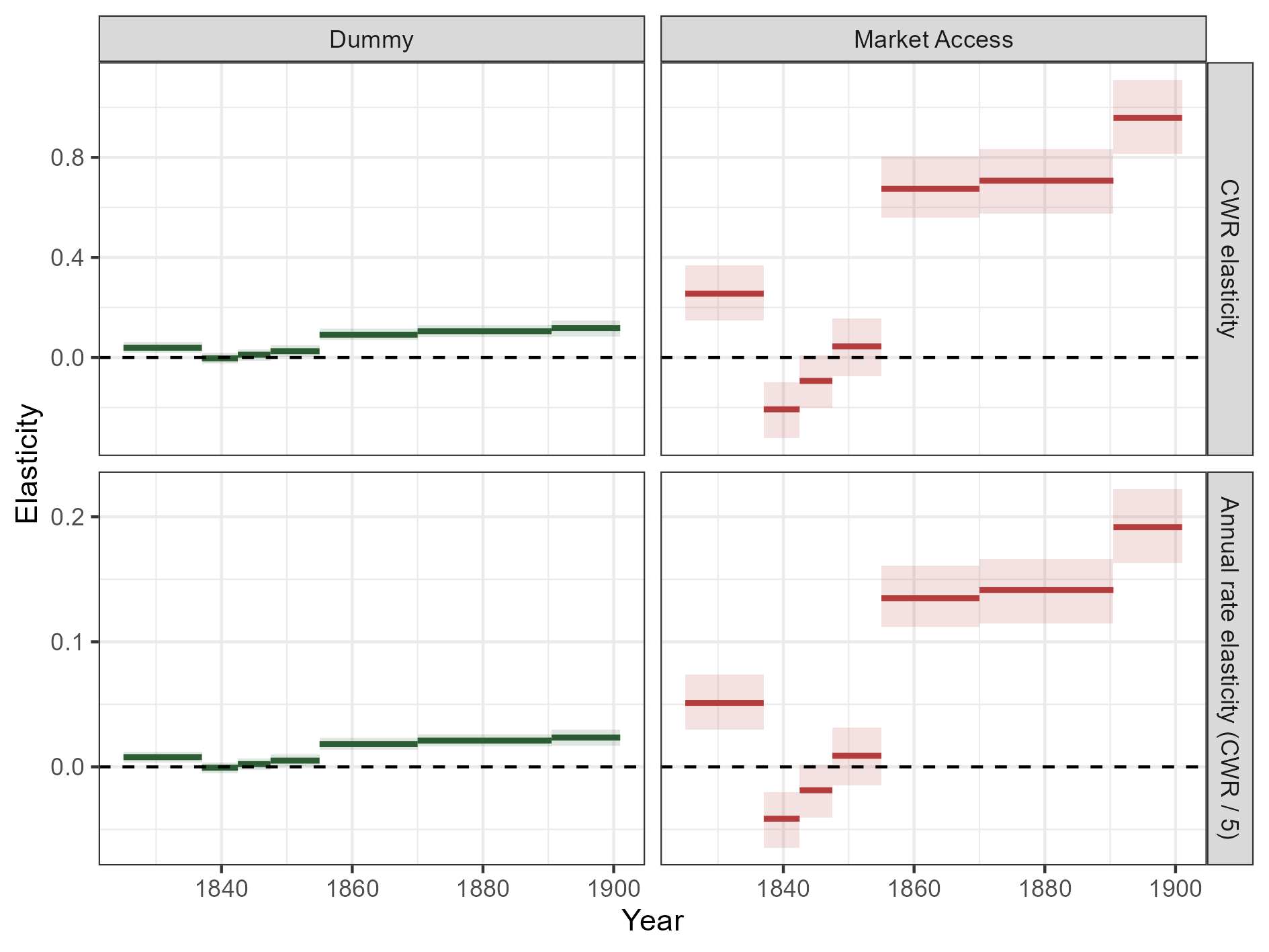}
    \parbox{0.9\textwidth}{\caption*{\footnotesize \textit{Notes:} $\hat{\beta}_f(t)$ is the event-study estimate of the effect of the channel on the child-women ratio in census year $t$, relative to 1801. Each estimate is held constant over its inter-census interval. Upper row: CWR effect. Lower row: annual rate effect (CWR$\,\div\,$5). Columns separate the market access and dummy approaches. Shaded bands are 95\% confidence intervals from 500 parish cluster bootstrap draws. \textit{Source: Danish census data.}}}
    \label{fig:boe_betaf}
\end{figure}

\begin{figure}[h!]
    \centering
    \caption{Annual fertility contribution to population effect}
    \includegraphics[width=\textwidth]{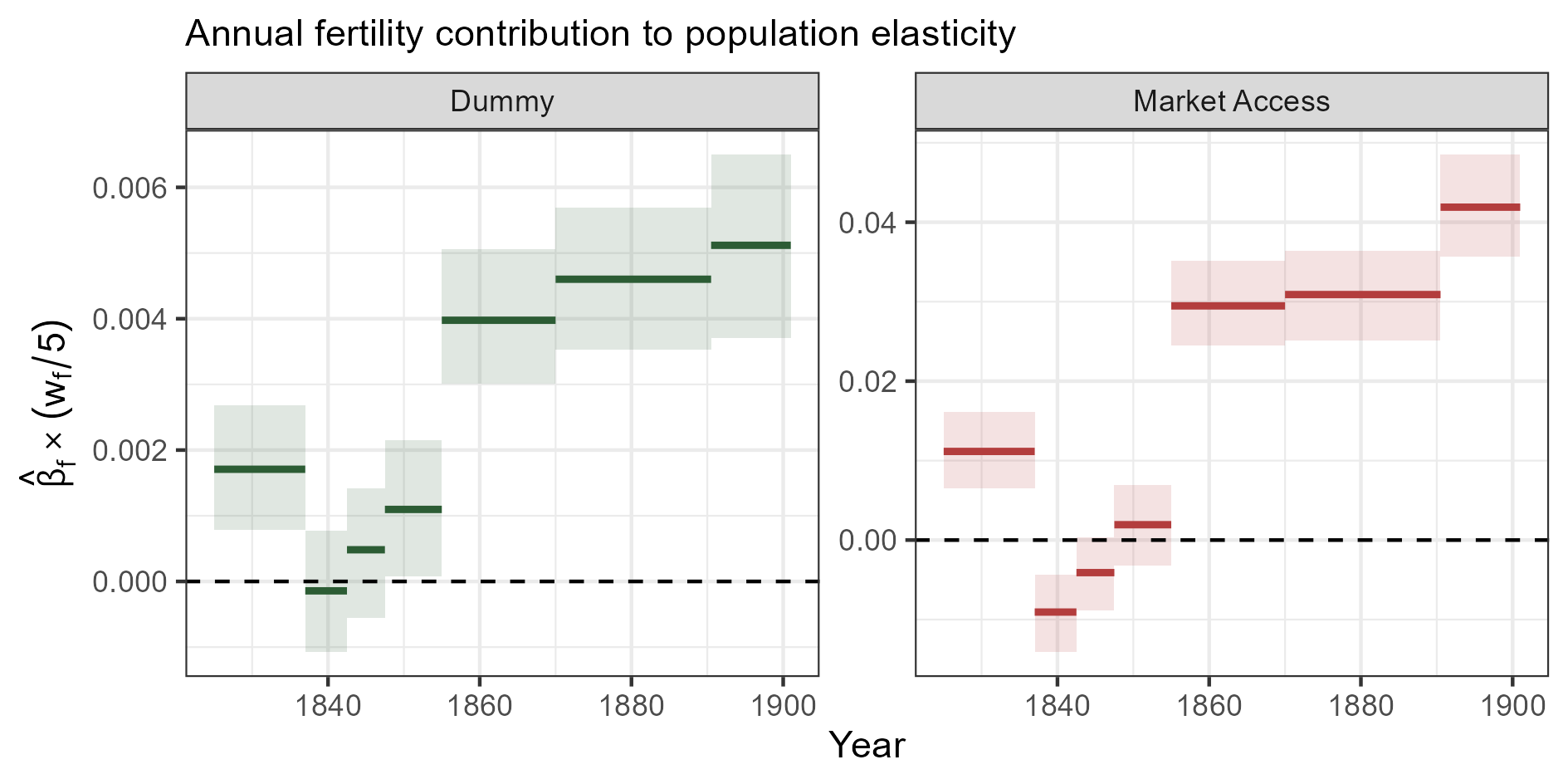}
    \parbox{0.9\textwidth}{\caption*{\footnotesize \textit{Notes:} Annual fertility contribution $= \hat{\beta}_f(t) \times (w_f/5)$, where $w_f$ is the 1801 share of women aged 15--44 in treated parishes. Shaded bands are 95\% bootstrap confidence intervals. \textit{Source: Danish census data.}}}
    \label{fig:boe_contrib}
\end{figure}

\begin{figure}[h!]
    \centering
    \caption{Cumulative fertility contribution vs.\ observed population effect}
    \includegraphics[width=\textwidth]{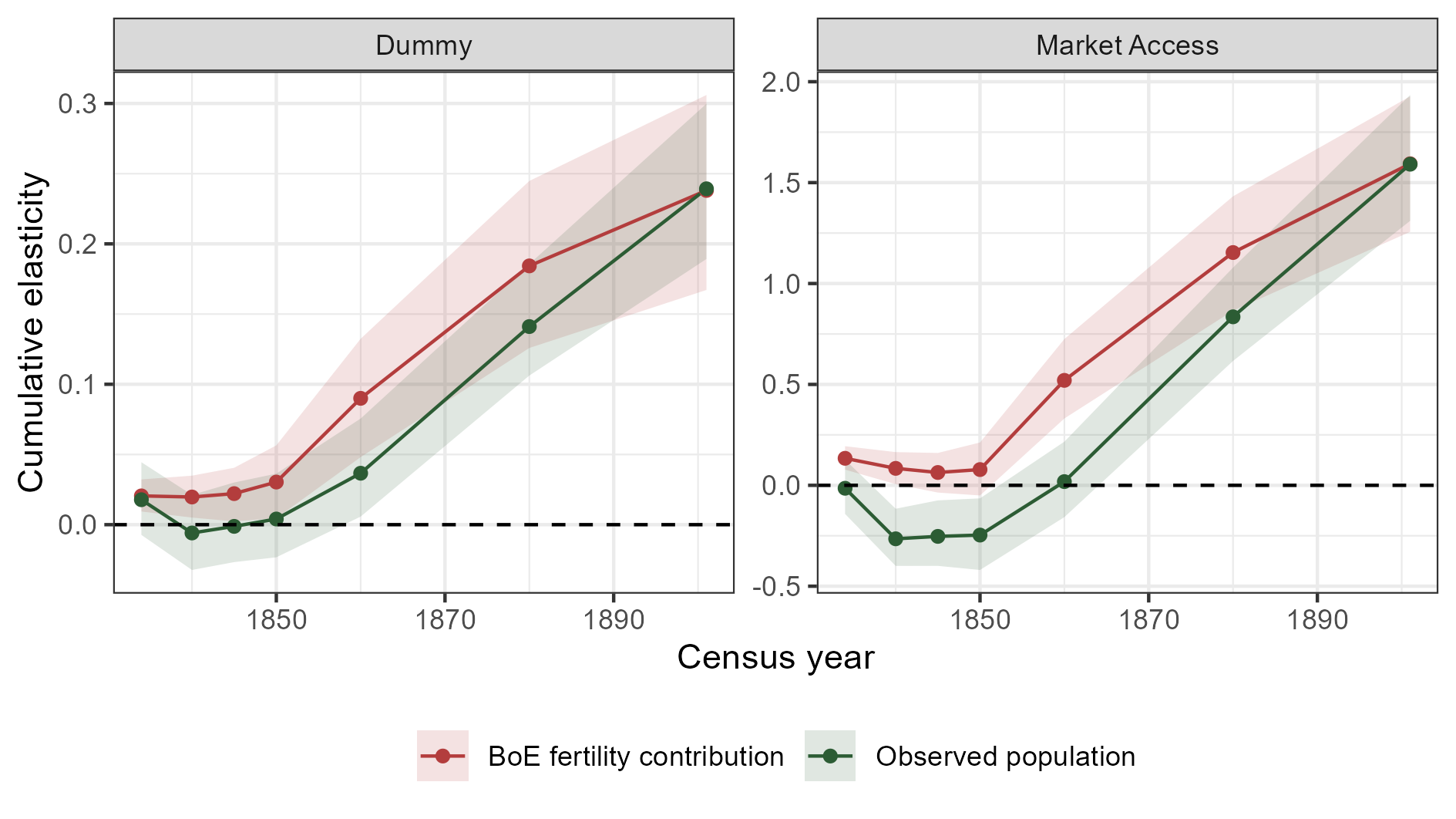}
    \parbox{0.9\textwidth}{\caption*{\footnotesize \textit{Notes:} Red: cumulative fertility contribution to the population stock, $(w_f/5)\sum_{s \leq t} \hat{\beta}_f(s)\,\Delta t_s$. Green: event-study estimate of the effect of the channel on log population. Shaded bands are 95\% bootstrap confidence intervals from 500 parish cluster bootstrap draws. \textit{Source: Danish census data.}}}
    \label{fig:boe_cumul}
\end{figure}
\FloatBarrier
\newpage
\section{Archaeological findings} \label{arch_appendix}

\subsection{Math note} \label{math_note}
This appendix outlines the methodology used to construct a panel of economic activity from archaeological data, serving as a proxy for medieval economic conditions.

The underlying data come from the Danish Registry of Archaeological Sites ('Fund og Fortidsminder'), which includes geo-referenced records of archaeological findings (e.g., coins and buildings) along with a dating range $[Y_{min}^c, Y_{max}^c]$ for each finding $c$. For each finding, the reported date range is interpreted as a uniform distribution so that the probability of the finding corresponding to a given year $t$ is defined as:
$$
P(t|c)=
\begin{cases}
\displaystyle \frac{1}{Y_{max}^c - Y_{min}^c}, & \text{if } Y_{min}^c \le t \le Y_{max}^c, \\
0, & \text{otherwise.}
\end{cases}
$$
This implies that each year within the interval is equally likely.

Since the goal is to estimate the probability that any finding (coin or building) was generated in a given parish at a specific year, the probabilities are aggregated over all findings in that parish. If a parish $i$ has $K_i$ findings, the probability that none of these findings are attributed to year $t$ is $\prod_{c=1}^{K_i} \left[ 1 - P(t|c) \right]$. Thus, the probability that at least one finding is generated at year $t$ is
$$
P_i(\{ \text{findings} \}|t)=1 - \prod_{c=1}^{K_i} \left[ 1-P(t|c) \right].
$$

This represents the extensive margin of archaeological activity. 

To estimate this probability for each parish-year pair, 1,000 Monte Carlo draws are performed for each finding's date from its reported interval, and the frequency of obtaining at least one finding for that parish and year is recorded. Inference is conducted using a clustered bootstrap procedure that resamples the Monte Carlo draws.

The derivation of this way of estimating now follows for a single parish $i$. To construct a panel this is simply repeated for all parishes. The derivation is based on coin findings. But it generalises to any kind of finding.    

The data is of the form: $Coin=c$ was generated in time interval $t\in [Y_{min}^c;Y_{max}^c]$ (Archaeologists report a coin finding and date it to a range.)

That is, for each coin we observe $P(t|c)\sim [Y_{min}^c;Y_{max}^c]$ 

We want to know the probability, that any coin, $c\in \{1, ..., K\}$, finding was generated at any particular point in time. This event is referred to as $\{coins\}|t$

\subsubsection{Probability of a single coin} 
We are interested in the probability that a any coin comes from a specific point in time. What is observed is a range provided by the archaeologist. We have to give this range some specific interpretation. This could be something like

\begin{equation}
P(t|c)=\begin{cases}
\frac{1}{Y_{max}^c - Y_{min}^c + 1}, & \text{if }Y_{min}^c\leq t \leq Y_{max}^c \\
0 & \text{otherwise}
\end{cases}.
\end{equation}

Or written with an indicator function

\begin{equation}
P(t|c)=1[t\in [Y_{min}^c;Y_{max}^c]]\frac{1}{Y_{max}^c - Y_{min}^c + 1}.
\end{equation}

I.e. it is equally likely that a coin truly originates at any particular point in time the range offered by the archaeologists. 

This distribution is an assumption. The archaeologists specify a range but not a distribution. How should this range be interpreted? A straightforward alternative is to interpret it as a 95 percent confidence interval of the normal distribution. This is also tested. 

\begin{equation}
\begin{split}
P(t|c)&=\mathcal{N}(\mu_c,\sigma_c)\\
\text{where:} \\
\mu_c&=0.5\times(Y_{max}^c + Y_{min}^c)\\
\sigma_c&\approx(Y_{max}^c - Y_{min}^c)/3.92
\end{split}
\end{equation}

\subsubsection{At the parish level}
The probability that at least one coin finding in the parish belongs to year $t$ is:

\begin{equation}
\begin{split}
P(t|\{coins\})&=1-\prod_{c=1}^K \left( 1 - P(t|c) \right) \\
P(t|\{coins\})&=1-\prod_{c=1}^K \left(1 - 1[t\in [Y_{min}^c;Y_{max}^c]] \left(\frac{1}{Y_{max}^c - Y_{min}^c}\right) \right)
\end{split}
\end{equation}

where $\{coins\}$ is the event of at least one coin $\{coins\}=\{1,...,c,...,K\}$. 

The intuition is this: The inner part of the expression ($1 - P(t|c)$) is the probability that a coin is \textit{not} associated with that particular point in time. Taking the product over all coins, generates the combined probability, that \textit{no coins at all} are associated with that particular point in time. The compliment of this is the probability we are interested in, in this step. It is the probability that \textit{any} coin is associated with a particular point in time. 

\subsubsection{Estimation}
The following loop (pseudocode) estimates $P_i(\{\text{findings}\}|t)$ for each parish-window pair. Sampled years are rounded to the nearest 50-year window $\tau$, so $\tau$ in the pseudocode below refers to a 50-year window rather than a single year.

\begin{verbatim}
    ```{pseudo code}
    B = 1000 # Number of Monte Carlo samples

    for b in 1 to B: # Loop of MC samples
    ...	# Generate samples from Y_min_c to Y_max_c and round to 50-year window:
    ...	for c in 1 to C:
    ...	...	t_c = round(sample_uniform(Y_min_c, Y_max_c) / 50) * 50
    ...	...	# Is t_c equal to tau?
    ...	...	coins_t[c] = t_c == tau

    ...	# Were there any coins associated with this window in this draw?
    ...	number_of_coins = sum(coins_t)
    ...	succces_t[b] = number_of_coins >= 1

    # Estimating the probability for each tau
    P_findings_given_tau = sum(succces_t) / B
    ```
\end{verbatim}

This code is then repeated for every parish and every 50-year window $\tau$. This gives a panel of size $N\times T$ containing the estimated probability that at least one finding was generated in a parish during a given 50-year window. This in turn can be used in econometric applications.

\newpage
\subsection{All parameter estimates} \label{all_arch_param}
Table \ref{tab:A_arch1} and \ref{tab:A_arch2} contain parameter estimates for all years for the regressions using archaeological findings. 

\begin{table}[H]
\centering
\footnotesize
\caption{All parameters of table 3 columns 1-4} \label{tab:A_arch1}
\begin{tabular}{lcccc}
   \tabularnewline \midrule \midrule
   Dependent Variable: & \multicolumn{4}{c}{activity}\\
   Model:                                           & (1)             & (2)             & (3)             & (4)\\  
   \midrule
   \emph{Variables}\\

   Year750 $\times$ Affected                        & 0.0962$^{***}$  & 0.0045          & -0.0600$^{**}$  & -0.0147$^{***}$\\   
                                                    & (0.0223)        & (0.0043)        & (0.0252)        & (0.0050)\\   
   Year800 $\times$ Affected                        & 0.0115          & -0.0032         & -0.0609$^{***}$ & -0.0087$^{**}$\\   
                                                    & (0.0207)        & (0.0041)        & (0.0208)        & (0.0042)\\   
   Year850 $\times$ Affected                        & 0.0176          & -0.0022         & -0.0583$^{***}$ & -0.0085$^{**}$\\   
                                                    & (0.0204)        & (0.0041)        & (0.0208)        & (0.0042)\\   
   Year900 $\times$ Affected                        & 0.0043          & -0.0025         & -0.0295$^{***}$ & -0.0044$^{**}$\\   
                                                    & (0.0180)        & (0.0037)        & (0.0105)        & (0.0020)\\   
   Year950 $\times$ Affected                        & -0.0058         & -0.0024         & -0.0046         & -0.0010\\   
                                                    & (0.0172)        & (0.0036)        & (0.0047)        & (0.0009)\\   
   Year1050 $\times$ Affected                       & -0.0143         & -0.0035         & -0.0378$^{***}$ & -0.0050$^{***}$\\   
                                                    & (0.0184)        & (0.0037)        & (0.0089)        & (0.0017)\\   
   Year1100 $\times$ Affected                       & 0.0487          & 0.0023          & -0.2048$^{***}$ & -0.0299$^{***}$\\   
                                                    & (0.0317)        & (0.0054)        & (0.0442)        & (0.0084)\\   
   Year1150 $\times$ Affected                       & 0.0509          & 0.0027          & -0.2123$^{***}$ & -0.0319$^{***}$\\   
                                                    & (0.0318)        & (0.0054)        & (0.0444)        & (0.0084)\\   
   Year1200 $\times$ Affected                       & -0.1343$^{***}$ & -0.0151$^{***}$ & -0.2043$^{***}$ & -0.0281$^{***}$\\   
                                                    & (0.0344)        & (0.0059)        & (0.0452)        & (0.0095)\\   
   Year1250 $\times$ Affected                       & -0.2905$^{***}$ & -0.0287$^{***}$ & -0.2283$^{***}$ & -0.0346$^{***}$\\   
                                                    & (0.0447)        & (0.0076)        & (0.0480)        & (0.0090)\\   
   Year1300 $\times$ Affected                       & -0.3398$^{***}$ & -0.0310$^{***}$ & -0.1980$^{***}$ & -0.0353$^{***}$\\   
                                                    & (0.0440)        & (0.0082)        & (0.0413)        & (0.0080)\\   
   Year1350 $\times$ Affected                       & -0.3412$^{***}$ & -0.0355$^{***}$ & -0.1446$^{***}$ & -0.0333$^{***}$\\   
                                                    & (0.0490)        & (0.0079)        & (0.0416)        & (0.0084)\\   
   Year1400 $\times$ Affected                       & -0.1622$^{***}$ & -0.0195$^{***}$ & -0.1190$^{***}$ & -0.0288$^{***}$\\   
                                                    & (0.0399)        & (0.0067)        & (0.0407)        & (0.0081)\\   
   Year1450 $\times$ Affected                       & 0.0275          & 0.0047          & -0.1081$^{***}$ & -0.0273$^{***}$\\   
                                                    & (0.0470)        & (0.0082)        & (0.0414)        & (0.0079)\\   
   Year1500 $\times$ Affected                       & 0.1120$^{***}$  & 0.0083          & -0.0827$^{**}$  & -0.0237$^{***}$\\   
                                                    & (0.0336)        & (0.0059)        & (0.0406)        & (0.0077)\\   
   \midrule
   \emph{Fit statistics}\\
   Observations                                     & 29,568          & 29,568          & 29,568          & 29,568\\  
   R$^2$                                            & 0.05303         & 0.03937         & 0.02100         & 0.02282\\  
   Adjusted R$^2$                                   & 0.05204         & 0.03732         & 0.01997         & 0.02073\\  
   \midrule \midrule
   \multicolumn{5}{l}{\emph{Custom standard-errors in parentheses}}\\
   \multicolumn{5}{l}{\emph{Signif. Codes: ***: 0.01, **: 0.05, *: 0.1}}\\
\end{tabular}

\end{table}

\begin{table}
\centering
\footnotesize
\caption{All parameters of table 3 columns 5-8} \label{tab:A_arch2}
\begin{tabular}{lcccc}
   \tabularnewline \midrule \midrule
   Dependent Variable: & \multicolumn{4}{c}{activity}\\
   Model:                      & (1)             & (2)             & (3)             & (4)\\  
   \midrule
   \emph{Variables}\\

   Year750 $\times$ Affected   & 0.0854$^{**}$   & 0.0085          & -0.0601         & -0.0126$^{**}$\\   
                               & (0.0422)        & (0.0062)        & (0.0423)        & (0.0063)\\   
   Year800 $\times$ Affected   & 0.0080          & -0.0023         & -0.1153$^{***}$ & -0.0143$^{**}$\\   
                               & (0.0429)        & (0.0067)        & (0.0380)        & (0.0059)\\   
   Year850 $\times$ Affected   & 0.0327          & 0.0021          & -0.1118$^{***}$ & -0.0137$^{**}$\\   
                               & (0.0383)        & (0.0057)        & (0.0379)        & (0.0059)\\   
   Year900 $\times$ Affected   & 0.0161          & 0.0010          & -0.0605$^{***}$ & -0.0080$^{***}$\\   
                               & (0.0355)        & (0.0052)        & (0.0193)        & (0.0030)\\   
   Year950 $\times$ Affected   & -0.0011         & -0.0001         & -0.0070         & -0.0011\\   
                               & (0.0349)        & (0.0050)        & (0.0075)        & (0.0012)\\   
   Year1050 $\times$ Affected  & -0.0125         & -0.0018         & -0.0098         & -0.0021\\   
                               & (0.0445)        & (0.0063)        & (0.0142)        & (0.0021)\\   
   Year1100 $\times$ Affected  & 0.1082$^{**}$   & 0.0155$^{**}$   & -0.0287         & -0.0120\\   
                               & (0.0505)        & (0.0074)        & (0.0729)        & (0.0107)\\   
   Year1150 $\times$ Affected  & 0.1104$^{**}$   & 0.0160$^{**}$   & -0.0494         & -0.0157\\   
                               & (0.0505)        & (0.0075)        & (0.0741)        & (0.0107)\\   
   Year1200 $\times$ Affected  & -0.0998$^{*}$   & -0.0098         & -0.0680         & -0.0182\\   
                               & (0.0557)        & (0.0083)        & (0.0764)        & (0.0120)\\   
   Year1250 $\times$ Affected  & -0.2566$^{***}$ & -0.0268$^{**}$  & -0.1187         & -0.0308$^{**}$\\   
                               & (0.0753)        & (0.0112)        & (0.0869)        & (0.0124)\\   
   Year1300 $\times$ Affected  & -0.2530$^{***}$ & -0.0241$^{**}$  & -0.1221         & -0.0295$^{***}$\\   
                               & (0.0698)        & (0.0111)        & (0.0746)        & (0.0111)\\   
   Year1350 $\times$ Affected  & -0.2320$^{***}$ & -0.0302$^{***}$ & -0.1112         & -0.0260$^{**}$\\   
                               & (0.0787)        & (0.0117)        & (0.0779)        & (0.0121)\\   
   Year1400 $\times$ Affected  & -0.0903         & -0.0158         & -0.0985         & -0.0224$^{*}$\\   
                               & (0.0695)        & (0.0098)        & (0.0788)        & (0.0117)\\   
   Year1450 $\times$ Affected  & 0.0685          & 0.0078          & -0.0863         & -0.0207$^{*}$\\   
                               & (0.0888)        & (0.0118)        & (0.0781)        & (0.0112)\\   
   Year1500 $\times$ Affected  & 0.1196$^{*}$    & 0.0133          & -0.0583         & -0.0170\\   
                               & (0.0655)        & (0.0091)        & (0.0779)        & (0.0110)\\   
   \midrule
   \emph{Fit statistics}\\
   Observations                & 6,912           & 6,912           & 6,912           & 6,912\\  
   R$^2$                       & 0.05097         & 0.03762         & 0.03129         & 0.02987\\  
   Adjusted R$^2$              & 0.04670         & 0.03328         & 0.02693         & 0.02550\\  
   \midrule \midrule
   \multicolumn{5}{l}{\emph{Custom standard-errors in parentheses}}\\
   \multicolumn{5}{l}{\emph{Signif. Codes: ***: 0.01, **: 0.05, *: 0.1}}\\
\end{tabular}
\end{table}

\FloatBarrier
\subsection{Normal distribution} \label{norm_arch}
Figure \ref{fig:arch_reg1}, \ref{fig:arch_reg_boot1}, \ref{fig:arch_reg2} and \ref{fig:arch_reg_boot2} show equivalent results to those presented in the paper. However, these are results based on assuming that the archaeological datings (e.g. coin finding dated to the years 1300-1495) represent a 95 percent confidence interval from a normal distribution rather than an uniform distribution. Figure \ref{fig:arch_reg1} shows results confidence intervals for all parameters using the full sample. Figure \ref{fig:arch_reg2} shows the same results using the matched sample. Figure \ref{fig:arch_reg_boot1} and \ref{fig:arch_reg_boot2} show all the bootstrap draws for 1350.

Note that the results are qualitatively the same as the main results. The standard deviation used here is $\sigma_c = (Y_{max}^c - Y_{min}^c)/3.92$, placing 95 percent of the distribution's mass within the archaeologists' reported date range.

\begin{figure}
    \centering
    \caption{Archaelogical results (full sample)}
    \begin{subfigure}[b]{0.45\textwidth}
        \centering
        \caption{Coins: Market access approach} \label{fig:arch1a_norm}
        \includegraphics[width=\textwidth]{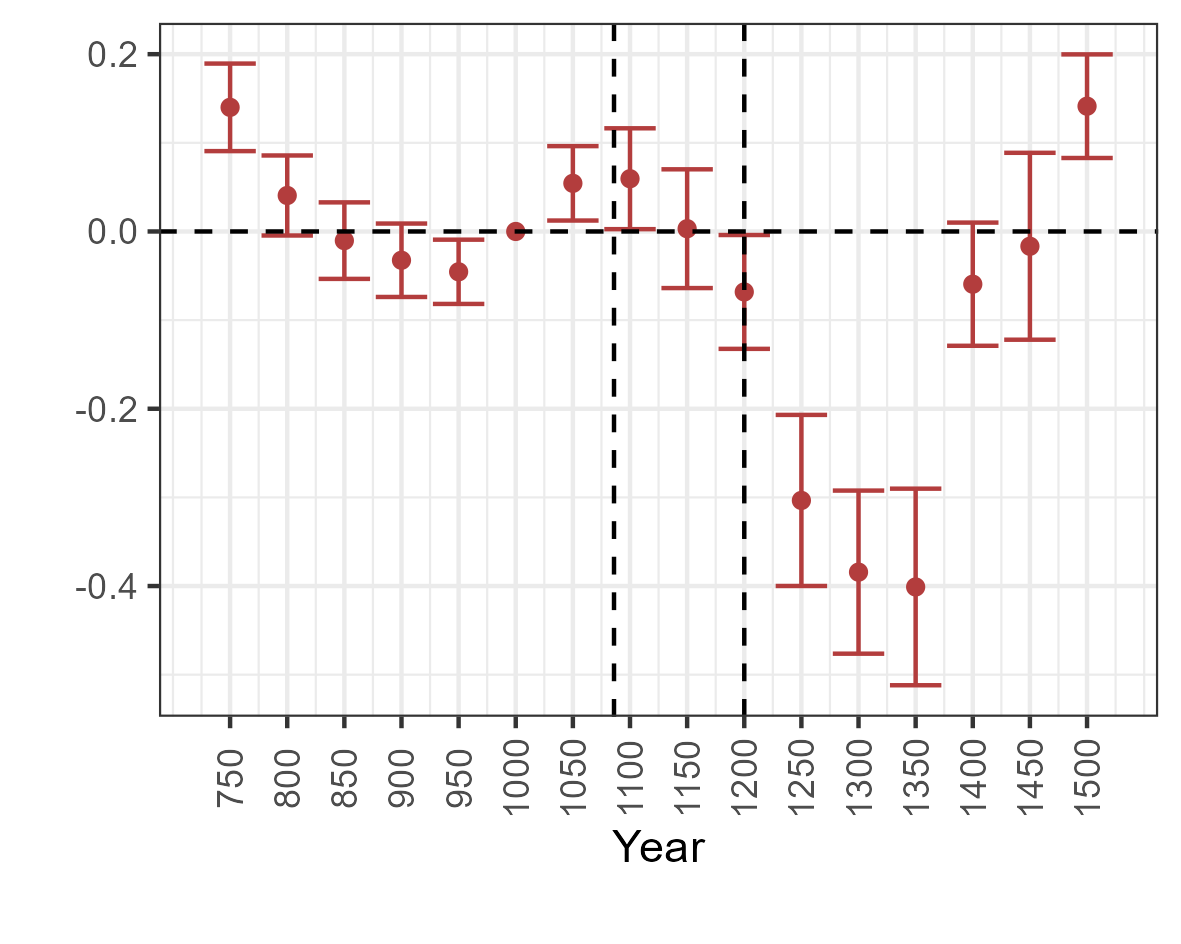}
    \end{subfigure}
    \hfill
    \begin{subfigure}[b]{0.45\textwidth}
        \centering
        \caption{Coins: Dummy approach} \label{fig:arch1b_norm}
        \includegraphics[width=\textwidth]{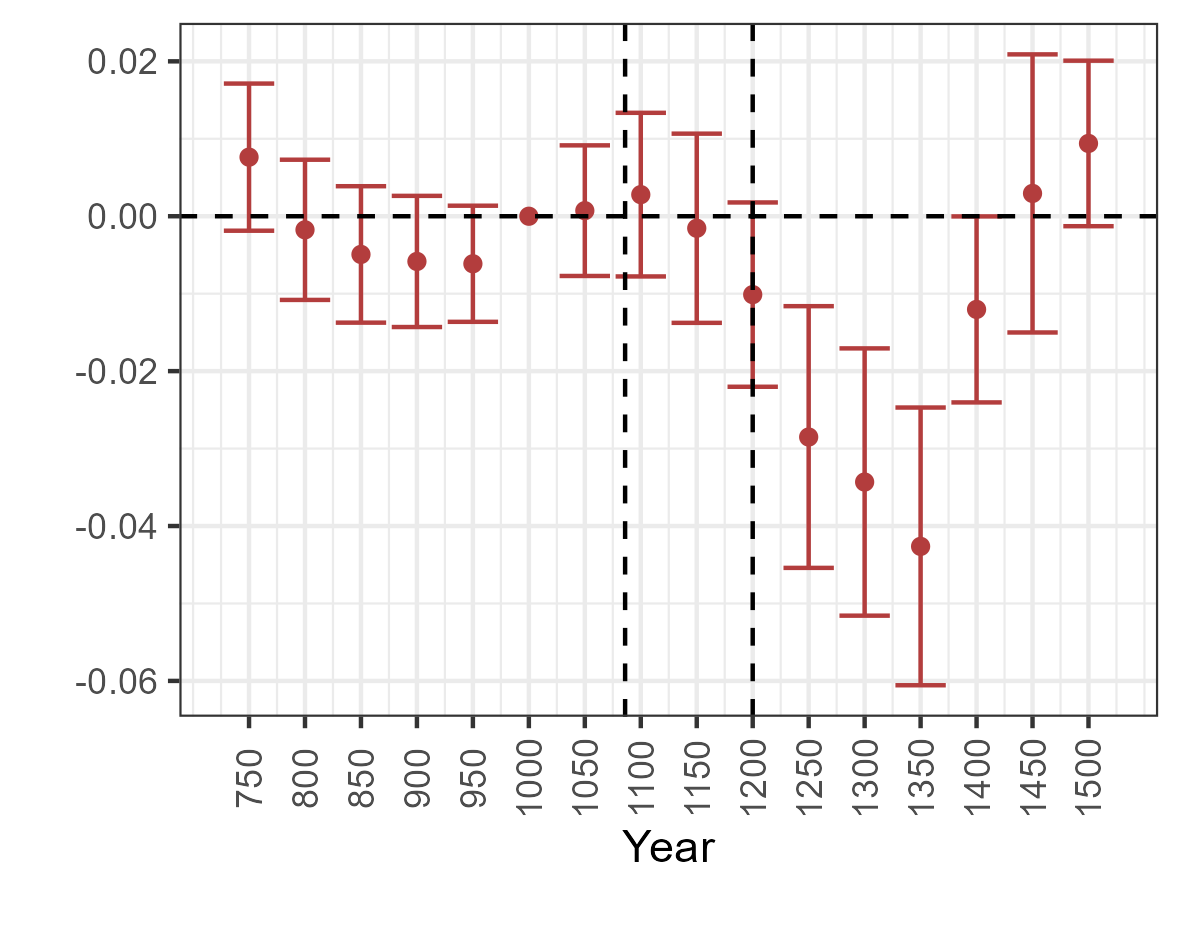}
    \end{subfigure}
    \vspace{0.45cm}
    \begin{subfigure}[b]{0.45\textwidth}
        \centering
        \caption{Buildings: Market access approach} \label{fig:arch1c_norm}
        \includegraphics[width=\textwidth]{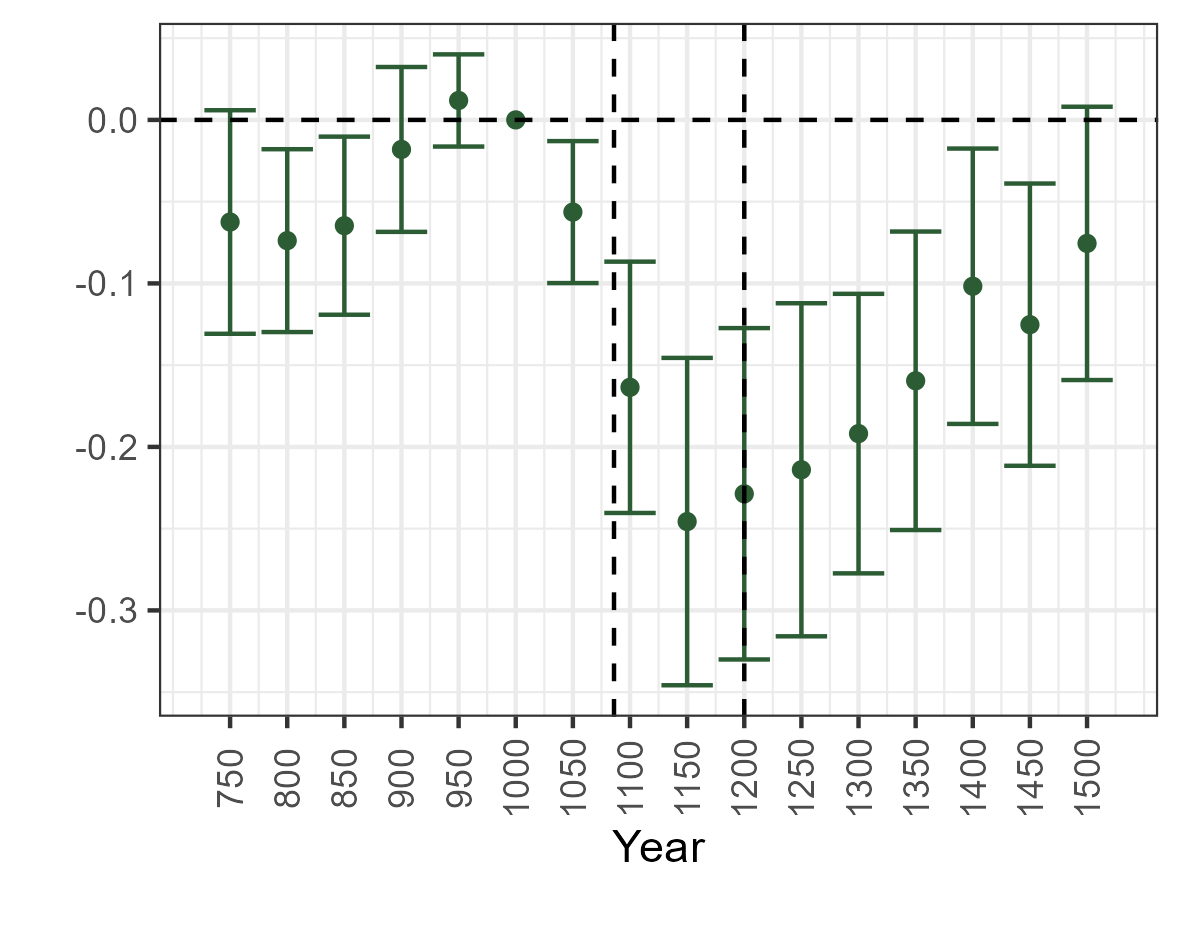}
    \end{subfigure}
    \hfill
    \begin{subfigure}[b]{0.45\textwidth}
        \centering
        \caption{Buildings: Dummy approach} \label{fig:arch1d_norm}
        \includegraphics[width=\textwidth]{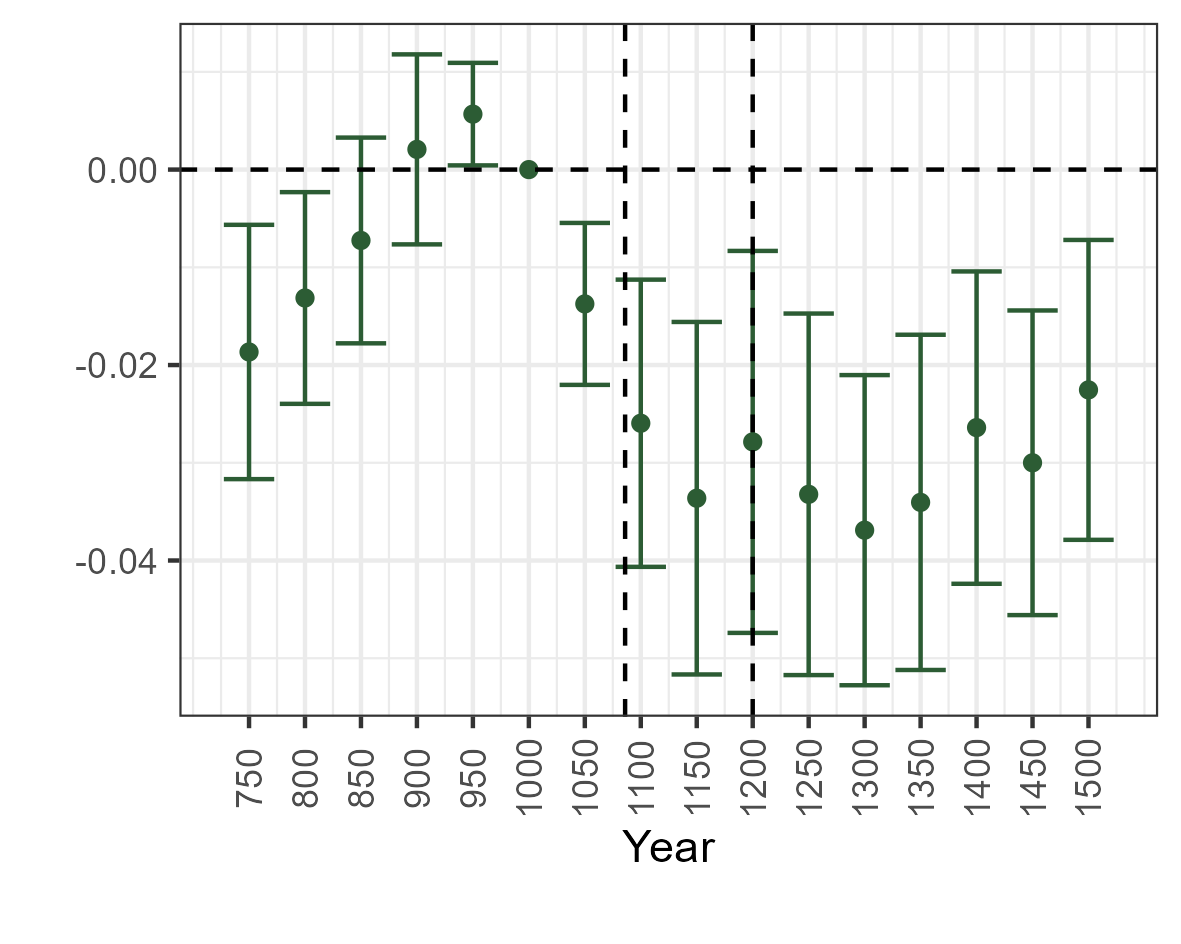}
    \end{subfigure}
    \label{fig:arch_reg1}
\end{figure}

\begin{figure}
    \centering
    \caption{Distribution of parameter estimates in 1350  (full sample)}
    \begin{subfigure}[b]{0.45\textwidth}
        \centering
        \caption{Coins: Market access approach} \label{fig:distri_a_norm}
        \includegraphics[width=\textwidth]{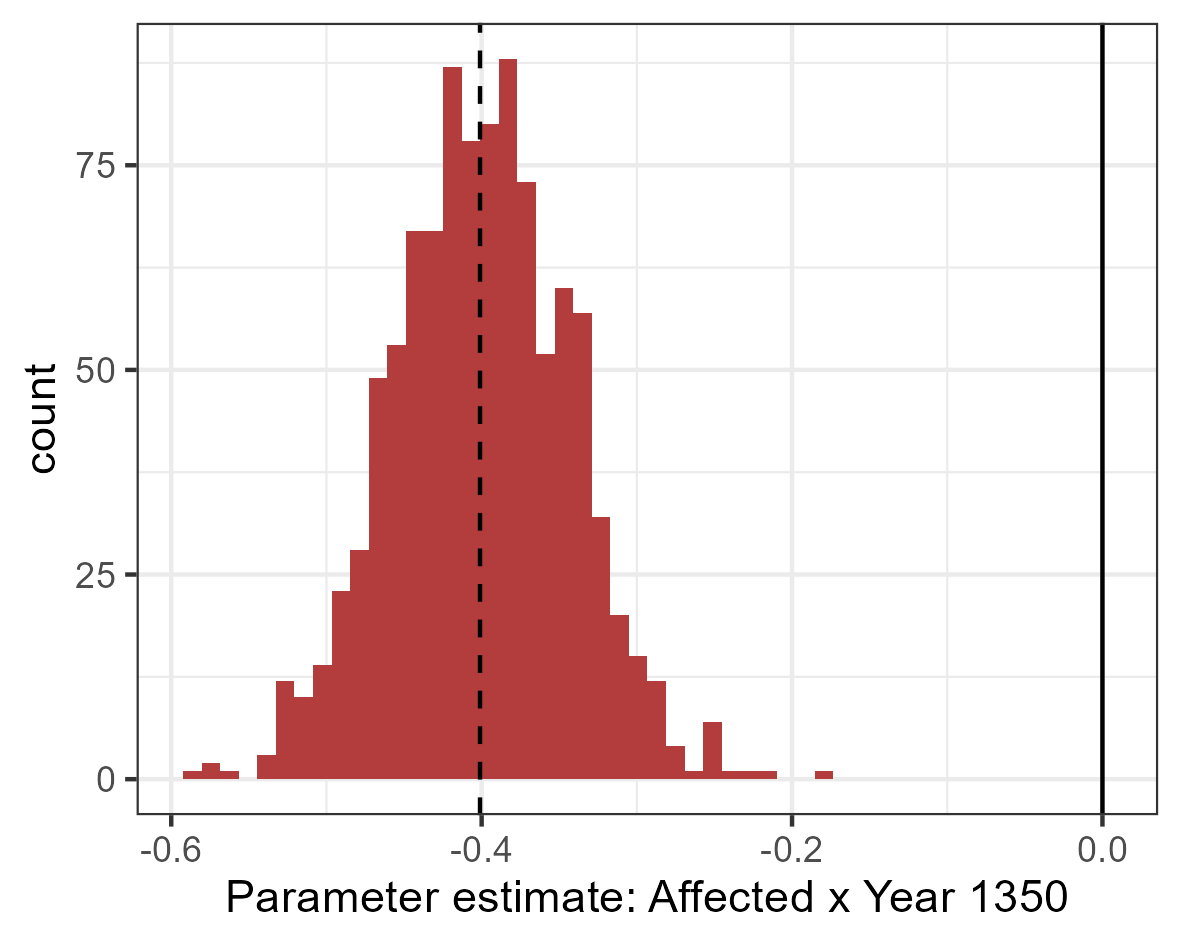}
    \end{subfigure}
    \hfill
    \begin{subfigure}[b]{0.45\textwidth}
        \centering
        \caption{Coins: Dummy approach} \label{fig:distri_b_norm}
        \includegraphics[width=\textwidth]{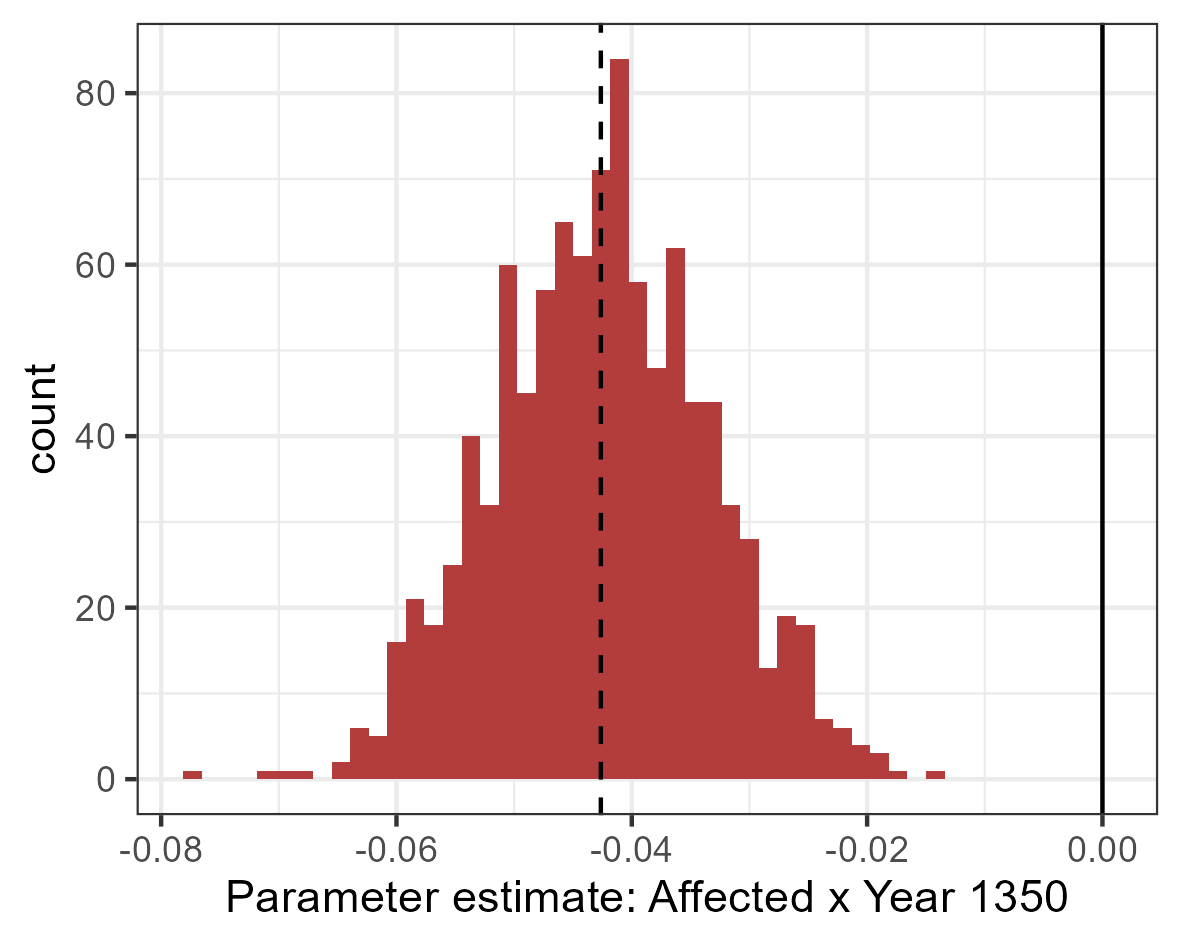}
    \end{subfigure}
    \vspace{0.45cm}
    \begin{subfigure}[b]{0.45\textwidth}
        \centering
        \caption{Buildings: Market access approach} \label{fig:distri_c_norm}
        \includegraphics[width=\textwidth]{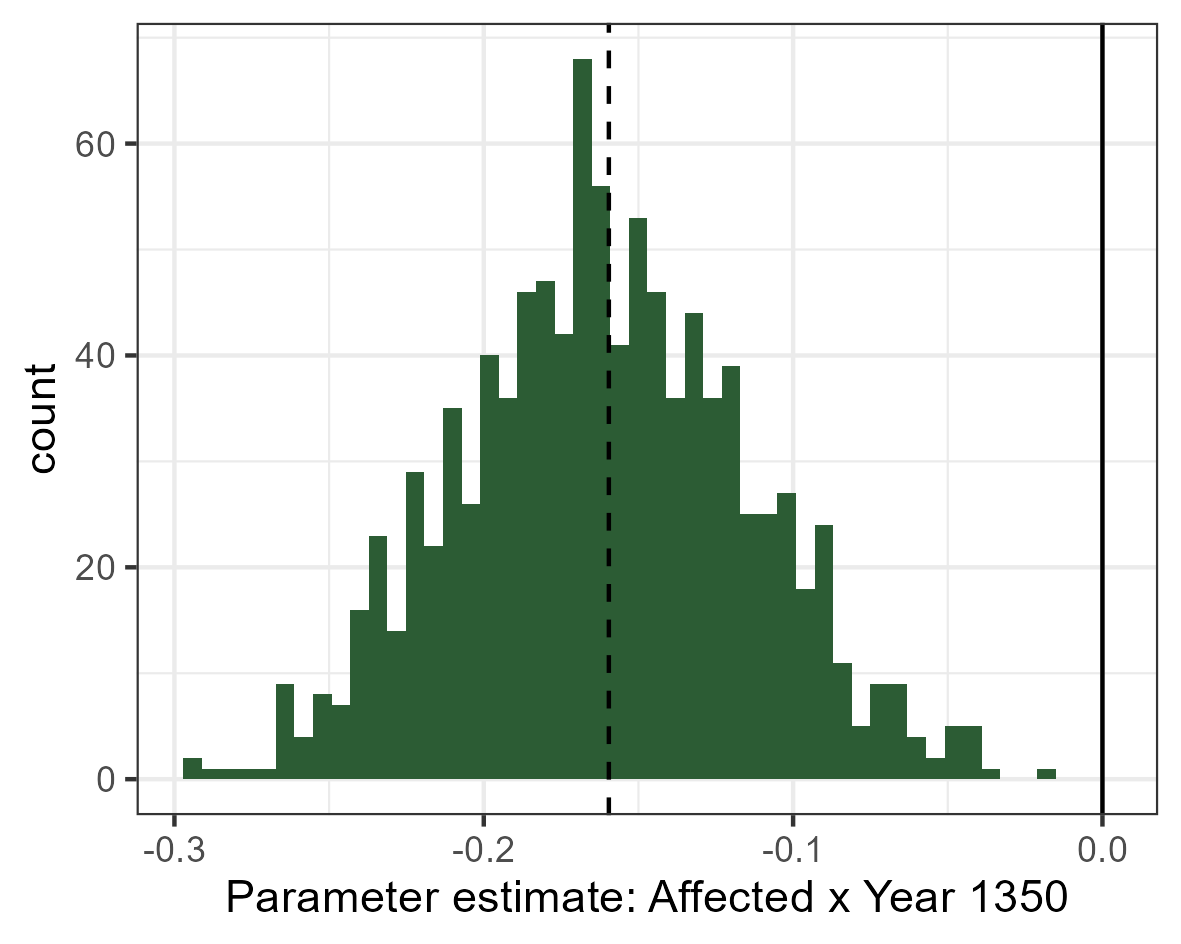}
    \end{subfigure}
    \hfill
    \begin{subfigure}[b]{0.45\textwidth}
        \centering
        \caption{Buildings: Dummy approach} \label{fig:distri_d_norm}
        \includegraphics[width=\textwidth]{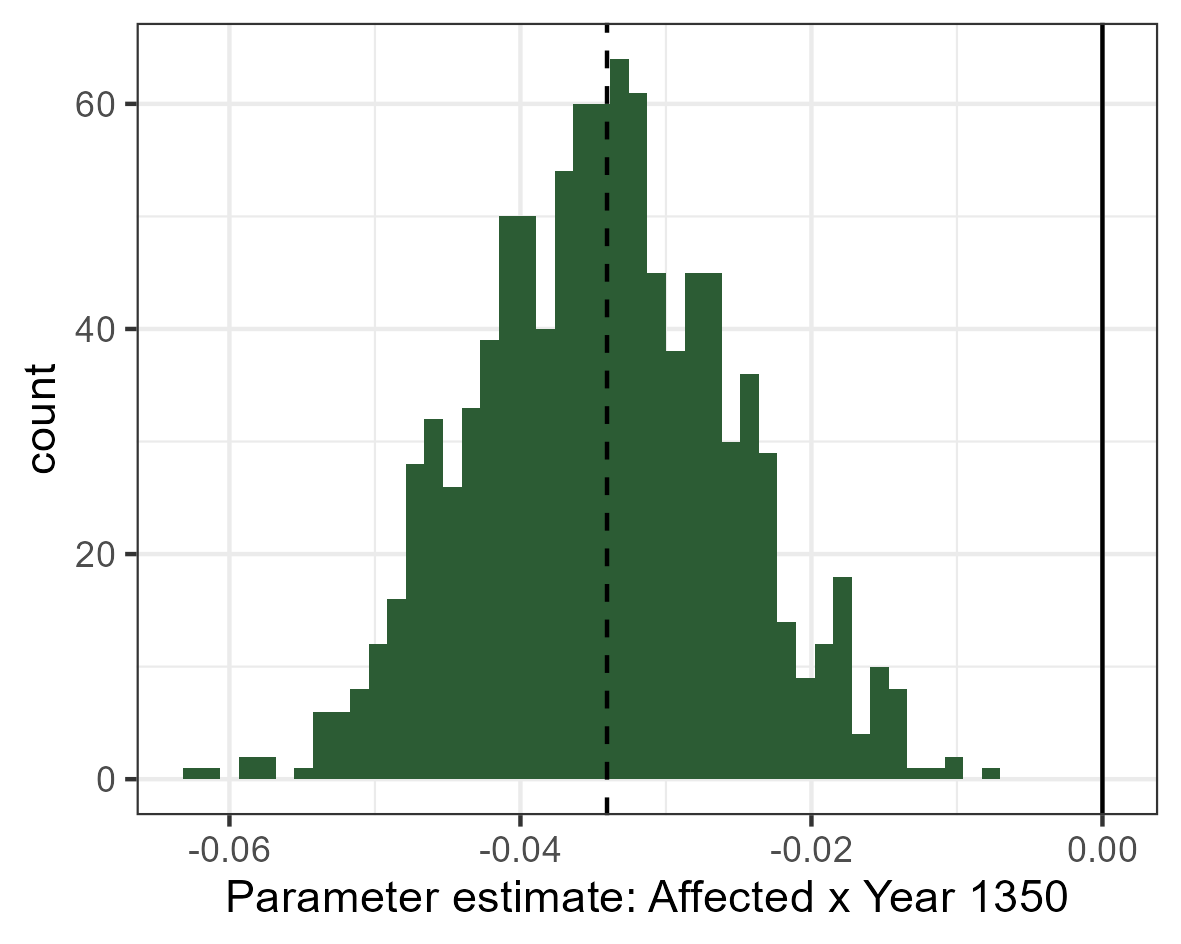}
    \end{subfigure}
    \label{fig:arch_reg_boot1}
\end{figure}

\begin{figure}
    \centering
    \caption{Archaelogical results (matched sample)}
    \begin{subfigure}[b]{0.45\textwidth}
        \centering
        \caption{Coins: Market access approach} \label{fig:arch1a_match_norm}
        \includegraphics[width=\textwidth]{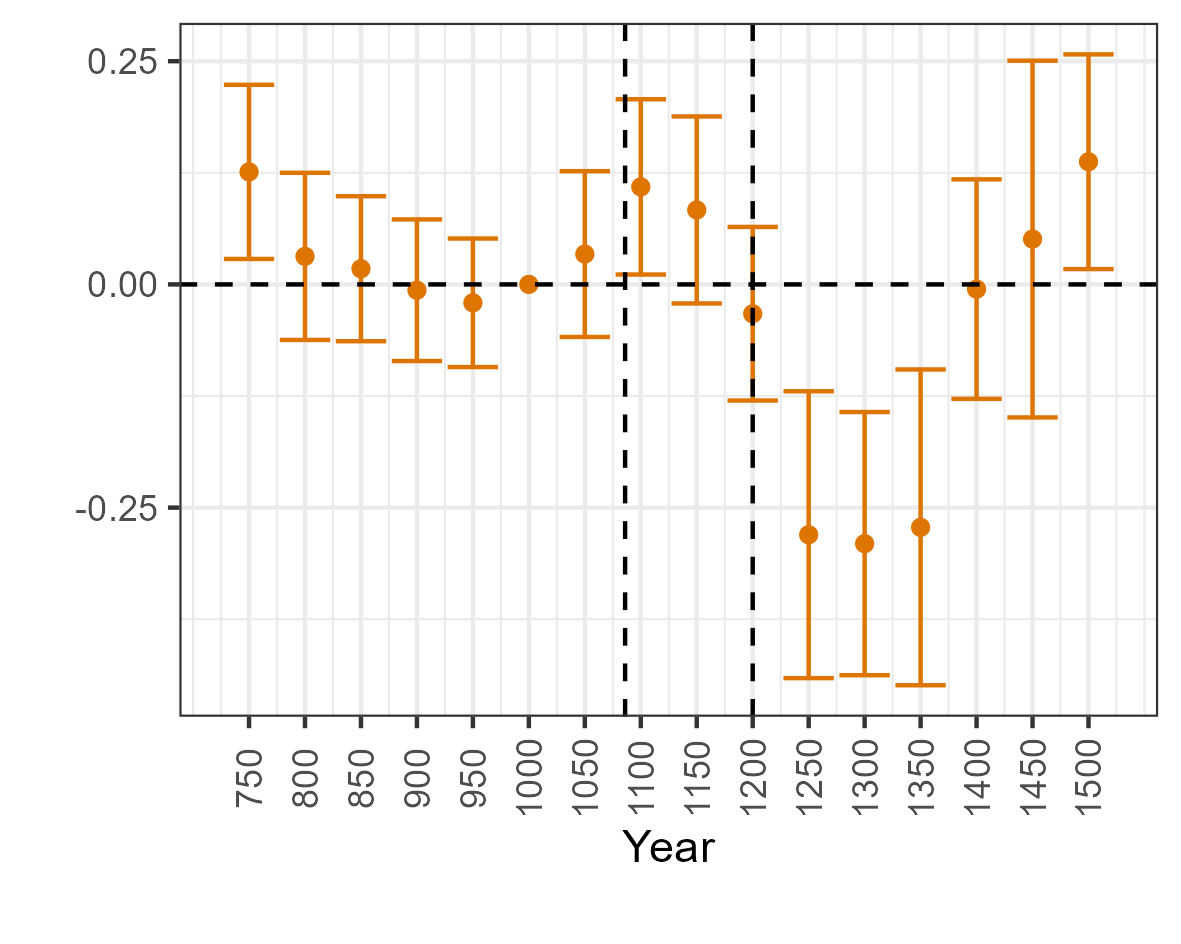}
    \end{subfigure}
    \hfill
    \begin{subfigure}[b]{0.45\textwidth}
        \centering
        \caption{Coins: Dummy approach} \label{fig:arch1b_match_norm}
        \includegraphics[width=\textwidth]{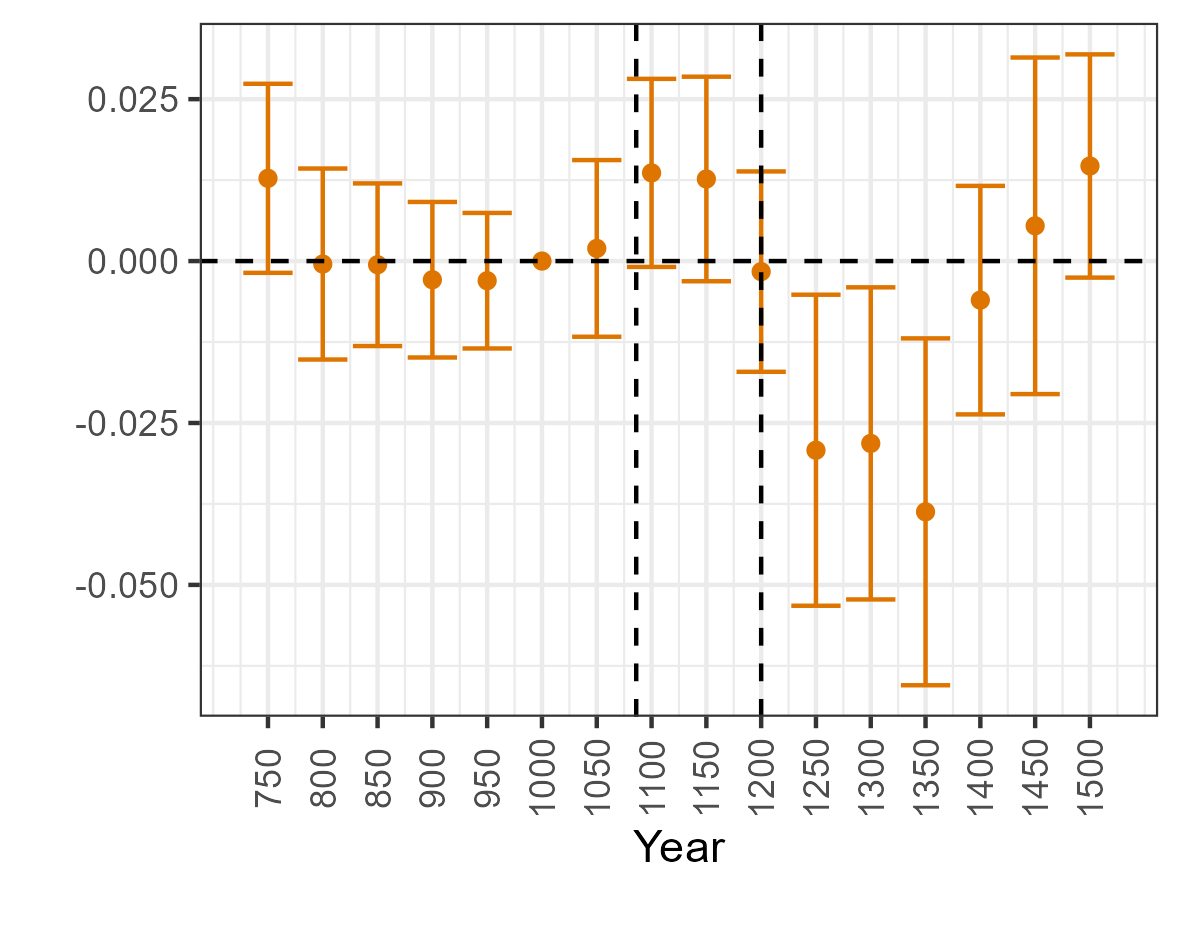}
    \end{subfigure}
    \vspace{0.45cm}
    \begin{subfigure}[b]{0.45\textwidth}
        \centering
        \caption{Buildings: Market access approach} \label{fig:arch1c_match_norm}
        \includegraphics[width=\textwidth]{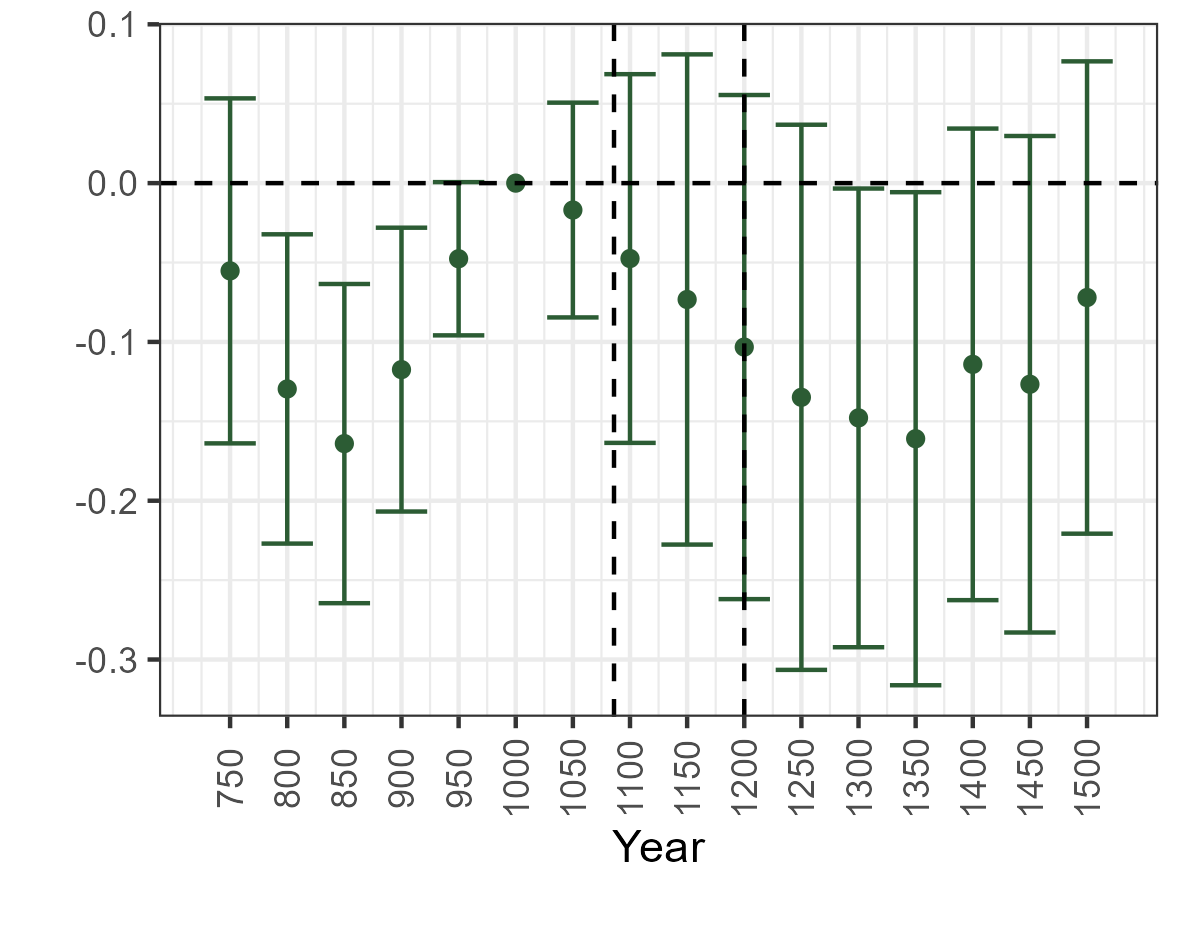}
    \end{subfigure}
    \hfill
    \begin{subfigure}[b]{0.45\textwidth}
        \centering
        \caption{Buildings: Dummy approach} \label{fig:arch1d_match_norm}
        \includegraphics[width=\textwidth]{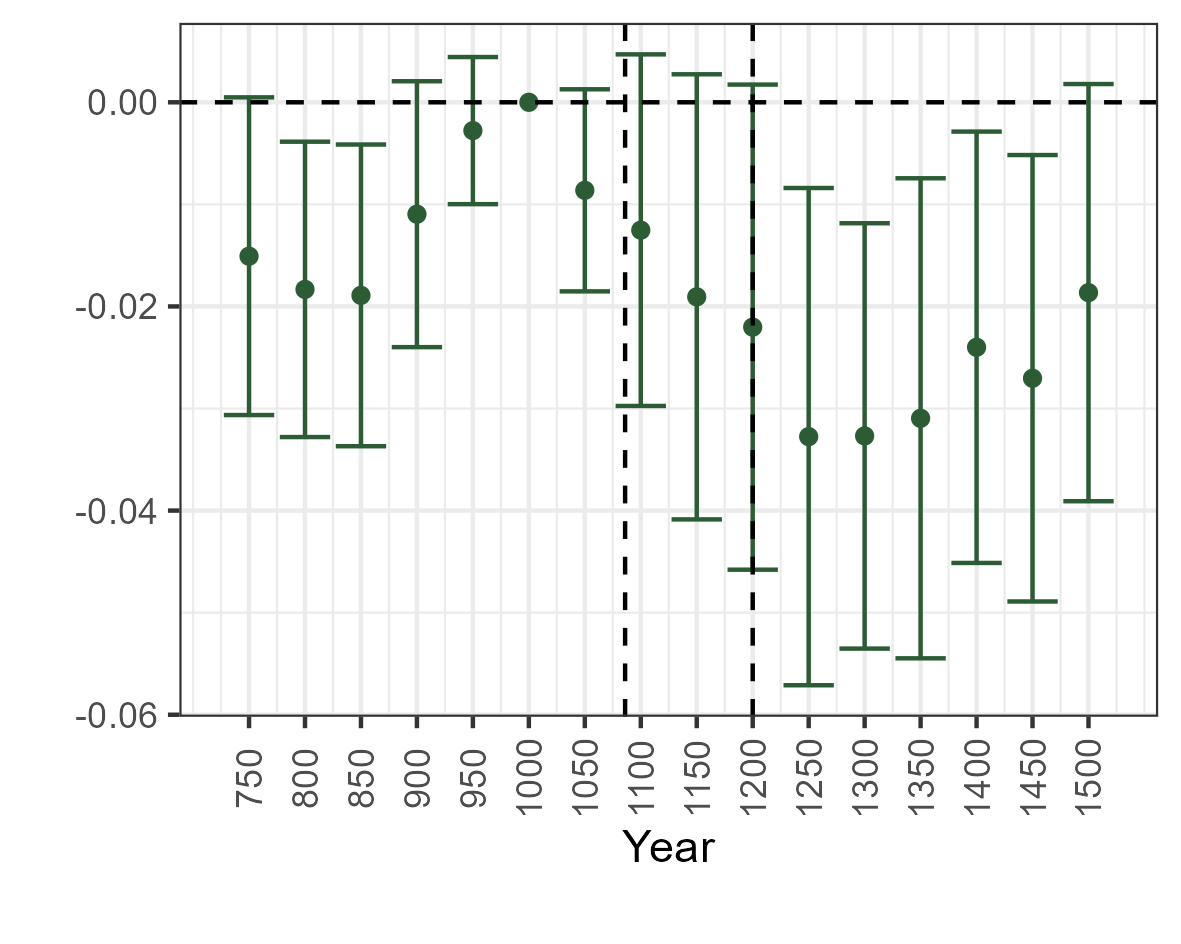}
    \end{subfigure}
    \label{fig:arch_reg2}
\end{figure}

\begin{figure}
    \centering
    \caption{Distribution of parameter estimates in 1350  (matched sample)}
    \begin{subfigure}[b]{0.45\textwidth}
        \centering
        \caption{Coins: Market access approach} \label{fig:distri_a_match_norm}
        \includegraphics[width=\textwidth]{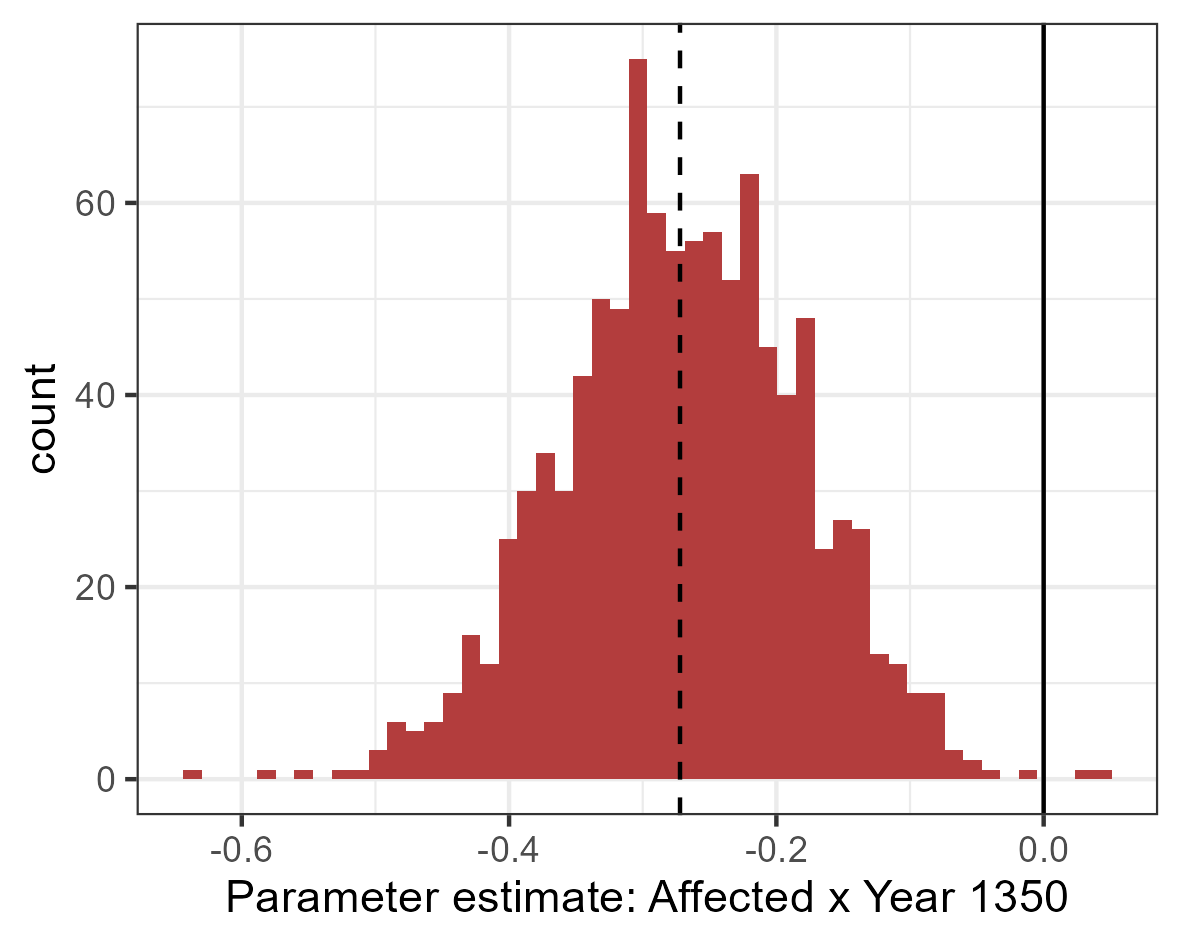}
    \end{subfigure}
    \hfill
    \begin{subfigure}[b]{0.45\textwidth}
        \centering
        \caption{Coins: Dummy approach} \label{fig:distri_b_match_norm}
        \includegraphics[width=\textwidth]{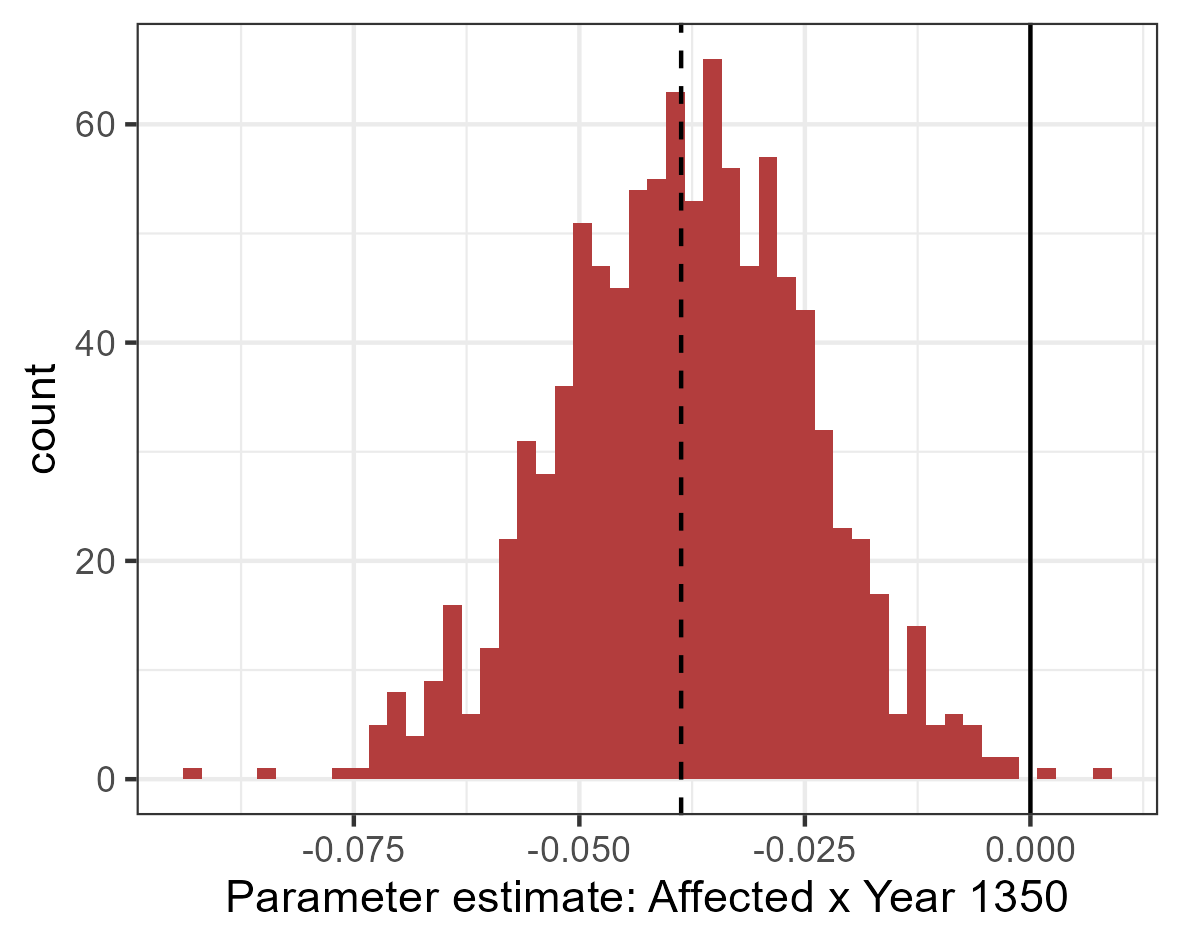}
    \end{subfigure}
    \vspace{0.45cm}
    \begin{subfigure}[b]{0.45\textwidth}
        \centering
        \caption{Buildings: Market access approach} \label{fig:distri_c_match_norm}
        \includegraphics[width=\textwidth]{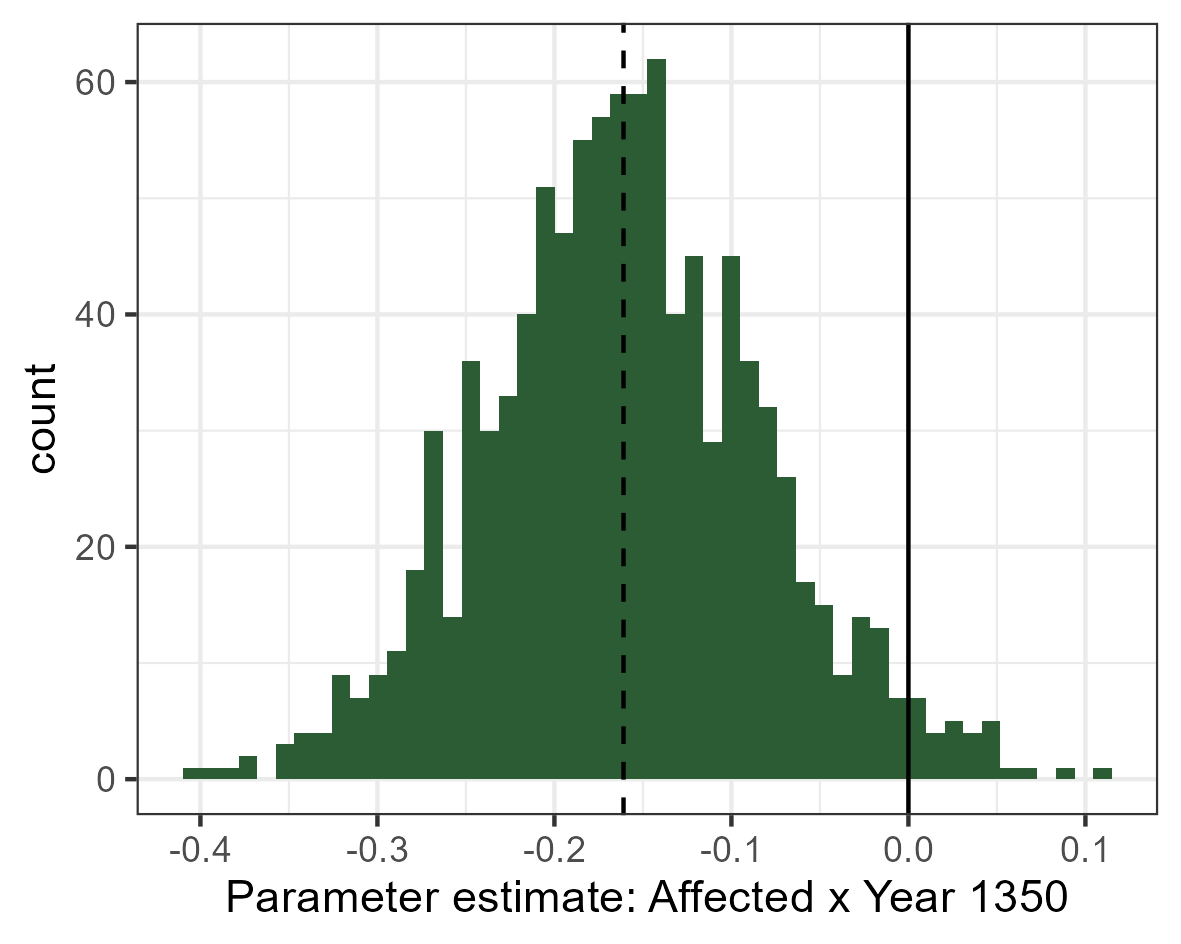}
    \end{subfigure}
    \hfill
    \begin{subfigure}[b]{0.45\textwidth}
        \centering
        \caption{Buildings: Dummy approach} \label{fig:distri_d_match_norm}
        \includegraphics[width=\textwidth]{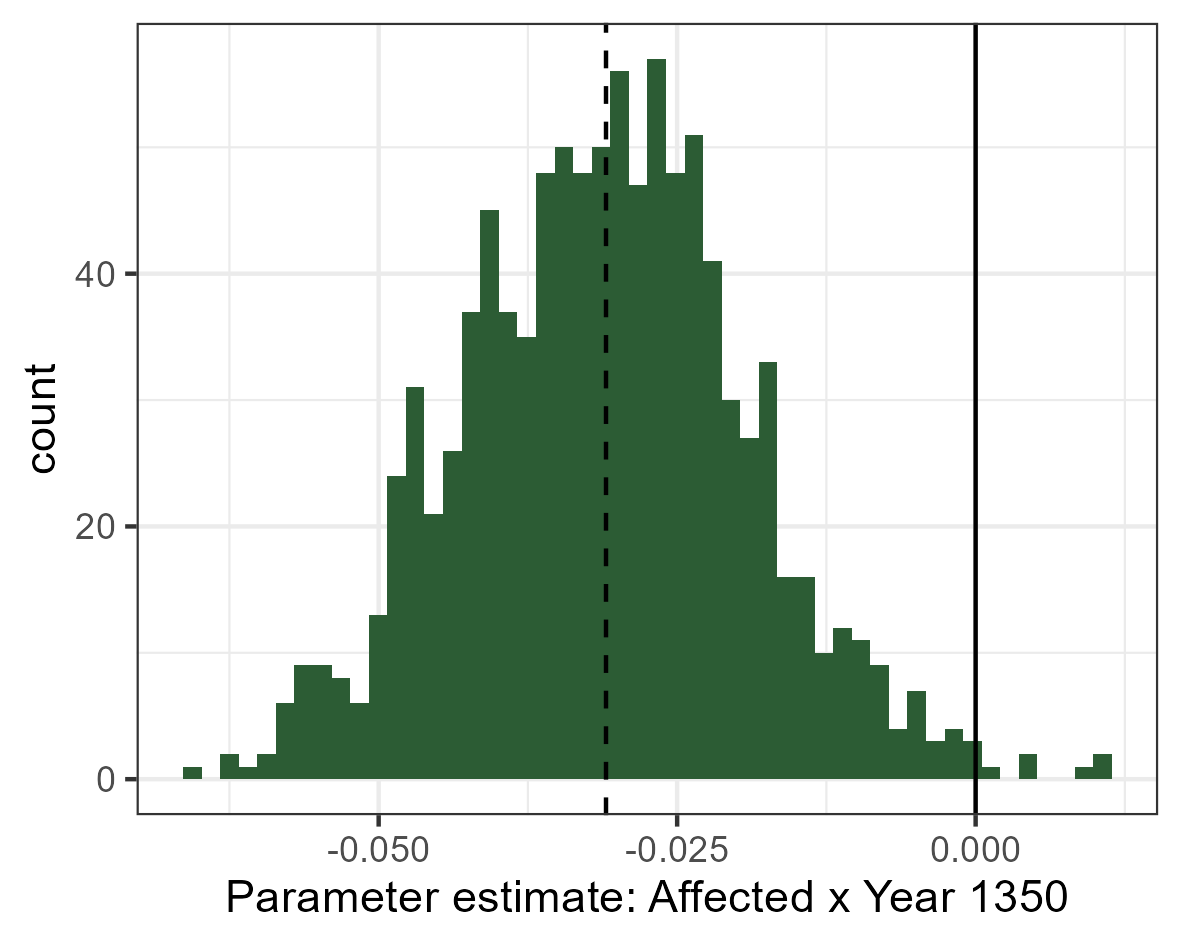}
    \end{subfigure}
    \label{fig:arch_reg_boot2}
\end{figure}

\end{document}